\documentclass[a4paper,fleqn,usenatbib]{mnras}

\usepackage[T1]{fontenc}
\usepackage{ae,aecompl}

\usepackage[pdftex]{graphicx}	
\usepackage{amsmath}	
\usepackage{rotating,txfonts,gensymb,graphicx,color,natbib,lscape,url,textcomp,longtable,pdflscape,times,mn2e-breakabs,multirow}

\newcommand{\ascasimple}{{\it ASCA}}
\newcommand{\swiftsimple}{{\it Swift}}
 
\newcommand{\ergs}{~erg s$^{-1}$}             

\newcommand{\etal}{et al.~}              

\newcommand{\kms}{~km s$^{-1}$}   

\newcommand{\msunsimple}{~M$_{\odot}$}
\newcommand{\msun}{~M$_{\odot}$\,}
\newcommand{\invmsun}{~$\rm{M_{\odot}}^{-1}$\,}

\newcommand{\zsun}{Z$_{\odot}$}
\newcommand{\msunpyr}{~M$_{\odot}~\rm{yr^{-1}}$}

\newcommand{\mdot}{${\rm M_{\odot}/yr}$}
\newcommand{\rosat}{{\it ROSAT }}
\newcommand{\rosatsimple}{{\it ROSAT}}
\newcommand{\chandra}{{\it Chandra~}}
\newcommand{\chandrasimple}{{\it Chandra}}
\newcommand{\xmm}{{\it XMM-Newton~}}
\newcommand{\xmmsimple}{{\it XMM-Newton}}

\newcommand{\Ha}{H$\alpha$~}

\newcommand{\meanAv}{$<A_{V}>$~}

\def\aaps{A\&AS}              
\def\aap{A\&A}                
\def\actaa{Acta Astron.}    
\def\aj{AJ}                   
\def\ao{Applied Optics}       
\def\apjl{ApJ}                
\def\apjs{ApJS}               
\def\apj{ApJ}                 
\def\araa{ARA\&A}             

\def\gca{Geochim.~Cosmochim.~Acta} 
\def\iaucirc{IAU Circ.}       

\def\mnras{MNRAS}             
\def\nar{New Astron.~Rev.}  
\def\nat{Nature}              
\def\pasp{PASP}               

\title[Star-formation history and XRB populations]{Star-formation history and X-ray binary populations: the case
of the Large Magellanic Cloud}

\author[V. Antoniou and A. Zezas]{
V. Antoniou$^{1}$\thanks{E-mail: vantoniou@cfa.harvard.edu} and A. Zezas$^{2,3,1}$
\\
$^{1}$Harvard-Smithsonian Center for Astrophysics, 60 Garden Street, Cambridge, MA 02138, USA\\
$^{2}$Physics Department \& Institute of Theoretical \& Computational Physics, University of Crete, 71003 Heraklion, Crete, Greece\\
$^{3}$Foundation for Research and Technology-Hellas, 71110 Heraklion, Crete, Greece
}

\date{Accepted 2016 January 17. Received 2015 December 14; in original form 2015 July 31}

\pubyear{2016}

\begin{document}
\label{firstpage}
\pagerange{\pageref{firstpage}--\pageref{lastpage}}
\maketitle

\begin{abstract}
In the present work we investigate the link between high-mass X-ray binaries (HMXBs) and star formation in the Large Magellanic Cloud (LMC), our nearest star-forming galaxy. Using optical photometric data, we identify the most likely counterpart of 44 X-ray sources. Among the 40 HMXBs classified in this work, we find 33 Be/X-ray binaries, and 4 supergiant XRBs. Using this census and the published spatially resolved star-formation history map of the LMC, we find that the HMXBs (and as expected the X-ray pulsars) are present in regions with star-formation bursts $\sim$6--25 Myr ago, in contrast to the Small Magellanic Cloud (SMC), for which this population peaks at later ages ($\sim$25--60 Myr ago). We also estimate the HMXB production rate to be equal to 1 system per $\sim43.5\times10^{-3}$\,\msunpyr\, or 1 system per $\sim$143\msun of stars formed during the associated star-formation episode. Therefore, the formation efficiency of HMXBs in the LMC is $\sim$17 times lower than that in the SMC. We attribute this difference primarily in the different ages and metallicity of the HMXB populations in the two galaxies. We also set limits on the kicks imparted on the neutron star during the supernova explosion. We find that the time elapsed since the supernova kick is $\sim$3 times shorter in the LMC than the SMC. This in combination with the average offsets of the HMXBs from their nearest star clusters results in $\sim$4 times faster transverse velocities for HMXBs in the LMC than in the SMC.
\end{abstract}




\section{Introduction}\label{Intro}

The production rate of X-ray binaries (XRBs) is a key parameter for understanding their formation and evolution. In the last few years, numerous theoretical and observational works (e.g., \citealt{Belczynski et al. 2008}, \citealt{2009ApJ...702.1387S}, \citealt{Linden et al. 2010}, \citealt{2013ApJ...764...41F}, \citealt{2013ApJ...774..136T}) have been performed, with the main focus in improving the realism in the population synthesis models as more observational constraints become available. The Magellanic Clouds are the only galactic environment so far that allow for a direct measurement of the production rate of HMXBs, systems consisting of a compact object (either a neutron star --NS-- or a black hole --BH-- and in few cases a white dwarf --WD) orbiting an early-type companion. The most numerous subclass of this type of objects up to now detected\footnote{There is an increasing number of Supergiant Fast X-ray Transients (SFXTs; \citealt{2006ESASP.604..165N}) being discovered in the Milky Way mainly with {\it INTEGRAL}. SFXTs have incredibly short outburst times (rising in tens of minutes and lasting few hours), and quiescent luminosities of $\sim10^{32}-10^{34}$\ergs\, \citep{2015arXiv151007681C}, making them even harder to detect than the transient Be-XRBs. There may be a significant population of these sources in the Magellanic Clouds that have yet to be discovered. For example, \citet{2014A&A...568A..76D} estimate the number of Galactic SFXTs at $37^{+53}_{-22}$, suggesting that the SFXTs constitute a large subclass of XRBs with supergiant companions.} is the so called Be/X-ray binaries (Be-XRBs) with O- or B-type donor stars that exhibit line emission over the photospheric spectrum (for a review see \citealt{2003PASP..115.1153P}).

In the Small Magellanic Cloud (SMC) several studies revealed a large number of HMXBs in low fluxes (e.g., \citealt{Antoniou et al. 2009b}, \citealt{Antoniou et al. 2010}, \citealt{2012A&A...545A.128H}, \citealt{2013A&A...558A...3S}, \citealt{2015arXiv151100445H}), which allowed us to investigate the link between HMXBs and star formation (e.g., \citealt{2005MNRAS.362..879S}, \citealt{Antoniou et al. 2010}). We found that the Be-XRBs and the X-ray pulsars (all but one of the $\sim70$ known X-ray pulsars have Oe- or Be-type companions, with the other system being a supergiant; \citealt{2015MNRAS.452..969C}) are observed in regions with star-formation bursts $\sim25-60$ Myr ago (strongly peaked at $\sim42$ Myr). On the other hand, we found that regions with strong but more recent star formation (e.g., the Wing) are deficient in Be-XRBs, in agreement with the \chandra survey of the SMC Wing (PI. M. Coe), which detected only 4 HMXBs in 20 observed fields\footnote{Though, as noted in this work, it is possible that a larger number of HMXBs exist in the SMC Wing, which nevertheless remain undetected due to the transient nature of the Be phase and the low probability of detecting them in outburst (for ${\rm L_{X}\sim10^{38}}$\ergs\, this probability is only $\sim$10\%; Fig. 4.62 in \citealt{GalachePhD}). In any case, based on the lower star-formation rate (SFR) of the SMC Wing when compared to that along the SMC Bar, we expect a smaller number of Be-XRBs than those identified in the SMC Bar  (\citealt{Antoniou et al. 2010}).} ~(\citealt{2008MNRAS.383..330M}) and the large \xmm survey of the SMC \citep[P.I. F. Haberl; ][]{2012A&A...545A.128H}, where only 6 HMXBs (3 confirmed and 3 candidate) have been identified in 10 fields in the South-East part of the SMC \citep{2013A&A...558A...3S}. By correlating the number of observed Be-XRBs (HMXBs) with the formation rate of their parent stellar populations (i.e., with the SFR of the regions that host these systems), \citet{Antoniou et al. 2010} derived a Be-XRB (HMXB) production rate of $\sim$1 system per 3 (2.5)$\times10^{-3}$\msunpyr. That was the {\it first} direct calibration of the XRB formation rate at $\sim40$ Myr by measuring the formation rate of the XRBs per unit SFR of their parent populations. 

Finally, the strong spatial correlation between the Be-XRBs in the SMC and their parent stellar populations provided strong evidence for relatively small supernova (SN) kicks during the formation of the compact object. We estimated an upper limit of $\sim$15-20 km s$^{-1}$, in agreement with measured velocities of Be-XRBs in the Galaxy ($15 \pm 6$\kms; \citealt{2000A&A...364..563V}) and estimated values in the SMC (\citealt{2005MNRAS.358.1379C} and \citealt{Antoniou et al. 2009b}). 

Motivated by the above work, we investigate here the link between HMXBs and star formation in the  Large Magellanic Cloud (LMC), our nearest star-forming galaxy (only at $\sim$50 kpc; \citealt{2006ApJ...652.1133M}). The LMC is a dwarf irregular galaxy in the Local Group. It has an extended circular shape with a prominent off-center bar, a nucleus, and irregular spiral arms (based on results from the DENIS survey; \citealt{2000A&A...358L...9C}). It experiences intense star-formation activity with a higher star formation rate than the SMC \citep{2009A&A...506.1137C}. Despite its moderate Galactic foreground absorption ($\rm{N_{H}}\simeq6.4\times10^{20} cm^{-2}$; \citealt{1990ARA&A..28..215D}) and closer distance, the LMC has received much less attention in the X-rays than the SMC, with most observations focusing so far on individual X-ray sources (such as LMC X-1, X-2, X-3, etc). 

Another reason for selecting the LMC as a target for this study is its sub-solar metallicity ($\sim$1/2.5~\zsun; \citealt{2005AJ....129.1465C}). Nowadays it is widely accepted that metallicity is one of the three main factors that determines the formation rate of young XRBs (the other two being the age of the parent stellar populations and the SFR). Even though the details of how  metallicity affects the HMXB populations are not well understood, there is a growing body of theoretical work indicating that low metallicity is associated with higher formation efficiency of HMXBs \citep{Dray 2006} and higher luminosity \citep{2013ApJ...764...41F}. For example, \cite{2015A&A...579A..44D} found that galaxies with metallicities below $\sim$1/5~\zsun, which is the typical value for the SMC \citep{1998AJ....115..605L}, have on average 10 times more HMXBs per unit SFR  than solar metallicity galaxies. On the other hand, \citet{Linden et al. 2010} find that for significantly sub-solar metallicity there is no dependence of the HMXB populations on the metallicity.

For such comparisons of course one needs an up-to-date tally of the number of HMXBs in this galaxy. Since \citet{2005A&A...442.1135L} published the list of known HMXBs in both Magellanic Clouds, several new such systems have been discovered in the LMC (mainly by the {\it Swift} and {\it INTEGRAL} satellites). As a by-product of this work, we compiled a list of all known HMXBs in the LMC, for which we also identified their optical counterparts. Until now, only 28 of such systems had a known counterpart in the literature. Our work provides the best candidates for spectroscopic follow-up programs that will identify the nature of the HMXBs in the LMC unambiguously. 

Based on this material and the spatially-resolved star-formation history map (\citealt{2009AJ....138.1243H}; hereafter [HZ09]), we are in the position to investigate the link between HMXBs and star formation in the LMC. In Section \ref{LMCHMXBs} we present an up-to-date tally of the young XRB population of the LMC and in Section \ref{SFH}  their optical properties. In particular, in Section \ref{OptLIKELY} we present their most likely optical counterparts along with estimations of the chance coincidence probability. The final classification of the XRBs is given in Section \ref{theCLASS}, while the metallicity of the young stellar populations in both Magellanic Clouds is discussed in Section \ref{MCmetal}. In Section \ref{globalSFH} we investigate the global star-formation history of this galaxy, and the link between stellar and XRB populations. In Section \ref{discussion} we study the properties of the HMXB population in the LMC. In particular, in Section \ref{SFH-youngXRBs} we discuss the star-formation history of young XRBs in the Magellanic Clouds, and in Section \ref{sectionFRs} the formation efficiency for the various XRB types. In Section \ref{Comparison} we present previous studies on this subject, and in Section \ref{SPTypedistribution} the spectral-type distribution of the LMC HMXBs. A discussion about the young parent stellar populations of the LMC X-ray pulsars follows in Section \ref{yngPULSARS}. The last part of this work discusses the supernova kick velocities of the LMC HMXBs (Section \ref{SNkicks}) and presents the "Corbet diagram" (\citealt{1984A&A...141...91C}) of HMXBs in the Magellanic Clouds and the Milky Way (Section \ref{Corbet}). Finally, in Section \ref{Summary} we summarize the main findings of this work.

Throughout this work, we adopt a distance modulus of $(m-M)_{V}=18.50 \pm 0.02$ mag (\citealt{2004NewAR..48..659A}). We also use $R_{V} = 3.41$ (\citealt{2003ApJ...594..279G}) in order to estimate the $E(B-V)$ ($=A_{V}/R_{V}$) reddening. A mean extinction value \meanAv for an  area with a radius of 12\arcmin\, around each HMXB is derived by using the ``LMC Extinction Retrieval Service" tool\footnote{Available from http://djuma.as.arizona.edu/$\sim$dennis/lmcext.html} \citep{2004AJ....128.1606Z}. We estimate an average value of  \meanAv $\sim$ 0.50 mag for early-type stars, thus resulting in $E(B-V)\sim0.15$ mag. Regarding the metallicity, we use the latest value of \zsun$=0.0134$ from the review of \citet{Asplund et al. 2009}, instead of the more widely-known canonical value of \zsun$=0.02$ (\citealt{1989GeCoA..53..197A}).

Since the completion of this work, 2 additional sources have been reported in the literature as HMXBs, and the spin period of a known source was derived. For completeness we list them here, and in Table \typeout{\ref{tableCensus}}1, but we do not include them in our analysis. \citet{2015A&A...579A.131C} identified an X-ray bright emission-line star as part of the VLT-FLAMES Tarantula Survey that could be the first HMXB identified within 30 Dor, if indeed XMMU J053833.9-691157 and CXOU J053833.4-691158 are associated with VFTS399. Swift J0549.7-6812  (reported in ATel \#5286; \citealt{2013ATel.5286....1K}) is located at (RA, Dec.)=(05:50:06.47, -68:14:55.7) with a 90\% confidence astrometric error ($r_{90}$) equal to 1.4\arcsec, and a 6.2-second pulse period (ATel \#5309; \citealt{2013ATel.5309....1K}). Although this source, also reported as LXP6.2, has a blue counterpart suggesting a neutron-star HMXB system in the hard state (ATel \#5293; \citealt{2013ATel.5293....1K}), it is not included in our analysis, because these results not been published in a refereed paper since they were reported in ATels 3 years ago, thus we are uncertain about its nature. A new X-ray pulsar that was recently detected in the LMC \xmm survey is reported by Vasilopoulos \etal 2016 (to be subm.; private communication) to have a pulse period of 38.55(1) s. This source (LXP38.55) has been previously detected and it is associated with sources IGR J05007-7047 and CXOU J050046.0-704436. With the addition of the latter 2 new pulsars, the total number of known systems in the LMC is now sixteen.

\section{The HMXB population of the LMC}\label{LMCHMXBs}

So far there has not been any complete X-ray survey of the LMC in the soft X-rays ($\sim$0.5-10 keV).  Shtykovskiy \& Gilfanov (2005) analyzed 23 \xmm archival observations (covering $\sim$3.8 sq. degrees down to $3\times10^{33}$~\ergs\, at 2$-$8 keV) and found $\sim$460 point sources. However, more than 94\% of those objects have been identified as background X-ray sources observed through the LMC, 
and only 9 sources as candidate HMXBs (with 19 additional objects of uncertain nature). The {\it INTEGRAL} observatory has also surveyed the LMC, thus allowing us to study for the first time the hard (15 keV$-$10 MeV) X-ray emission of a handful of sources in this galaxy (\citealt{2006A&A...448..873G}). In particular, the {\it INTEGRAL} survey of the LMC discovered 5 new faint high-energy sources, although only 2 of them are potentially located in this galaxy. More recently, \citet{2013MNRAS.428...50G} presented the results from the ultra deep ($\sim$7 Ms) {\it INTEGRAL} survey of the LMC. They detected 7 known HMXBs, and 2 new hard X-ray sources; the nature of the latter is unknown. 

Forty five LMC HMXBs are known to date (confirmed and candidate systems, however in a few cases discussed below, we question this nature based on more recent photometric surveys). Only 16 systems have been confirmed as Be-XRBs (one of these sources is most likely a WD/Be-XRB; \citealt{2006A&A...458..285K}), while 5 systems have been classified as candidate Be-XRBs.  In the compilation of \citet{2005A&A...442.1135L}, one finds 2 LMC sources listed as supergiant XRBs (SG-XRBs): RX J0532.5-6551 (\citealt{1995A&A...303L..49H}) and RX J0541.4-6936 (\citealt{2000A&AS..143..391S}), though in the original publications these systems are only identified as candidate SG-XRBs. Moreover, there are 2 black-hole HMXBs (LMC X-1: \citealt{1983ApJ...275L..43H}, \citealt{1984ApJ...281..354W}; LMC X-3: \citealt{Cowley et al. 1983}). The remaining 20 sources are classified as NS/HMXBs based on their hard X-ray spectra, and in some cases, their association with a massive star showing \Ha emission. These systems have been classified  in this work as either candidate Be-XRB or SG-XRBs, except for 4 cases where the classification was revised to non-HMXB systems (see below Section \ref{theCLASS} and Table \typeout{\ref{finalclass}}5). 

In Table \typeout{\ref{tableCensus}}1 we present a list of all known HMXBs in the LMC compiled from the literature (as of Dec. 2014\footnote{Two additional sources published in the meantime are reported at the end of the previous section. With the exception of source IGR J05007-7047, which is already included in our HMXBs list, we do list these 2 sources in Table \typeout{\ref{tableCensus}}1 for completeness, but we do not include them in our present work. Similarly, we added in this table the pulse period of  source IGR J05007-7047, but we also do not include this source among the known LMC X-ray pulsars in our analysis ( i.e., we only include it in the list of  known HMXBs).}). In Columns 1 and 2, we give the X-ray source ID and source name, respectively. In Columns 3 and 4, we list the Right Ascension and Declination (J2000.0), and in Column 5 the positional uncertainty of the X-ray source (followed by the references in parenthesis). In cases of several detections reported in the literature, the coordinates with the smallest positional uncertainty are presented. Column 6 gives the source ID from the catalog of \citet{2005A&A...442.1135L}, which is the most recent compilation of HMXBs in the LMC prior to this work. Column 7 gives the pulse period in seconds followed by the orbital period in days (with corresponding references given in parenthesis). Finally, the spectral type of the optical counterpart and the XRB type, as presented in the literature, are given in Columns 8 and 9, respectively, with the reference list as updated as possible.

\section{The optical properties of the young X-ray source population of the LMC}\label{SFH}

There are two major photometric surveys of the LMC in the optical band: the Magellanic Clouds Photometric Survey (MCPS; \citealt{2004AJ....128.1606Z}) and the Optical Gravitational Lensing Experiment (OGLE-III; \citealt{2008AcA....58...89U}). The MCPS is the largest area survey of the LMC (with a pixel scale of $\sim$0.7\arcsec/pixel) covering $\sim$8.5\degr$\times$7.5\degr\, with observations in the $U$, $B$, $V$, and $I$\, filters reaching a limiting $V$-band magnitude of 20--21 (depending on the local crowding). The final LMC photometric catalog contains over 24 million objects, the vast majority of which are stars in the LMC (\citealt{2004AJ....128.1606Z}). All but one (LMC X-3) of the LMC HMXBs presented in Table \typeout{\ref{tableCensus}}1 are covered by the MCPS survey. On the other hand, the OGLE-III survey is a significant extension of the OGLE-II maps \citep{2000AcA....50..307U} that covered only the central regions of the LMC. This survey covers $\sim$40 sq. degrees of the LMC and contains $V$ and $I$ photometry and astrometry for about 35 million stars observed during 7 observing seasons. However, since only 27 out of the 45 X-ray sources studied here are covered even by the extended OGLE-III survey, in the present work we chose to use exclusively the MCPS catalog.

\subsection{Optical counterparts of HMXBs in the LMC}\label{OptLIKELY}

As discussed in Section \ref{Intro}, only 28 systems had a known counterpart in the literature prior to this work (and these are clearly marked in Tables  \typeout{\ref{tableOpt}}2 -- \typeout{\ref{tableMatchesOut}}4). In this work we perform a systematic cross correlation of the known X-ray sources in the LMC with the MCPS catalog. For consistency we extend this search to the 28 systems with known counterpart. The vast majority of the sources reported in Table \typeout{\ref{tableCensus}}1 have been identified with X-ray telescopes with typical astrometric errors of $\sim$5\arcsec\, (e.g., \xmmsimple, \swiftsimple, \rosatsimple) or worse (e.g., \ascasimple). A few (5) sources have sub-arcsecond positions. Therefore, we adopt a search radius of 5\arcsec, which in the case of \xmm and  \rosat also includes the boresight error (e.g., \citealt{2007ApJS..172..353B}, \citealt{ROSATboresight}, respectively).

 The resulting matches are presented in Table \typeout{\ref{tableOpt}}2. In Column 1, we give the X-ray source ID (same as in Column 1 of Table \typeout{\ref{tableCensus}}1). In Columns 2 and 3, we give the Right Ascension and Declination (J2000.0) of the MCPS counterparts, while their photometric data are listed in Columns 4--13 (these data are taken directly from the original catalogs without applying any reddening correction): apparent magnitudes in the $U$, $B$, $V$, and $I$ bands (Columns 4, 6, 8, and 10), the $B-V$ color (Column 12), and their errors (Columns 5, 7, 9, 11 and 13, respectively). The distance (in arcseconds) of the counterpart to the X-ray source is given in Column 14, while in Column 15 "n" indicates a newly identified counterpart from this work, "k" a known counterpart in the literature, and "A" those counterparts that are discussed in the Appendix \ref{appendixA}. There are only 10 X-ray sources with unique optical matches within 5\arcsec, while there is one more source without any match at this radius. For completeness, we also repeated this exercise for larger radii. This is particularly important for sources detected with \rosat PSPC or \ascasimple, which have error circles larger than 5\arcsec\, (see Table \typeout{\ref{tableCensus}}1). The resulting matches are presented in Tables \typeout{\ref{tableBright}}3 and \typeout{\ref{tableMatchesOut}}4, which follow the structure of Table \typeout{\ref{tableOpt}}2. However, we emphasize that in the case of sources with sub-arcsecond positional uncertainties, we do not identify any counterpart at distances larger than $\sim$1.7\arcsec\, (consistent with \chandrasimple's off-axis averaged astrometric error following the prescription of \citealt{2005ApJ...635..907H}).

Based on thorough Monte Carlo simulations presented in Section \ref{cc}, we propose here as the most likely counterpart of sources with more than one optical matches within the search radius, the brightest one (and among sources of similar brightness the bluest one), which is indicated in bold in Table \typeout{\ref{tableOpt}}2. Following Section \ref{cc} we give preference to sources with $M_{V_{o}}<-2$ and $(B-V)_{o}\leq 0.25$ or  $M_{V_{o}}<-3$ and $(B-V)_{o}> 0.25$; these are referred to as "unambiguous" matches because of their very low chance coincidence. For only 26 out of the 44 X-ray sources with MCPS coverage there is an unambiguous counterpart within 5\arcsec, i.e. within the 1$\sigma$ search radius. For other 6 X-ray sources, this unambiguous match is a bright source ($V \lesssim 20$ mag and $B-V \lesssim 0.2$ mag) that lies within the 5\arcsec\, to 10\arcsec\, annulus (presented at Table \typeout{\ref{tableBright}}3), while 2 additional X-ray sources have an unambiguous match at a distance larger than 10\arcsec\, (these sources have large X-ray positional uncertainties; Table \typeout{\ref{tableMatchesOut}}4). 

\begin{figure*}
\includegraphics[width=1.0\textwidth]{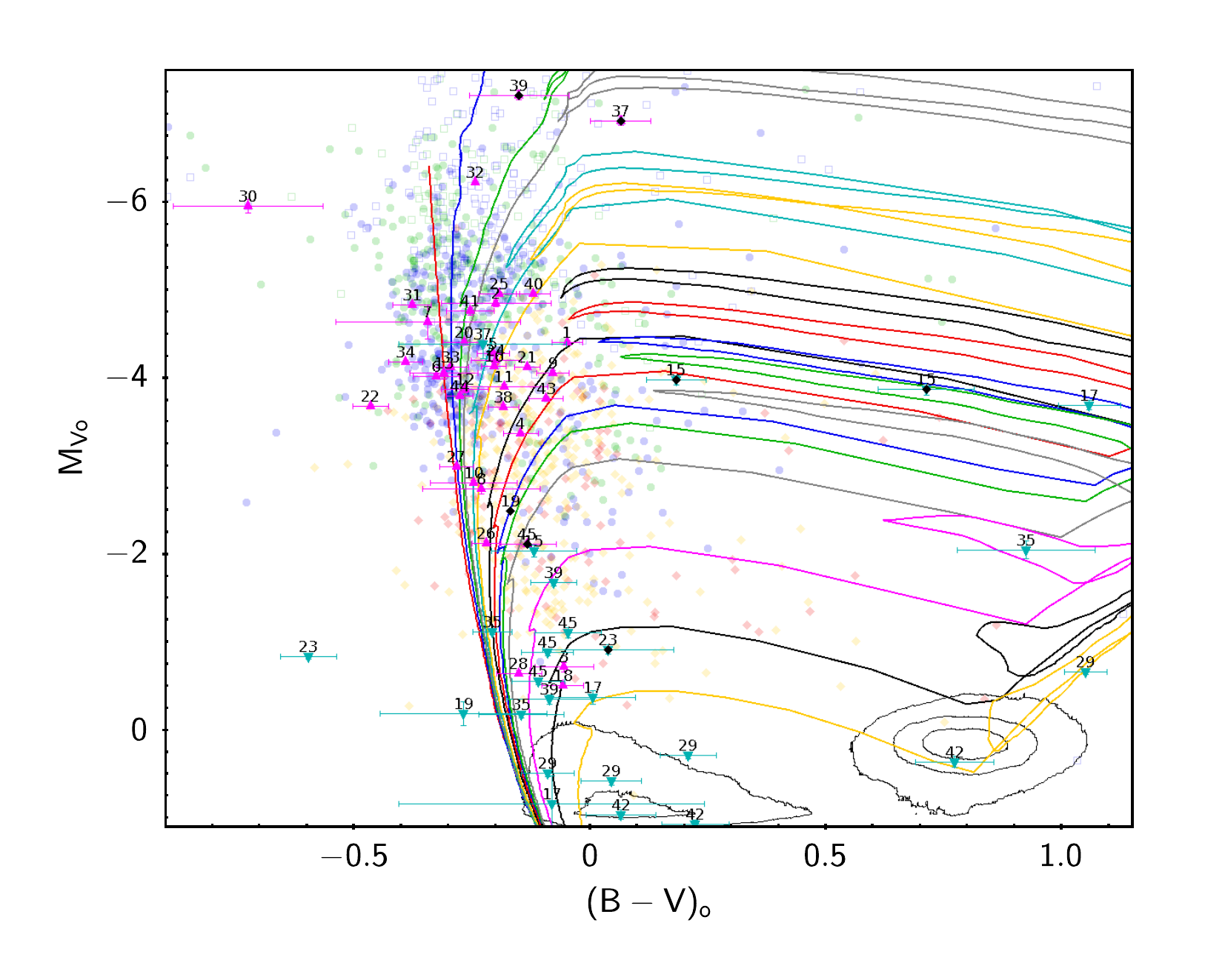}
\begin{minipage}{170mm}
\caption{A reddening-corrected $V,B-V$ color-magnitude diagram (presented as the absolute $V$-magnitude $M_{V_{o}}$ versus the $(B-V)_{o}$ color) of the optical counterparts of the LMC HMXBs (marked with their IDs; Column 1 of Table \typeout{\ref{tableCensus}}1).  Only the matches shown in bold in Tables \typeout{\ref{tableOpt}}2 -- \typeout{\ref{tableMatchesOut}}4 are presented. The counterpart of LXP 8.04 (source ID \# 14) is not shown in this figure due its very blue $B-V$ color (see discussion in Appendix \ref{appendixTOOblue}). For sources with more than one optical match, we select as the most likely counterpart the one that is located in the region of least chance coincidence probability (c.f. Section \ref{cc}). Magenta symbols indicate unambiguous matches and cyan symbols show the remaining sources for which such a choice is not feasible (in few cases, and based on arguments discussed in Appendix \ref{appendixA}, we mark with black diamonds the most likely matches among similarly likely associations). We also show {\it (i)} 500,000 randomly selected MCPS stars (listed in the catalog of \citet{2004AJ....128.1606Z}; shown with black contours); {\it (ii)} the Be and B[e] III-V stars from the spectroscopic study of \citet{2012MNRAS.425..355R} (shown in dark yellow and red filled circles, respectively); {\it (iii)} the O and B III-V stars (shown in green and blue filled circles, respectively) and O and B I-II stars (shown in green and blue open circles, respectively) from the census of \citet{2009AJ....138.1003B}; {\it (iv)} the isochrones from the Geneva database \citep{Lejeune et al. 2001} for Z$=$0.008 and ages ranging from 1 Myr to $\sim$490 Myr  -- from top to bottom: 1 Myr (red), 5 Myr (blue), 10 Myr (green), 12.3 Myr (gray), 20 Myr (cyan), 25 Myr (yellow), 44.7 Myr (black), 56.2 Myr (red), 69.2 Myr (blue), 79.4 Myr (green), 100 Myr (gray), 177.8 Myr (magenta), 316.2 Myr (black), 489.8 Myr (yellow).}\label{VBVall} 
\end{minipage}
\end{figure*}

In Fig. \ref{VBVall} we show a $V, B-V$ color-magnitude diagram constructed from 500,000 randomly selected MCPS stars (black contours) listed in the catalog of \citet{2004AJ....128.1606Z}. Shown with dark yellow and red filled circles are the Be and B[e] III-V stars, respectively, from the spectroscopic study of \citet{2012MNRAS.425..355R}. The O and B stars from the census of \citet{2009AJ....138.1003B} are shown in green and blue circles, respectively, with luminosity class III-V objects shown with filled symbols, and luminosity class I-II objects shown with open symbols. Since \citet{2012MNRAS.425..355R} do not list the MCPS photometry of their objects, we cross-correlated their coordinates with the MCPS catalog using a search radius of 1\arcsec. All these different datasets constitute the background of this color-magnitude diagram, while in the foreground we show the identified optical counterparts to the X-ray sources marked by the X-ray source ID (Column 1 of Table \typeout{\ref{tableCensus}}1). Magenta symbols indicate unambiguous matches (i.e. the counterpart of each source that falls in the area of lowest chance coincidence probability; c.f. Section \ref{cc}). All the remaining sources are shown with cyan symbols, and the most likely matches among those are marked with black diamonds (if possible to select one; see detailed discussion in the Appendix \ref{appendixA}). We also overplot the Geneva isochrones \citep{Lejeune et al. 2001} for metallicity Z$=$0.008 and for various ages ranging from 1 Myr to $\sim$490 Myr.

From the comparison of the $V, B-V$ color-magnitude diagram in the LMC (Fig. \ref{VBVall}), and the SMC (Fig. 2 of \citealt{Antoniou et al. 2009b}), we see that the counterparts of HMXBs in the LMC are bluer than those in the SMC. Most of the matches in the LMC are clustered around $(B-V)_{o}\sim-0.2$ mag (corrected for reddening), while in the SMC they are found around $(B-V)_{o}\sim-0.1$ mag. Similarly, from the overlaid Geneva isochrones (\citealt{Lejeune et al. 2001}), we find counterparts as young as $\sim3$ Myr in the LMC with their majority being younger than $\sim45-55$ Myr, while in contrast in the SMC their estimated age is $\sim15-85$ Myr (\citealt{Antoniou et al. 2009b}).

\subsubsection{Chance coincidence probability}\label{cc}

In order to estimate the chance coincidence probability of identifying spurious matches from the MCPS catalog as the optical counterparts of the HMXBs, we performed extensive Monte Carlo simulations showing that the brightest match is the most likely optical counterpart.

\begin{figure}
\centering
\rotatebox{270}{\includegraphics[width=5.7cm]{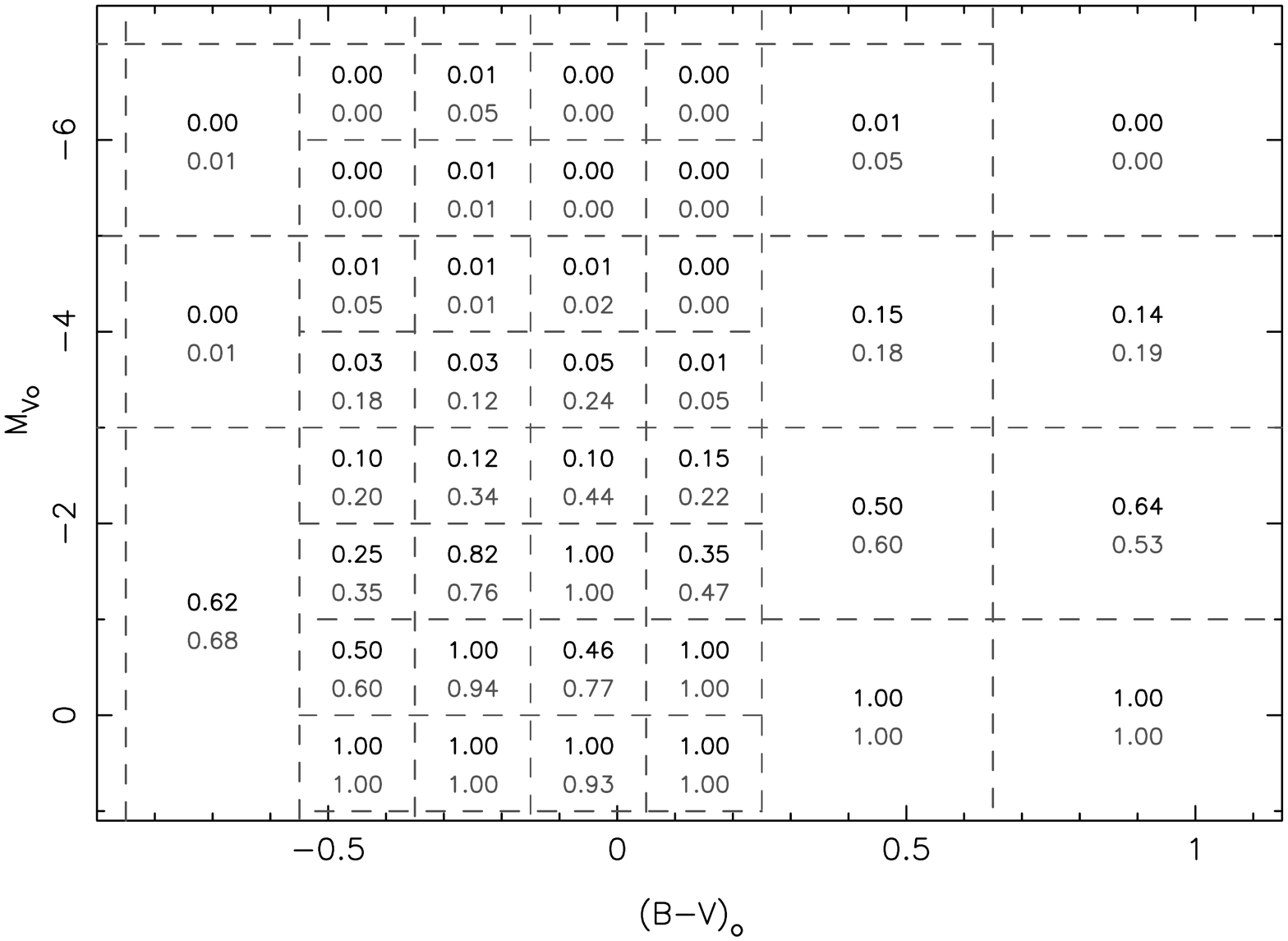}}
\begin{minipage}{80mm}
\caption{Chance coincidence probability (optical matches per X-ray source) in different regions of the reddening-corrected $V,B-V$ color-magnitude diagram. In each of these regions (i.e. cells identified by the grid shown with dashed lines) we present the probability of detecting one or more spurious matches for a source, for 2 different search radii (5\arcsec\, shown in black at the top part of each cell, and 10\arcsec\, shown in grey at the bottom part of each cell). For example, for sources with $-4 < M_{V_{o}}\leq-3$ magnitudes and $-0.55 < (B-V)_{o}\leq-0.35$ colors (corresponding to $15<V\leq16$  and $-0.4<B-V\leq-0.2$, respectively) there is a 3\% chance coincidence probability for the matches found within 5\arcsec\, from the X-ray source position, while this probability increases to 18\% when found within 10\arcsec.\label{fig_gridCC}}
\end{minipage}
\end{figure}

Following the same procedure as in \citet{Antoniou et al. 2009b}, we looked for matches of the X-ray sources after applying a random positional offset in Right Ascension and Declination. In particular, we created 1,000 random such samples with offsets drawn from a uniform distribution, taking care that the new position is outside the search radius of each source, and then we cross-correlated each of these samples with the MCPS catalog in the same way as the observed data. The chance coincidence probability was estimated for 2 different search radii (5\arcsec\, and 10\arcsec, corresponding to 1 and 2 times our typical search radius, respectively). In order to account for the varying density of stars in different regions of the color-magnitude diagram, we performed the cross correlations for a grid of magnitudes and colors, in regions of the $V,B-V$ color-magnitude diagram that have a range of 1 mag in the $V$ band and 0.2 mag in the $B-V$ color, in all but the cases where the stellar density of these regions was small. In these cases, we increased this range accordingly so as to obtain  meaningful results from the cross correlations. The results are given in Fig. \ref{fig_gridCC}. It is clear from this figure that for X-ray sources with more than one optical matches, the brightest one has the lowest probability of being a spurious match, while for equally bright matches, the one with the smallest chance coincidence probability is the bluest source. For sources with fainter optical counterparts the chance coincidence increases significantly (see also \citealt{Antoniou et al. 2009b}).

\begin{figure}
\centering
\rotatebox{270}{\includegraphics[width=5.7cm]{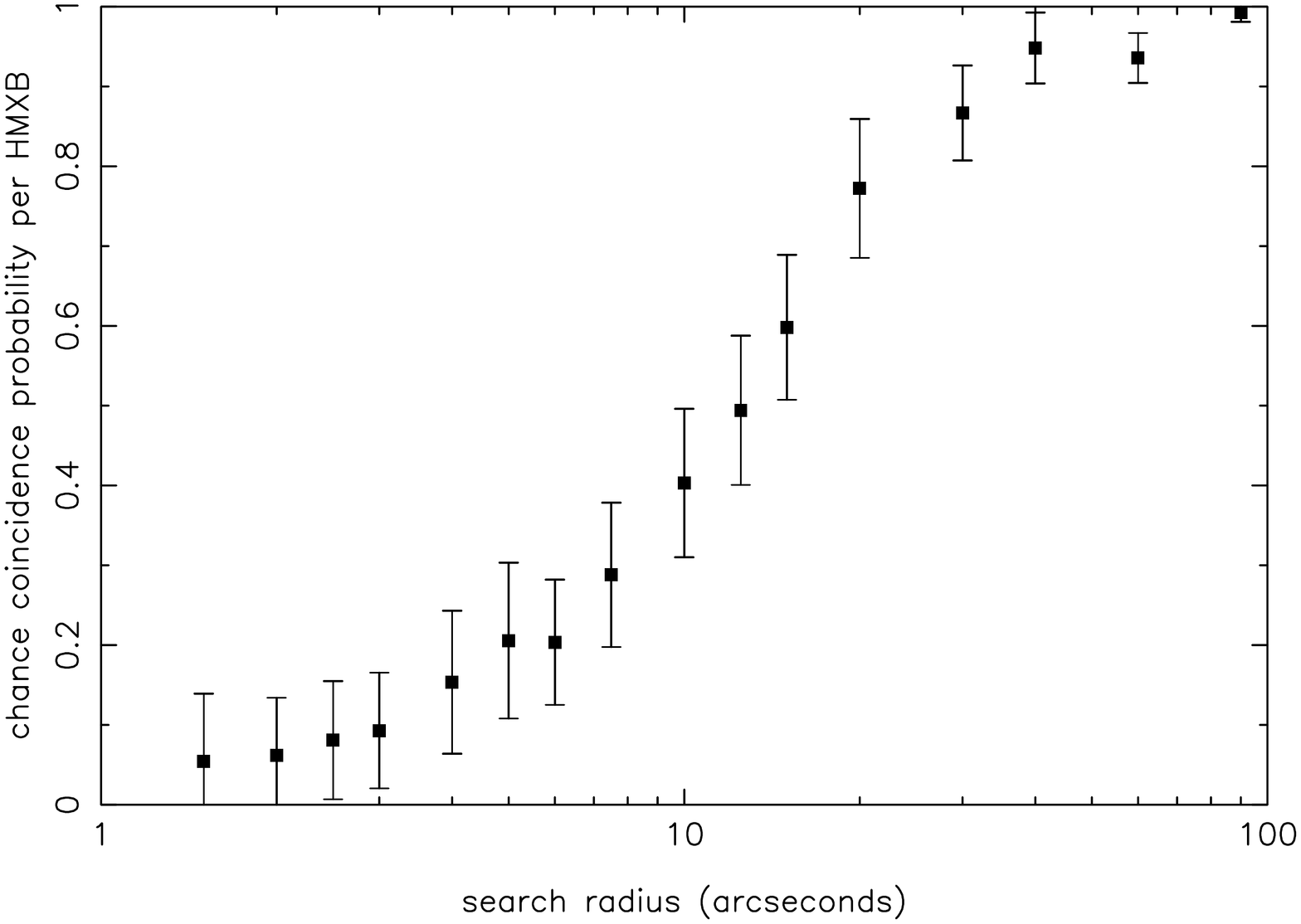}}
\begin{minipage}{80mm}
\caption{Chance coincidence probability for optical sources in the locus of OB stars as a function of the search radius (for a comparison with the SMC refer to Fig. 4 of \citealt{Antoniou et al. 2009b}).\label{figCCtests}}
\end{minipage}
\end{figure}

We also studied the chance coincidence probability as a function of search radius for stars in the locus of OB spectral types ($M_{V_{o}}\leq-0.75$ and $(B-V)_{o}\leq 0.20$). These results indicate that for a search radius of 5\arcsec, about $21\pm10\%$ of the bright blue ($V\lesssim 18.25$ mag and $(B-V)\lesssim 0.35$ mag) matches are spurious associations (this probability increases to $\sim29\pm9\%$ and $\sim40\pm9\%$ for search radii of 7.5\arcsec\, and 10\arcsec, respectively). By comparing these results with Fig. \ref{fig_gridCC}, we conclude that this relatively large chance coincidence probability is mainly driven by the fainter, more numerous stars in the OB-star locus. Instead for the earlier spectral types the chance coincidence probability is considerably lower. In Fig. \ref{figCCtests} we present the normalized chance coincidence probability for OB stars as a function of the search radius.

\subsection{Final source classification}\label{theCLASS}

In Table \typeout{\ref{finalclass}}5 we present the final classification of the X-ray sources listed in Table \typeout{\ref{tableCensus}}1. In particular, we list the X-ray source ID, name and XRB classification found in the literature, followed by the classification from this work (Columns 1 to 4, respectively\footnote{Columns 1, 2, and 3 of Table \typeout{\ref{finalclass}}5 are identical to Columns 1, 2, and 9 of Table \typeout{\ref{tableCensus}}1.}). In Column 5 we give additional notes for each source. In summary, we classify the X-ray sources listed in the literature as HMXBs based on the photometric properties of their optical counterparts identified in this work. In total we revise the classification of 4 systems from being a member of the HMXB class to being an X-ray source with a late-type companion (their most likely counterpart falls on or above the Red Giant Branch on the $V,B-V$ color-magnitude diagram). For other 12 X-ray sources, which are listed as confirmed or candidate HXMBs in the list of \citet{2005A&A...442.1135L} and for which we revise their classification, we propose a candidate Be-XRB nature, while for one additional source we are able to only classify it as an HMXB. Most importantly, we classify 2 sources (identified until now only as HMXBs) as supergiant XRBs (SG-XRBs), bringing the total number of known SG-XRBs in the LMC to four (i.e. doubling the currently known number), which makes them significant enough as a class to be studied further on their own. These sources are discussed in detail in Appendix \ref{appendixA}. In Table \typeout{\ref{tblPOP}}6 (Column 1) we summarize our results by presenting the number of sources in the LMC for the different XRB types (Be-, SG-, pulsar-, e.t.c.), found in the literature and revised in the present work (Columns 2 and 3, respectively).

\subsection{The metallicity of young stellar populations in the Magellanic Clouds}\label{MCmetal}

 Metallicity is a key parameter that determines the evolution and physical parameters of stars and binary stellar systems (e.g., \citealt{Belczynski et al. 2008}, \citealt{Linden et al. 2010}, \citealt{2013ApJ...764...41F}). The generally used values of the Magellanic Clouds' metallicities are ${\rm Z_{LMC}\sim1/2.5}$~\zsun\, and ${\rm Z_{SMC}\sim1/5}$~\zsun. For \zsun$=0.0134$ \citep{Asplund et al. 2009}, these correspond to ${\rm Z_{LMC} \sim 0.005}$ and ${\rm Z_{SMC} \sim 0.003}$  (while assuming \zsun$=0.02$, they would be ${\rm Z_{LMC} \sim 0.008}$ and ${\rm Z_{SMC} \sim 0.004}$).

As a by-product of this work, we compiled a list of metallicities for young stars in the Magellanic Clouds. We first examined the metallicity of B-type stars\footnote{\label{samefootnote}Ideally, of course, we would like to have information about the metallicity of the particular companion stars -- in other words, of the known counterparts of the studied HMXBs -- or of regions as close as possible to their location. However, the vast majority of the companions are Be-stars, which as broad-line stars are not used in metallicity studies.} (i.e. stars with ages up to $\sim$50 Myr), and then we investigated the metallicity of young ($<100$ Myr) clusters.

In Table \typeout{\ref{tblFeH}}7 we present the Fe abundances found in the literature for B-type stars in the LMC, compared to those for the same type of  objects in the SMC, SMC Wing, and the Magellanic Bridge areas (Column 1). To the best of our knowledge, the studies used here are the most comprehensive ones in which an Fe abundance is presented for these spectral-types. Systems for which there were no available Fe abundances are excluded from Table \typeout{\ref{tblFeH}}7. In Column 2 we list the name of the B-type star studied, in Columns 3 and 4 its Right Ascension and Declination (J2000.0), and in Columns 5 and 6 its spectral type and [Fe/H] abundance (followed by the corresponding references in parenthesis), respectively. The Z/\zsun\, and Z values are derived using Z/\zsun ${\rm =10^{[Fe/H]}}$ (e.g., \citealt{1992ApJ...384..508R}) and \zsun$=0.0134$ \citep{Asplund et al. 2009} (Columns 7 and 8, respectively). The estimated mean Z values are: $\sim0.007\pm0.003$ for the LMC, $\sim0.003\pm0.002$ for the SMC, $\sim0.001$ for the SMC Wing, and $\sim10^{-4}$ for the Magellanic Bridge. We note though the following caveats that apply to this compilation: {\it (a)} these are measurements for B-type stars, i.e. non Be-stars$^{\ref{samefootnote}}$, {\it (b)} most of them are supergiant stars (for example, in the LMC only 3 out of 17 stars presented in Table \typeout{\ref{tblFeH}}7 are III-V luminosity class stars), in contrast to the majority of the companions of the HMXBs in the Magellanic Clouds which are main-sequence stars (see Table \typeout{\ref{tblPOP}}6), and {\it (c)} most of them are stars in clusters of particularly young ages (only few Myr old), while the formation of the HMXBs, in the SMC at least, shows its peak at ages as old as $\sim25-60$ Myr (this is less of a concern for the LMC; see Section \ref{SFH-XRB}).

In order to obtain a more complete picture of stellar metallicities in a broader age range, in Table \typeout{\ref{tblAgeFeH}}8 we present the ages and Fe abundances of young ($<$100 Myr) star clusters in the Magellanic Clouds compiled from an extensive literature search. Since there are multiple, often contradictory, values for the age and metallicity of star clusters in the Magellanic Clouds reported in the literature, mainly due to the different methods used for their derivation, we opted to use the metallicity and age of the clusters listed only in [HZ09], which is the largest homogeneous compilation of cluster parameters (only the data for the NGC2100 and NGC1818 clusters are taken from different works). This list contains 85 star clusters, which is an adequate number for a comparison of the average  metallicities of young stellar populations in the Magellanic Clouds. In Columns 1 and 2 we list the galaxy and star cluster names (24 in total for the LMC and 9 in the SMC). The age, [Fe/H], Z/\zsun\, and Z values are given in Columns 3, 5, 7, and 9, respectively (followed by their errors in Columns 4, 6, 8, and 10, respectively). The reference from where most data are taken is given in Column 11. When the Z/\zsun\, (and/or Z) values were not available in the literature, we used the Z/\zsun ${\rm =10^{[Fe/H]}}$ relation and \zsun$=0.0134$ (\citealt{Asplund et al. 2009}). From the data of Table \typeout{\ref{tblAgeFeH}}8 we estimate mean Z values of: $\sim0.006\pm0.002$ for the LMC, and $\sim0.003$ (with $<0.001$ standard deviation) for the SMC, in good agreement within the errors with the values derived from the B-type stars above.

\section{The star-formation history of the LMC}\label{globalSFH}

[HZ09] presented the detailed spatially resolved star-formation history map of the whole galaxy (8.5\degree\, $\times$ 7.5\degree) in 24\arcmin\, $\times$ 24\arcmin\, cells (each cell was further subdivided into a $2\times2$ grid of subregions with $>$25,000 stars)
 by utilizing the MCPS survey (\citealt{2004AJ....128.1606Z}). This study focused on the structure of the LMC, and on the recent episodes of enhanced star-formation activity (younger than $\sim$100 Myr). More recently, also \citet{2011A&A...535A.115I} studied the recent star-formation history of both Magellanic Clouds, and found peaks in the star-formation activity of the LMC at 0--10 Myr and 90--100 Myr, and at 0--10 Myr and 50--60 Myr in the SMC, in broad agreement with [HZ09] and  [HZ04], respectively. At $\sim$40 Myr ago, which is the age of the regions in the SMC that host the largest number of HMXBs (\citealt{Antoniou et al. 2010}), the SFR for the entire LMC is $\sim0.25_{-0.10}^{+0.15}$\msunpyr ~(derived from its integrated star-formation history as presented in Fig.~11 of [HZ09]). At the same epoch the SFR of the SMC  is $\sim0.30_{-0.15}^{+0.55}$\msunpyr ~(Fig. 19 in [HZ09]; note that based on this figure, the maximum SFR value in the SMC is at least 0.85\msunpyr). In order to derive the star-formation history of the Magellanic Clouds, [HZ09] and [HZ04] have fitted the color-magnitude diagrams with stellar populations of 4 and 3 different metallicities for the LMC and the SMC, respectively. Throughout this work we are using the star-formation history for the highest metallicity populations (${\rm Z=0.008}$), which is the most relevant for the young stellar populations, and in agreement with the average metallicity of LMC stars discussed in the previous section. Since the above SFRs in the two Magellanic Clouds are comparable (within the errors), one would expect to find not too different numbers for the same types of XRBs in the two galaxies.

\subsection{Star-formation history and young XRB populations in the LMC}\label{SFH-XRB}

In order to investigate the link between stellar and XRB populations, we follow the procedure described in \citet{Antoniou et al. 2010}: 
we calculate the average star-formation history for the MCPS regions
($\sim12\arcmin\times12\arcmin$; [HZ09]) that host the 15 confirmed Be-XRBs\footnote{We chose to show the star-formation history of this XRB population --and not that of all LMC HMXBs for example-- for a direct comparison with the SMC (\citealt{Antoniou et al. 2010}). As a reminder, the SMC has an HMXB population consisting almost exclusively of NS/Be-XRBs, while the LMC has also 2 BH-HMXBs and 4 SG-XRBs among the 40 HMXBs studied in this work.} listed in Table \typeout{\ref{finalclass}}5. Since the HMXBs presented in this work have been identified by many different X-ray observatories (from {\it ROSAT} to \xmmsimple, {\it Swift} and {\it INTEGRAL}), without any of them having performed a complete survey of the whole galaxy so far, we opted to derive the star-formation history only for the MCPS region that contains each HMXB, and not for the whole field of view of each X-ray satellite in each case. The SFR errors are derived based on the upper and lower confidence intervals given by [HZ09], while when more than one HMXBs fall in the same MCPS region, we weight the SFR by the number of encompassed X-ray sources in each region.

We repeat this exercise for regions in the LMC hosting each of the following source classes: 14 X-ray pulsars, 4 SG-XRBs, 1 BH-HMXB (the only known within the area covered by MCPS), 40 HMXBs, 17 candidate Be-XRBs, and 1 WD/Be-XRB that have been so far identified (Table \typeout{\ref{finalclass}}5). For comparison, we also derive the star-formation history of the MCPS region in the SMC (which also has a $\sim12\arcmin\times12\arcmin$ size; [HZ04]) that hosts the 1 candidate WD/Be-XRB.

\begin{figure}
\centering
\rotatebox{270}{\includegraphics[height=8.5cm]{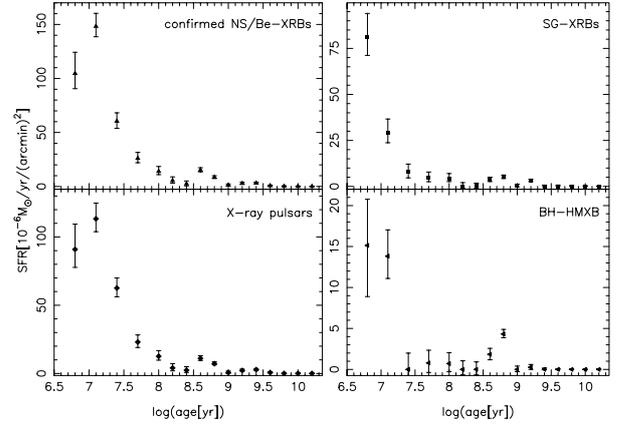}}
\caption{Average star-formation history (using data from [HZ09]) of regions in the LMC with: confirmed NS/Be-XRBs {\it (top left)}; X-ray pulsars {\it (bottom left)}; SG-XRBs {\it (top right)}; and LMC X-1, the only BH-HMXB  with MCPS coverage {\it (bottom right)}.  For a comparison with the SMC, refer to Fig. 1 of \citet{Antoniou et al. 2010}.}\label{fig1SFHmean}
\end{figure}

\begin{figure}
\centering
\rotatebox{270}{\includegraphics[height=8.5cm]{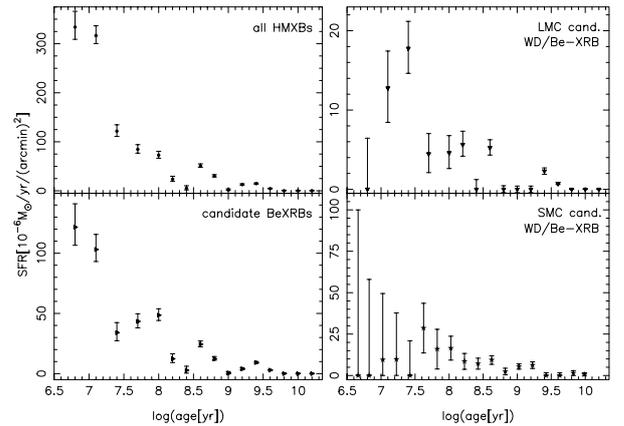}}
\caption{Average star-formation history (using data from [HZ09]) of regions in the LMC with: all HMXBs {\it (top left)}; candidate Be-XRBs, i.e. systems without an optical spectroscopic identification {\it (bottom left)}; candidate WD/Be-XRBs in the LMC and SMC {\it (top and bottom right, respectively)}. For a comparison with the SMC, refer to Fig. 1 of \citet{Antoniou et al. 2010}.}\label{fig2SFHmean}
\end{figure}

In Fig. \ref{fig1SFHmean} we present the average star-formation history of regions in the LMC with: confirmed Be-XRBs {\it (top left)}; X-ray pulsars {\it (bottom left)}; SG-XRBs {\it (top right)}; and LMC X-1 (the only BH-HMXB with MCPS coverage; {\it bottom right}). Similarly, in Fig. \ref{fig2SFHmean} we present the average star-formation history of regions in the LMC with: all HMXBs {\it (top left)}; candidate Be-XRBs (i.e. based on photometric properties, but still lacking an optical spectroscopic identification; {\it bottom left}); candidate WD/Be-XRBs in the LMC and SMC {\it (top and bottom right panels, respectively)}.

We find that the star-formation history in regions associated with the different XRB types is strongly peaked at the following ages:\\
\indent
 {\it (a)} $\sim$6.3 Myr for all known HMXBs, HMXBs without a BH or WD compact object, the candidate Be-XRBs, the SG-XRBs, and the BH-HMXB;\\
\indent
 {\it (b)} $\sim$12.6 Myr for all the confirmed Be-XRBs, the confirmed Be-XRBs with a NS compact object, and the X-ray pulsars; and\\
\indent
 {\it (c)} $\sim$25.1 Myr for the candidate WD/Be-XRB in the LMC. In contrast the candidate WD/Be-XRB in the SMC lies in a region with a star-formation peak at $\sim$42.2 Myr.

A comparison of the total SFR for the Magellanic Clouds at the above ages shows that at $\sim$10 Myr the SFRs are $\sim 0.43_{-0.15}^{+0.30}$\msunpyr ~and $\sim 0.33_{-0.18}^{+0.52}$\msunpyr ~for the LMC and SMC, respectively, while at $\sim$25 Myr they become $\sim 0.30_{-0.12}^{+0.18}$\msunpyr ~and $\sim 0.15_{-0.10}^{+0.70}$\msunpyr, respectively. Although the upper bound of the SMC SFR at both ages  is very large (i.e. the maximum SFR value is at least 0.85\msunpyr), we see that at $\sim$10 Myr and $\sim$25 Myr the SFR is higher in the LMC, while for older star-formation episodes ($\sim$40 Myr) it is higher in the SMC.

\section{DISCUSSION}\label{discussion}

In the previous sections we presented a comprehensive compilation of all known X-ray binaries associated in the LMC. By comparing them with stellar catalogs of the LMC we updated and completed their optical classifications. Based on these results we identified the main star-formation episodes in the history of the LMC that are associated with its HMXB populations. In this section we discuss these results in the context of the formation efficiency of HMXBs, and their nature in comparison with the SMC and our Galaxy.

\subsection{Star-formation history of young XRBs in the Magellanic Clouds}\label{SFH-youngXRBs}

Below we discuss the major findings of this work and we compare them with the results for the young XRBs in the SMC (c.f. \citealt{Antoniou et al. 2010}):\\
\indent
 {\it (a)} In contrast to the SMC, the confirmed Be-XRBs and (as expected) XRB pulsars in the LMC are related to a major star-formation burst $\sim6-25$ Myr ago (as a reminder, the SMC Be-XRBs are present in regions with star-formation bursts at $\sim25-60$ Myr ago), indicating younger Be-star populations in the LMC than in the SMC regions (also obvious from the position of the counterparts on the $V,B-V$ color-magnitude diagram as discussed earlier in Section \ref{OptLIKELY}).\\
\indent
 {\it (b)}  The SFR of the major burst  associated with the confirmed Be-XRBs and the X-ray pulsars in the LMC  is almost 3 and 2 times, respectively, higher than that in the SMC.\\
\indent
 {\it (c)} the candidate Be-XRBs show a star-formation episode that started earlier than 12.6 Myr ago, which is similar to the star-formation episode for the confirmed Be-XRBs (although the latter is more centered around $\sim$ 12 Myr). The fact that the peak of the star-formation episode associated with the candidate Be-XRBs  is at $\sim$ 6 Myr suggests that this population may include misclassified SG-XRBs, since the latter are expected to be abundant at these young ages due to their massive companions. This is the first hint for a significant SG-XRB population in the LMC (we already know 4 such systems), at least in comparison to the SMC where only two systems are known to date: SMC X-1 (\citealt{1972NPhS..240R.183W}) that is not covered by the MCPS survey, and source CXOU J005409.57-724143.5 (\citealt{2014MNRAS.438.2005M}; although the nature of this source is debated; \citealt{2013A&A...560A..10C}). From Fig. \ref{fig2SFHmean} we also see that there is another star-formation episode  $\sim$100 Myr ago at the regions associated with candidate Be-XRBs. However, the association of this peak with the observed XRB populations is ruled out by the location of their optical counterparts in the OB locus of the color-magnitude diagram (Fig. \ref{VBVall}), since the least massive B-type stars have a main-sequence lifetime of less than 100 Myr.\\
\indent
 {\it (d)} the five regions that host the 4 known SG-XRB and LMC X-1 (the one BH-HMXB in the LMC with MCPS coverage) show a peak in their star-formation history at even younger ages ($\lesssim$10 Myr), as expected based on the more massive companions (i.e. shorter lifetimes) of these XRB types. This is also supported by the location of their counterparts on the color-magnitude diagram (Fig. \ref{VBVall}; they are systematically younger than $\lesssim$ 10 Myr). As a note, no BH-HMXB has been found in the SMC yet.\\
\indent
 {\it (e)} We also examined the star-formation history of the regions in which two of the most promising detections of WD/Be-XRBs have been made. For the LMC WD/Be-XRB (source XMMU J052016.0-692505 with ${\rm L_{X}>10^{36}}$\ergs; \citealt{2006A&A...458..285K}), the peak in its star-formation history is shifted to older ages ($\sim$11--40 Myr) with respect to the NS/Be-XRB population in this galaxy. We observe a similar trend in the SMC (source XMMU J010147.5-715550 with ${\rm L_{X}\sim3\times10^{34}}$\ergs; \citealt{2012A&A...537A..76S}), though the star-formation burst in this case is wide (it appears at $\sim$32--141 Myr ago). We note that in this analysis we have not included the second discovered WD/Be-XRB in the SMC (MAXI J0158-744; \citealt{2012ApJ...761...99L}), since this system is located in the SMC Wing, which is not covered by the MCPS survey, thus there is no information on its star-formation history.

\subsection{Formation efficiency for the various XRB types}\label{sectionFRs}

In order to measure the formation efficiency of XRBs we need first to associate them with an individual star-formation episode. The location of the optical counterparts of the X-ray sources on the color-magnitude diagram clearly links them with stellar populations younger than $\sim$50 Myr, which associates them with the star-formation episode peaking at $\sim$10--30 Myr, depending on the exact star-formation history at the location of different XRB populations (Figs. \ref{fig1SFHmean} and \ref{fig2SFHmean}). In the case of SG-XRBs in particular, we see from Fig. \ref{VBVall} that they are associated with even younger stellar populations, which clearly links them with an earlier star-formation episode ($\lesssim10$ Myr; Fig. \ref{fig1SFHmean}).

The number of systems in each class of XRBs is given in Table \typeout{\ref{FRates}}9: Column 1 lists the XRB types considered in each calculation, while Column 2 gives the number of objects. Column 3 lists the age of the peak of their maximum SFR\footnote{For the cases with a significant secondary star-formation burst, e.g., for all known HMXBs, the HMXBs without a BH or WD compact object, and the candidate Be-XRBs listed in Table \typeout{\ref{finalclass}}5, we also provide the same information for this secondary episode, although it is not associated with the production of the XRBs under consideration (Section \ref{SFH-youngXRBs}).}, while Column 4 presents its estimated duration (in this we take into account that the star-formation episodes started before reaching their maximum intensity at the time given in Column 3), and Column 5 gives the SFR at the peak of each episode. In Column 6 we list the number of MCPS subregions (each 12\arcmin$\times$12\arcmin\, [HZ09]) used for the derivation of the average star-formation history of regions with different XRB types (as these are given in Columns 1 and 2). In the same table, we also list the formation efficiency for each class of objects with respect to the SFR  at the peak of the star-formation episode they are associated with. In particular, in Column 7 we list the number of systems we observe today per unit SFR at the peak of the SF episode, and in Column 8 the required SFR for the production of 1 system for each class. We note that we derive a formation efficiency only for those XRB populations for which we can establish a strong correlation with a particular star-formation episode. In addition, we estimate the number of XRBs normalized by the stellar mass M$_{\star}$ formed during the star-formation episode that produced them. We calculate the total M$_{\star}$ by integrating the binned star-formation histories shown in Figs. \ref{fig1SFHmean}, \ref{fig2SFHmean}. The formation rate of XRBs per unit stellar mass formed during their respective SF episode is given in Column 9. The errors in this quantity reflect the upper and lower limits of the SFR in Figs. \ref{fig1SFHmean}, \ref{fig2SFHmean}. For the SMC we have found that we need a SFR of $\sim3\,\,\,(2.5)\times10^{-3}$\msunpyr ~for the production of one Be-XRB (HMXB) at $\sim$40 Myr after a star-formation episode (\citealt{Antoniou et al. 2010}). In the present work we find that:\\
\indent
  {\it (a)} the production of NS/Be-XRBs and HMXBs in the LMC ($\sim21.4\times10^{-3}$\msunpyr\, and $\sim43.5\times10^{-3}$\msunpyr, respectively; Table \typeout{\ref{FRates}}9) is $\sim7$ and $\sim17$ times less efficient than the production of the same populations in the SMC (with respect to their parent star-formation episodes, i.e. at $\sim$12.6 Myr and $\sim$42.2 Myr, respectively).\\
 \indent
{\it (b)} based on the SFR of the SMC at the age of $\sim$10 Myr ($\sim 0.33_{-0.18}^{+0.52}$\msunpyr; Section \ref{SFH-XRB}) and the formation efficiency of LMC SG-XRBs at this age ($\sim85.3_{-44.0}^{+44.7}$ systems/(\mdot); Table \typeout{\ref{FRates}}9), we estimate a large number of $28_{-21}^{+47}$ SG-XRBs in the SMC (or equivalently, about 7 to 75 such systems). This large number is in stark contrast with the only 2 known SG-XRBs in the SMC (Section \ref{SFH-youngXRBs}). On the other hand, recently, we have identified the elusive population of SG-XRBs in our Galaxy: (a) the heavily obscured SG-XRBs (a small known population exhibiting low luminosities, e.g., IG J16318-4848: \citealt{2003MNRAS.341L..13M}, \citealt{2009A&A...508.1275B}; CI Cam: \citealt{1999ApJ...527..345B}, \citealt{2013MNRAS.429.1213B}; GX301-2: \citealt{1973ApJ...184..237R}, \citealt{2014MNRAS.441.2539I}), and (b) the SFXTs, which have short (less than few hrs long) outbursts and low-level quiescent X-ray emission (${\rm L_{X}}\sim 10^{32}-10^{34}$\ergs; Section \ref{Intro}). However, this population is not expected to be easily detectable in \chandra and \xmm surveys of the Magellanic Clouds. Given that this calculation is based on rescaling the numbers of SG-XRBs detected in the LMC to the SMC, both of which have been surveyed with similar strategies, we would not expect that the discrepancy between observed and estimated number of SG-XRBs in the SMC is the result of selection effects or missing populations as these would influence the SMC and the LMC in similar ways. Therefore, we consider that this discrepancy could be the result of a strong metallicity dependence of the XRB formation efficiency at least in the case of SG-XRBs.\\
\indent
 {\it  (c)} Based on the same argument, the SFR of the SMC at 25 Myr ($\sim 0.15_{-0.10}^{+0.70}$\msunpyr; Section \ref{SFH-XRB}), and the formation efficiency of the LMC WD/Be-XRBs at this age (390.3 systems/(\mdot); Table \typeout{\ref{FRates}}9), we predict $\sim$ 59 SMC WD/Be-XRBs. However, thus far only one system has been identified as a candidate WD/Be-XRB in the LMC (Section \ref{SFH-XRB}), also pointing to a strong metallicity dependence.

\subsection{Previous studies}\label{Comparison}

Two other works discuss in some detail the link between young X-ray source populations and recent star formation: 

 -- \citet{2005A&A...431..597S} used archival \xmm data to study the X-ray source population of the LMC. They found significant field-to-field variations of the HMXBs in the LMC, which are not correlated with differences in their FIR or ${\rm H}_\alpha$ emission, hence they suggest that these differences are due to the age dependence of the HMXB population. This is in good agreement with our work, which also shows differences in the types of XRBs and their formation efficiency as a function of age. Furthermore, they estimated that the number of HMXBs per total stellar mass to less than $10^{-4}$\invmsun at ages $\sim$1--2 Myr, and  $(1.00\pm0.45)\times10^{-4}$\invmsun at $\sim$10--12 Myr. This is a factor of $\sim$70 lower than the formation efficiency listed in Table  \typeout{\ref{FRates}}9. This difference however is the result of the more complete census of HMXBs used in the present work, and the different assumptions in the calculation of the total stellar mass in the two works.
 
-- On the other hand, \citet{2013ApJ...772...12W} found that the preferred age of HMXBs in NGC 300 and NGC 2403 is 40--55 Myr, which is the {\it same} age in which HMXBs are observed in the SMC. Although these galaxies have metallicities higher than that of the SMC, the remarkable similarity of the ages of the stellar populations in which the HMXBs are observed in the SMC, NGC 300 and NGC 2403, make the argument for increased HMXB formation efficiency at an age of $\sim$40 Myr even stronger.

\subsection{Spectral-type distribution of LMC HMXBs}\label{SPTypedistribution}

\begin{figure}
\centering
\includegraphics[width=0.5\textwidth]{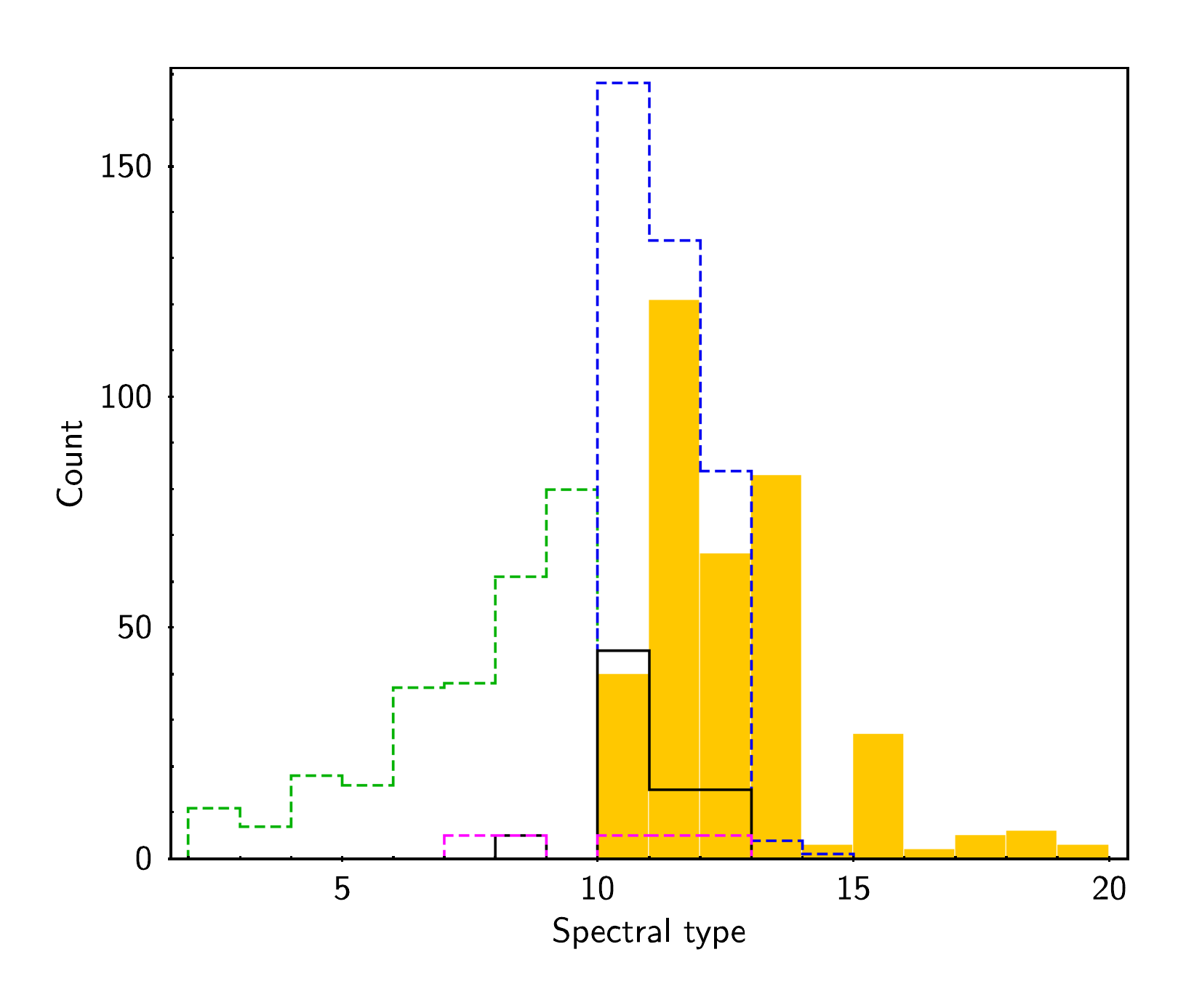}
\begin{minipage}{80mm}
\caption{Spectral-type distribution for {\it (a)} Be field starts (filled dark yellow histogram); {\it (b)} O-type field stars (green dashed line); {\it (c)} B-type field stars (blue dashed line); {\it (d)} HMXBs with Oe or Be stars as the optical counterparts (black solid line); and {\it (e)} HMXBs with O- or B-type stars as the optical counterparts (magenta dashed line). The x-axis shows the spectral type, starting from O2 stars (corresponding to number 2) up to A0 stars (corresponding to 20), in steps of one spectral subtype. All sources have III-V luminosity classes and are taken from {\it (a)}  the spectroscopic study of \citet{2012MNRAS.425..355R},  {\it (b,c)} the census of \citet{2009AJ....138.1003B}, and {\it (d,e)} the literature (see Table \typeout{\ref{tableCensus}}1), respectively. Due to the small number of known HMXBs, the distributions {\it (d,e)} have been rescaled by a factor of 5.}\label{figSPTypedistribution}
\end{minipage}
\end{figure}

Moreover, we investigated differences in the spectral-type distribution of Be-XRBs, HMXBs with OB companions (i.e. non-Be stars), and Be stars and OB stars not associated with XRBs in the LMC. In Fig. \ref{figSPTypedistribution} we show this distribution for Be starts (filled dark yellow histogram; \citealt{2012MNRAS.425..355R}), and O- and B-type stars (green and blue solid lines, respectively; \citealt{2009AJ....138.1003B}), all of luminosity class III-V. The HMXBs associated with Oe or Be stars are shown with a black solid line, while those with O- or B-type optical counterparts are presented with a magenta dashed line (all data are taken from Table \typeout{\ref{tableCensus}}1). In  this figure, spectral type IDs from 0 to 9 correspond to O-type stars, and IDs from 10 to 19 correspond to B-type stars. We find that the peak of the spectral-type distribution of the LMC Be-XRBs is around B0 and that of LMC Be stars around B1, in agreement with previous studies which have used smaller sample sizes (e.g., \citealt{2002A&A...385..517N}, \citealt{Antoniou et al. 2009a}).

For completeness, we note here that the peak of the spectral-type distribution of the SMC Be-XRBs is instead around B1, but based on the two-sample Kolmogorov-Smirnov (K-S) test, it is not possible to definitely say if the SMC Be-XRBs follow a different spectral-type distribution
from Galactic and LMC Be-XRBs \citep{Antoniou et al. 2009a}.  This is consistent with \citet{2008MNRAS.388.1198M}, who note that there is indication for similar distributions in the SMC and the Milky Way. They find a spectral distribution cutoff for the SMC Be-XRBs around B2, but \citet{2014MNRAS.438.2005M} find that the spectral-type distribution is skewed towards later spectral types (actually, they identify several systems within the B2 to B5 spectral-type range). On the other hand, the SMC Be-star population peaks at earlier spectral types (i.e. at B0), in contrast to the LMC, where it peaks at later spectral types ($\sim$B1) than the Be-XRBs ($\sim$B0). The main caveats in these comparisons are the (large in many cases) uncertainties in the spectral-type classification, the small size of the LMC samples, and a possible selection effect towards brighter targets in the optical, thus earlier spectral types.

\subsection{The young parent stellar populations of the LMC X-ray pulsars}\label{yngPULSARS}

From the above analysis, perhaps one would find surprising that the LMC X-ray pulsars  are associated with a star-formation episode at such a young age, i.e. at only $\sim$12.6 Myr (if not younger; see Fig. \ref{fig1SFHmean} and Table \typeout{\ref{FRates}}9). The obvious question then is: could an X-ray pulsar form in such a short period? The answer to this question is not straightforward. Nevertheless, we approach this based on simple arguments regarding the age of the optical companions of these systems. Out of the 14 X-ray pulsars, 9 have known spectral types (Table \typeout{\ref{tableCensus}}1). Of those, 8 have a late-type Oe or early-type Be star companion (only LMC X-4 with a pulse period of 13.5 s has an O7III-V/O8III counterpart, i.e. it has not shown  \Ha in emission). On the other hand, we know that the LMC HMXBs with Be-star companions show a peak in their spectral-type distribution at B0 (Fig. \ref{figSPTypedistribution}), and that the main-sequence lifetime can be simply estimated as:

\begin{center}
${\rm t_{MS}/t_{\odot}=(M/M_{\odot})^{-2.5}}$ \,\,\,\,\,\,\,\,\,\,\,(1)
\end{center}

\noindent
using ${\rm t_{\odot}\sim10}$ Gyr. Thus, a B0 V star (having a mass of 14.6 \msun; see Table \typeout{\ref{tblLIFE}}10) has ${\rm t_{MS}\sim12.3}$ Myr, while the lifetime of a blue supergiant is $\lesssim1$ Myr (Table 4.1 from the PhD thesis of \citealt{ChitaPhD}). This main-sequence lifetime of 12.3 Myr is very similar  to the age of the peak of the star-formation history of the stellar populations associated with Be-XRBs and the X-ray pulsars in the LMC (see Figs. \ref{fig1SFHmean} and \ref{fig2SFHmean}). 

We note that the Be stars can appear at ages even younger than $\lesssim10$ Myr. They have been observed in  Magellanic Clouds clusters --NGC 346 and NGC 371 in the SMC and LH 72 and NGC 1858 in the LMC-- with ages as young as $\sim$5--8 Myr (\citealt{2007ApJ...671.2040W}). Thus, although one would expect to mainly find SG-XRBs or BH-XRBs at that young ages based simply on their more massive companions\footnote{This is true for the 4 SG-XRBs and the one BH-HMXB in the LMC, for which we can derive their star-formation histories (Fig. \ref{fig1SFHmean}, right panels).}, there is evidence that the Be-XRB pulsars in the LMC can be as young as $\sim$10 Myr.

\citet{Belczynski et al. 2008} find that a NS forms from stars with zero-age main sequence (ZAMS) masses between $\sim$7.5 \msun and $\sim$21 \msun (though we note that these estimations are based on single stellar evolution and solar metallicity). From Table \typeout{\ref{tblLIFE}}10 we see that this mass range corresponds to a main-sequence lifetime of $\sim$5 Myr to $\sim$100 Myr. Therefore, XRB pulsars can indeed form at such young ages.

\subsection{Supernova kick velocities of LMC HMXBs}\label{SNkicks}

\begin{figure}
\centering
\includegraphics[width=0.5\textwidth]{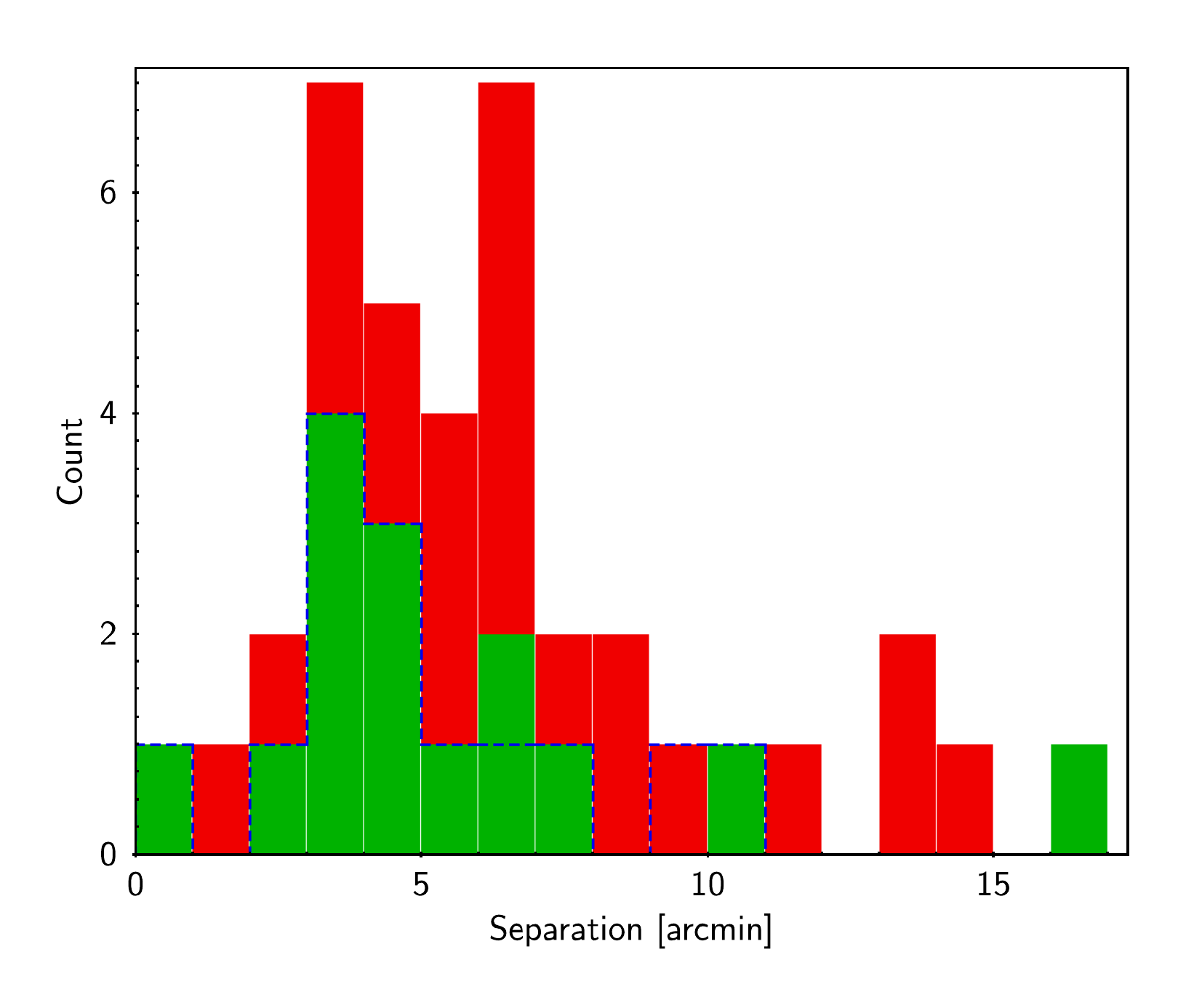}
\caption{Distribution of the distance (in arcminutes) between XRBs in the LMC and their nearest star cluster from the catalog of \citet{2011AJ....142...48W}. The 38 HMXBs without BH or WD compact objects are shown with a red solid histogram, the 15 confirmed NS/Be-XRBs are shown with a green solid histogram, and the 14 X-ray pulsars are shown with a blue dashed histogram.}\label{figDISTCLUS}
\end{figure}

 In order to estimate the kick velocity imparted onto the compact object at the time of its formation as a result of an asymmetry in the SN explosion, we follow the approach of \citet{2005MNRAS.358.1379C} that was applied to the HMXB population of the SMC.
We first derive the traveled distance of the HMXBs since their birth by using the mean offset between each X-ray source from Table \typeout{\ref{tableCensus}}1 and its nearest star cluster. The catalog of \citet{2011AJ....142...48W} (hereafter [WZ11]) contains 1,066 clusters from the MCPS survey. All known LMC HMXBs but LMC X-3 are also covered by this survey. For the 38 non-BH or -WD HMXBs of this work (see Table \typeout{\ref{finalclass}}5), we find a mean offset of $6.4\arcmin \pm 3.6\arcmin$ ($\sim 92.0 \pm 52.4$\,pc assuming a distance to the LMC equal to $\sim$50 kpc; \citealt{2006ApJ...652.1133M}), with the minimum distance  equal to 1\arcmin\, and the maximum equal to 16.7\arcmin\, (in the cases of HMXBs and HMXBs with neutron stars exclusively these distributions are bimodal; Fig. \ref{figDISTCLUS}). All reported error values are the  standard deviations of the distribution of offsets. On the other hand, the 15 confirmed Be-XRBs and the 14 confirmed X-ray pulsars have a mean offset of $5.5\arcmin \pm3.7\arcmin$ and $5.0\arcmin \pm 2.5\arcmin$, respectively (we consider the two classes separately since not all Be-XRBs have been identified as pulsars, and not all pulsars have been associated with spectroscopically confirmed Be stars). 

 From these offsets we can calculate the kick velocities imparted on the neutron star if we have an estimate of the travel time (e.g., \citealt{2002ApJ...577..726Z}; \citealt{2005MNRAS.358.1379C}). The travel time is equal to the time since the supernova explosion. The spectral-type distribution of our stars peaks at B0 (Section \ref{SPTypedistribution}), which corresponds to a star with a mass of 14.6 \msun and thus a main-sequence lifetime of $\sim$ 12.3 Myr (Table \typeout{\ref{tblLIFE}}10; \citealt{WilliamsPhD}). The earliest time that pulsars are produced is given by the most massive stars that can produce compact objects with masses $\lesssim$ 3 \msun. These are $\sim$ 21 \msun stars \citep{Belczynski et al. 2008}, which have a main-sequence lifetime of $\sim$ 5 Myr (Section \ref{yngPULSARS}). Therefore, the minimum travel time is $\sim$ 7.3 Myr. Here we take into account the fact that the Be phenomenon appears towards the end of the main-sequence lifetime of a B-type star (e.g., \citealt{2003ASPC..292...65F}). If Be stars are younger, then the estimated velocities are even larger. Combining this with the mean offset of $6.4\arcmin \pm 3.6\arcmin$ found above, we estimate a transverse velocity of $\sim12.4\pm7.0$\kms\, (or a space velocity $\sqrt{2}$ times larger) for the HMXBs in the LMC (excluding the BH or WD systems). The confirmed NS/Be-XRB population in particular travels distances equal to $5.5\arcmin \pm3.7\arcmin$ at the same time, resulting in a  transverse velocity of $\sim10.8\pm7.3$\kms. Similarly, the LMC X-ray pulsars travel $5.0\arcmin \pm 2.5\arcmin$ with a transverse velocity of $\sim9.7\pm4.9$\kms.

We also investigate whether the positions of HMXBs in the LMC are significantly different from a uniform distribution. For this reason, we have estimated the minimum distance between 100,000 random points drawn from a uniform distribution in R.A., Dec., and in each case, the nearest star cluster (similarly to \citealt{2005MNRAS.358.1379C}). The mean value was found to be equal to $4.4\arcmin \pm 1.4\arcmin$. Then we used the Kolmogorov-Smirnov (K-S) test to quantify the probability that the distributions of the offsets between the HMXBs and the real clusters in one case, and the random positions in the LMC and the clusters in the second case are different. We find that the two distributions are different at >99.95\% confidence level. However, for the smaller samples (X-ray pulsars and Be-XRBs) the results from the K-S test are inconclusive.

In order to compare these results with those for the SMC, we follow the same approach. \citet{2005MNRAS.358.1379C} has estimated a mean distance of 3.85\arcmin\, between the HMXBs in the SMC and their nearest clusters, and assumed a maximum lifetime of the Be star after the formation of the neutron star of $\sim$ 5 Myr. We thus estimate a transverse velocity of $\sim13.1$\kms\, (by using that 1\arcmin\, corresponds to 17.4 pc at the 60 kpc distance to the SMC; \citealt{2005MNRAS.357..304H}). On the other hand, if we take into account the fact that the peak of the spectral-type distribution of the companions of SMC HMXBs appears at B1 type that corresponds to an 11\msun mass and a main-sequence lifetime of $\sim$ 24.9 Myr, we estimate a minimum travel time of  $\sim$19.9 Myr. In this case, the SMC HMXBs have a transverse velocity of  $3.3$\kms, or equivalently the LMC HMXBs  travel with up to $\sim$4 times larger velocities than their SMC counterparts.

On the other hand, the Milky Way Be-XRBs have low runaway velocities (mean peculiar tangential velocity equal to $15 \pm 6$\kms), indicating low masses ejected into the supernovae that formed the neutron stars ($\lesssim$1-2 \msun) and kick velocities received by their neutron stars at birth in the range of $60-250$\kms\, \citep{2000A&A...364..563V}. These velocities are in broad agreement with the transverse velocities for the LMC neutron star accreting binaries estimated above. However, the SMC systems seem to have even smaller velocities, which is consistent with enhanced fraction of electron-capture supernovae that impart very small kicks, as predicted by \citet{Linden et al. 2009} for the SMC metallicity. 

\begin{figure}
\centering
\includegraphics[width=0.5\textwidth]{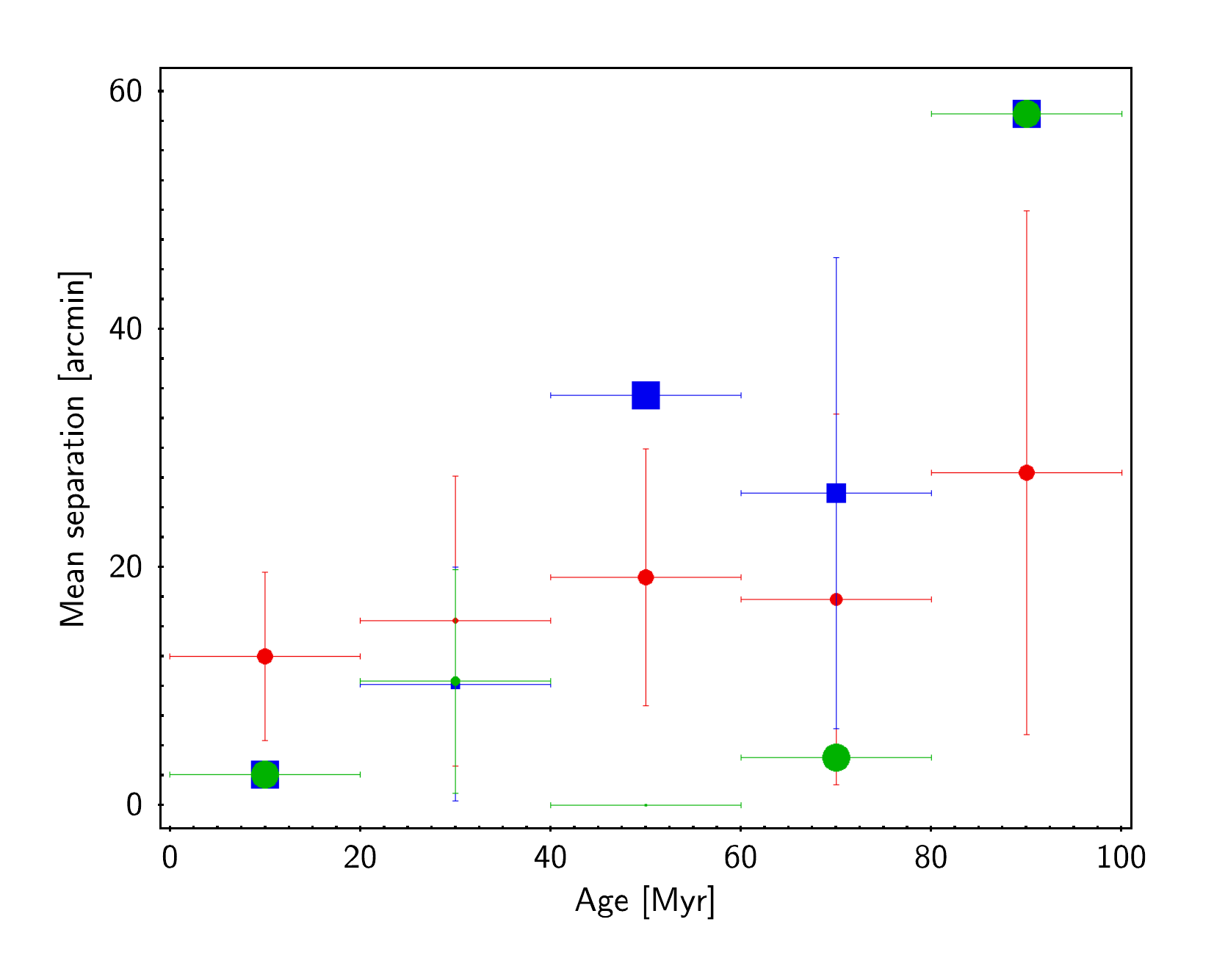}
\vspace{-0.5cm}
\includegraphics[width=0.5\textwidth]{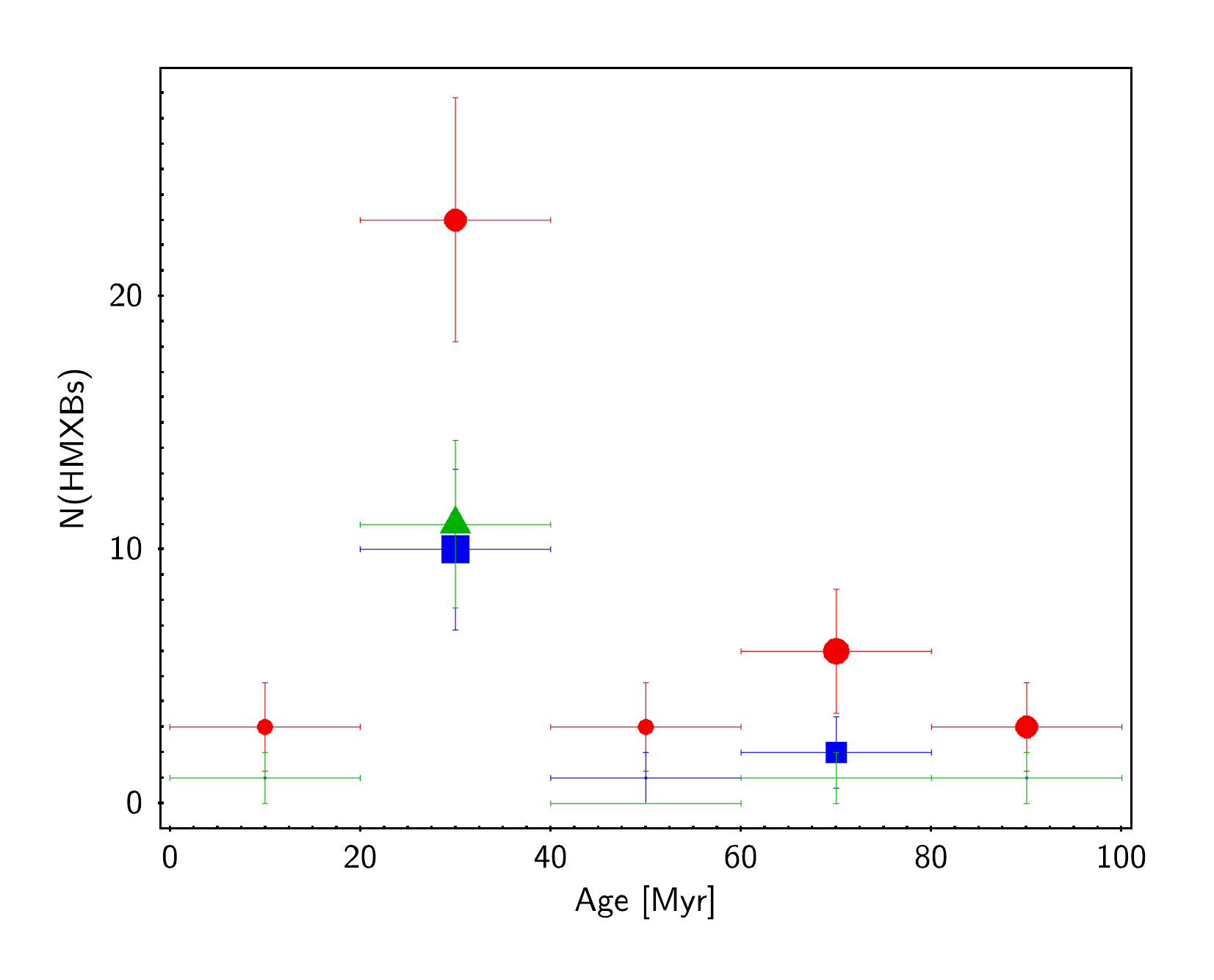}
\caption{{\it (a)} Mean separation (in arcminutes) of all non-BH or -WD HMXBs (red circles), only confirmed NS/HMXBs (blue squares) and  X-ray pulsars (green triangles; color scheme similar to Fig. \ref{figDISTCLUS}) from the nearest young star cluster (with an age $\leq$100 Myr) listed in \citet{Baumgardt et al. 2013} as a function of age. The data have been binned in the following age ranges: ${\rm (0,20]\,\,Myr, (20,40]\,\,Myr, (40,60]\,\,Myr, (60,80]\,\,Myr, (80,100]\,\,Myr}$, which are shown as error bars in the x-axis. The size of the data points is proportional to the relative error on the number of X-ray sources with matches in the cluster catalog within each age bin. The y-axis error bar shows the standard deviation of the mean separation distribution of sources within each age bin. {\it (b)} Similar to panel {\it (a)} but this time in the y-axis is the number of X-ray sources with matches in the cluster catalog within each age bin, while the size of the data points is proportional to the relative error on the separation (in arcminutes). The y-axis error bar corresponds to  $\sqrt{N}$, where N is the number of sources in each bin.}\label{figOFFCLUSAGE}
\end{figure}

Furthermore, we also examine the mean offset of the identified HMXBs from the nearest young star cluster as a function of the cluster's age. In this comparison we used the catalog of \citet{Baumgardt et al. 2013} (hereafter [B13]), which provides age information for a large number of star clusters in the LMC. About one third of the 320 clusters in the [B13] catalog has ages younger than or equal to 100 Myr (with a mean age of $52\pm28$ Myr). If we focus on these 99 young clusters, we find that the 38 non-BH or -WD HMXBs have a mean offset of $16.8\arcmin \pm 13.9\arcmin$, which is more than double the distance derived using the [WZ11] catalog. The large standard deviation is the result of few systems having very large offsets (which is expected since this catalog is sparser than [WZ11]). The other two subsets, the confirmed NS/Be-XRBs and the X-ray pulsars, have offsets of  $16.6\arcmin \pm 17.5\arcmin$ and $12.8\arcmin \pm 15.3\arcmin$, respectively. In Fig. \ref{figOFFCLUSAGE}(a) we present the mean separation (in arcminutes) of all HMXBs (excluding those systems with BH or WD compact objects), the confirmed NS/Be-XRBs and only the X-ray pulsars (red circles, blue squares, and green triangles, respectively) from the nearest young star cluster (with an age $\leq$100 Myr) listed in [B13] catalog as a function of the cluster's age. The size of each data point is proportional to the relative error on the number of the identified matches in each age bin, while the x-axis error-bars are equal to half the age bin. Similarly, in Fig. \ref{figOFFCLUSAGE}(b) we give the number of the identified matches as a function of the age of the cluster, following the color scheme of Fig. \ref{figOFFCLUSAGE}(a). In this case, the data points have sizes proportional to the relative error on the separation. We find that the majority of the X-ray sources  reside nearby a cluster with age (20, 40] Myr ($\sim$61\%, $\sim$67\% and $\sim$79\%, for the HMXBs without BH or WD compact objects, the confirmed NS/Be-XRBs and the X-ray pulsars, respectively). Although the star-formation history of the regions associated with X-ray pulsars shows its peak at younger ages ($\sim$12.6 Myr; see Table \typeout{\ref{FRates}}9) than the 30 Myr peak in Fig. \ref{figOFFCLUSAGE}, the major star-formation burst has a duration of $\sim$32 Myr, well within the (20,40] Myr age range. The mean separation of the aforementioned 3 populations at the (20,40] Myr age range is $15.5\arcmin \pm 12.2\arcmin$, $10.1\arcmin \pm 9.8\arcmin$, and $10.4\arcmin \pm 9.4\arcmin$, respectively. 
 
 \subsection{Corbet diagram}\label{Corbet}

\begin{figure}
\centering
\includegraphics[width=9.1cm]{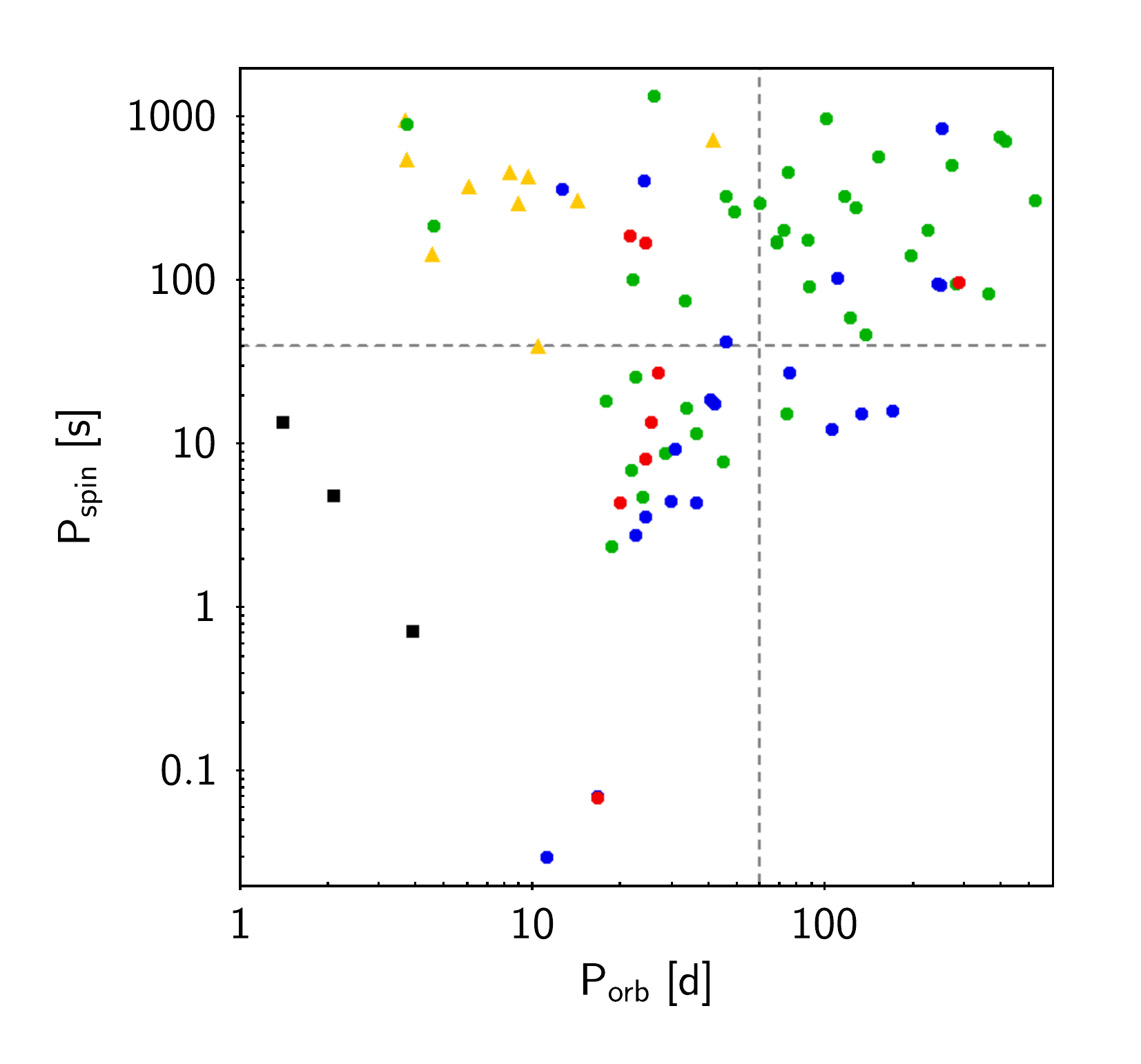}
\begin{minipage}{80mm}
\caption{Spin period versus orbital period --the so called "Corbet diagram"-- for LMC, SMC and Milky Way NS/Be-XRBs (shown in red, green and blue filled circles, respectively). We also show the Galactic wind-fed SG-XRBs (in yellow triangles) and 3 RLOF SG-XRB systems (Cen X-3, LMC X-4, and SMC X-1 shown in black squares).}\label{figCorbet}
\end{minipage}
\end{figure}

An updated version of the "Corbet diagram" (\citealt{1984A&A...141...91C}), i.e. a log-log plot of the spin period versus the orbital period, of X-ray pulsars with Be and SG stars  donors is shown in Fig. \ref{figCorbet}. The position of sources in this diagram depends on the accretion torques exerted into the NS, resulting in the division of the sources in 3 groups corresponding to 3 different mass-transfer mechanisms \citep{2009IAUS..256..361C}. When the primary is a Be star, the accretion occurs through the star's decretion disk, but when it is a SG star, the accretion can happen either through stellar winds (wind-fed systems) or through Roche-Lobe overflow (RLOF) if the star fills its Roche Lobe. The 8 LMC, 38 SMC and 20 Milky Way Be-XRBs with known ${\rm P_{spin}}$ and ${\rm P_{orb}}$ values are shown in red, green and blue circles, respectively, while the 10 Milky Way wind-fed SG-XRBs in yellow triangles. There are also 3 RLOF systems (Cen X-3, LMC X-4, and SMC X-1) shown in black squares. The data for the LMC are taken from the references listed in Table \typeout{\ref{tableCensus}}1, while those for the SMC and the Milky Way from the compilations of \citet{2014MNRAS.437.3863K} and \citet{2011MNRAS.416.1556T}, respectively. We see that all but one of the LMC Be-XRBs have orbital periods of less than $\sim$30 days. Only LXP96.1 (source RX J0544.1-7100) has a much larger orbital period of 286 days. Similarly, all but one of the LMC Be-XRBs have  spin periods longer than $\sim$4 sec. Only LXP0.07 (source RX J0535.6-6651) is known to have a 70 millisecond spin period (in the Milky Way there are 2 systems with such short spin periods). Although the number of known LMC Be-XRB systems is small, all identified members fall within the locus of the SMC and Milky Way Be-XRBs. In the same plot we also show the location of the apparent dips in the bimodal log(${\rm P_{spin}}$) and log(${\rm P_{orb}}$) distributions of Be-XRBs (dashed lines), as identified by \citet{2011Natur.479..372K}, and subsequently confirmed by \citet{2015MNRAS.452..969C}. These authors associate those sub-populations (also distinct in the log(${\rm P_{spin}}$)-eccentricity plane) with 2 types of supernovae: electron-capture supernovae systems are thought to produce short spin and orbital periods and low eccentricity (imparting smaller kick velocities to the neutron stars), while iron core-collapse supernovae are thought to be responsible for higher eccentricity systems with slightly higher neutron star masses. We find that most of the known LMC Be-XRBs fall on the bottom left part of this diagram, i.e. pointing to electron-capture supernovae, in agreement with the small estimated center of the mass velocities for these systems. However, the lack of LMC X-ray pulsars at the top right part of the "Corbet diagram" could be the result of a selection effect: it is easier to detect orbital periods of a few weeks than several months or years, and spin periods of a few seconds than $\sim$ 100 sec.

In contrast, \citet{Linden et al. 2009} have found that electron-capture supernovae are highly associated with the production of Be-XRBs for the SMC metallicity, so if the argument of \citet{2011Natur.479..372K} holds then one would expect to find the majority of the SMC Be-XRBs on the lower-left quadrant of the "Corbet diagram". Instead, Fig. \ref{figCorbet} shows that less than half of the systems  are located in the lower-left quadrant with respect to long spin and orbital period systems. On the other hand, \citet{2014ApJ...786..128C} proposed that the bimodal ${\rm P_{spin}}$ distribution is most likely due to the difference in the accretion modes of the NSs in the Be-XRB systems. They also note that the two types of supernovae can affect the spin evolution of the NS through the configuration of the decretion disk of the Be star. Recently, \citet{2015arXiv151100445H} used the most complete census of HMXBs in the SMC to investigate which mechanism is responsible for the bimodal ${\rm P_{spin}}$ distribution. They found larger long-term X-ray variability for sources with short spin periods, thus favoring the accretion model of  \citet{2014ApJ...786..128C} as an explanation of the bimodal ${\rm P_{spin}}$ distribution of Be-XRBs.

\section{SUMMARY}\label{Summary}

In this work, we present an up-to-date list of HMXBs in the LMC. Using the MCPS survey \citep{2004AJ....128.1606Z} and extensive Monte Carlo simulations, we identify the most likely counterpart of X-ray sources identified in previous X-ray surveys of the LMC and estimate their chance coincidence probability. In particular, we classify 40 such systems, further divided in 33 Be-XRBs, 4 SG-XRBs and 3 sources for which we could only assign a broad HMXB class. In addition, we revise the classification of 4 X-ray sources listed in the literature as (candidate) HMXBs to non-HMXB systems. As a by-product of this work, we compiled a list of metallicities available for young OB-type stars in the Magellanic Clouds. 

Using the spatially resolved star-formation history map of the LMC by [HZ09], we investigate the link of star formation and young XRBs. We find that the star-formation history is strongly peaked at $\sim$6.3 Myr for regions hosting HMXBs, SG-XRBs, and the one BH-HMXB (LMC X-1) with MCPS coverage; at $\sim$12.6 Myr for the NS/Be-XRBs and X-ray pulsars; and at $\sim$25.1 Myr for the candidate WD/Be-XRB. We find that the age of the stellar populations in regions associated with pulsars in the LMC is much younger than the age of the same population in the SMC ($25-60$ Myr; \citealt{Antoniou et al. 2010}). Similarly, the age of the stellar populations associated with the only WD/Be-XRB in the LMC is younger than the age of the same population in the SMC ($\sim$42.2 Myr). Thus, this study (in combination with previous works; e.g., \citealt{2005A&A...431..597S}, \citealt{Antoniou et al. 2010}, \citealt{2013ApJ...772...12W}) reinforces the idea that the HMXBs are associated with young stellar populations of ages $\sim$ 10--40 Myr.

Comparing the SFR at the peak of the star-formation episode associated with the NS/Be-XRBs (HMXBs) sources in the LMC and the SMC, we find that their production rate in the LMC is almost 7 (17) times lower than that in the SMC. In particular, we find a production rate of 1 system per $\sim21.4\times10^{-3}$\msunpyr\, and $\sim43.5\times10^{-3}$\msunpyr\, for the NS/Be-XRBs and HMXBs in the LMC, respectively, or 1 system per $\sim$333\msun ($\sim$143\msunsimple) of stars formed during the star-formation episode that is associated with the Be-XRBs (HMXBs). On the other hand, based on the SFR of the SMC at the age of $\sim$10 Myr ($\sim 0.33_{-0.18}^{+0.52}$\msunpyr; Section \ref{SFH-XRB}) and the formation efficiency of LMC SG-XRBs at this age ($\sim85.3_{-44.0}^{+44.7}$ systems/(\mdot); Table \typeout{\ref{FRates}}9), we estimate a large number of $28_{-21}^{+47}$ SG-XRBs in the SMC, which is in stark contrast with the only 2 known such systems in the SMC. Based on similar arguments, the SFR of the SMC ($\sim 0.15_{-0.10}^{+0.70}$\msunpyr; Section \ref{SFH-XRB}), and the formation efficiency of the LMC WD/Be-XRBs (390.3 systems/(\mdot); Table \typeout{\ref{FRates}}9), both at the age of $\sim$25 Myr, we predict $\sim$ 59 SMC WD/Be-XRBs. Although the fact that nowadays we know only one candidate WD/Be-XRB in the LMC points to a strong metallicity dependence, in order to investigate the formation channels  of the WD/Be-XRB systems and to constrain their evolutionary models, many more such systems need to be identified first.

Furthermore, we find a peak at the spectral-type distribution of the LMC Be-XRBs at around B0, in agreement with previous studies (e.g., \citealt{2002A&A...385..517N}, \citealt{Antoniou et al. 2009a}), which have used smaller sample sizes.

We have also examined the mean offset of the identified HMXBs from the nearest young star cluster as a function of the cluster's age (using the [B13] catalog; \citealt{Baumgardt et al. 2013}). We find that the majority of the {\it (a)} HMXBs without BH or WD compact objects, {\it (b)} the confirmed NS/Be-XRBs and {\it (c)} the X-ray pulsars reside near a cluster with age between $20-40$ Myr, while we find almost equal numbers of X-ray pulsars in the $20-30$ Myr and $30-40$ Myr age bins. 

In addition, using the [WZ11] catalog \citep{2011AJ....142...48W}, we find that the 15 confirmed NS/Be-XRBs (14 confirmed X-ray pulsars) have a mean offset of $5.5\arcmin \pm3.7\arcmin$ ($5.0\arcmin \pm 2.5\arcmin$) from their nearest cluster, respectively. Assuming a minimum pulsar birth timescale of $\sim$5 Myr after the burst and an elapsed time of $\sim$7.3 Myr since the kick imparted onto the neutron star  during the supernova explosion for a B0 star, we estimate a transverse velocity of $\sim10.8\pm7.3$\kms\, ($\sim9.7\pm4.9$\kms) for the NS/Be-XRBs (X-ray pulsars) in the LMC. In addition, the 38 HMXBs without BH or WD compact objects have a mean offset of $6.4\arcmin \pm 3.6\arcmin$, thus resulting in transverse velocities of $\sim12.4\pm7.0$\kms. Following the same approach in the case of the SMC, we find that SMC binaries have up to $\sim$4 times smaller transverse velocities (estimated assuming an elapsed time of $\sim$19.9 Myr for a B1 star and the same minimum pulsar birth timescale).

\section*{Acknowledgements}

We thank the anonymous referee and Georgios Vasilopoulos for their insightful comments, which have improved the paper. 
This material is based upon work supported by the National Aeronautics
and Space Administration under Grant No. NNX10AH47G
issued through the Astrophysics Data Analysis Program. VA also acknowledges financial support from NASA/Chandra grants  GO3-14051X  and AR4-15003X. AZ acknowledges financial support from NASA/ADAP grant NNX12AN05G and funding from the European Research
Council under the European Union's Seventh Framework Programme
(FP/2007-2013)/ERC Grant Agreement n. 617001. This research has made
use of the NASA/IPAC Extragalactic Database (NED) which is operated by
the Jet Propulsion Laboratory, California Institute of Technology,
under contract with the National Aeronautics and Space Administration;
the VizieR catalogue access tool, CDS, Strasbourg, France;  the SIMBAD
database, operated at CDS, Strasbourg, France; NASA's Astrophysics
Data System; the Tool for OPerations on Catalogues And Tables (TOPCAT)
software package \citep{2005ASPC..347...29T}. 





\clearpage
\begin{sidewaystable*}\scriptsize
\hspace{-8.5cm}{\bf Table 1.} Compilation of LMC HMXBs${\triangle}$\label{tableCensus}\\
\centering
  \begin{tabular}{@{}llccccccc@{}}
\hline
\multicolumn{2}{c}{X-ray Source}  &  R.A.                   & Dec.              & Pos.Unc.  & Liu's & ${\rm P_{pulse}/P_{orb}}$ & Spectral                 & XRB$\dag$\\
  ID                     &                   Name  &  \multicolumn{2}{c}{(J2000.0)} & (\arcsec)    & ID     &                                                & \multicolumn{2}{c}{Type}\\
   $[1]$                    &                   $[2]$        &       $[3]$                  &  $[4]$                 &  $[5]$             &  $[6]$    &                        $[7]$                    &    $[8]$             &      $[9]$\\  
\hline
 1  & Swift J045106.8-694803, 				 & 04  51  06.8   &  -69  48  03.2   & 3.5 (10)  &    ...    	        & {\color{black} 187.07$\pm$0.04 (70)} /     & B0-1 III-Ve (70) 	& HMXB/Be-XRB? (10), \\[0.05cm]
     &  LXP187                                                             & {\color{black} 04 51 07.0}  &  {\color{black} -69 48 03.1}  &  {\color{black} 3.6 (70)}         &                   &   {\color{black} 21.631$\pm$0.005 (70)}    &                                 & Be-XRB (70) \\[0.05cm]
\hline
2 &   Swift J045558.9-702001                         & 04 55 58.9     &   -70 20 01        &  2.2 (72)        &    ...          &     ...   	$/$    ...         &  B1-2e (72)                   & Be-XRB (72)\\[0.05cm]
\hline
 3$\star$  & RX J0456.9-6824	& 04  56  54.1   &  -68  24  35.0   & 11.1 (38) & 93    		        &    ...   	$/$    ...         &    ...    			& HMXB? (37,38)\\[0.05cm]
\hline
 4$\star$  & RX J0457.2-6612     & 04  57  12.4   &  -66  12  10.0   & 14.5 (38)  & 94   		        &    ...    $/$    ...          &    ...    			& HMXB? (37,38)\\[0.05cm]
\hline
5  & IGR J05007-7047,                   	                   &  05 00  46.08 &  -70  44  36.0   &  0.6 (20) &    ...    	        &    38.55$\pm$0.01$\lhd$ (44)  /   & B2 III (39),   & HMXB {\color{black} (39)}, \\[0.05cm]
     & CXOU J050046.0-704436 (20)                    &                          &                            &                 &                   &    30.77$\pm$0.01 {\color{black} (11)}                  & Be? (24)     & Be-XRB? (24) \\[0.05cm]
\hline
 6  & RX J0501.6-7034,  CAL 9, 		          & 05  01  23.9   &  -70  33  33.0   &  2.9 (2)    &  95   	       	         &    ...    $/$    ...           & B0 Ve (1),                 & Be-XRB (62)\\[0.05cm]
     & 2E 0501.8-7038, 1E 0501.8-7036               &                         &                            &                  &                              &                                            & O8e (42), B0e (43) &                        \\[0.05cm]
\hline
 7  & RX J0502.9-6626, CAL E,                              & 05  02  51.6   &  -66  26  25.0   &  1.2 (2)    & 96   			& 4.0635 (35) $/$    ...     & O9.5-B0 IIIe / B0 Ve (1),  & Be-XRB (62) \\[0.05cm]
     &   LXP4.10                                                           &                         &                            &                  &              		&                                             & O9 V (42),  B0 III(e) (43) &                      \\[0.05cm]
\hline
8$\star$  & RX J0507.6-6847, RX J050736-6847.8      & 05  07  37.9   &  -68  47  49.0   & 7.1 (38)   &  97    			&    ...    $/$    ...    &    ...    	           & HMXB/Be-XRB? (36) \\[0.05cm]
\hline
9       & LXP169,                   &  05 07 55.38           &   -68 25 04.6          &   0.17 (73)          &    ...    		& 168.777$\pm$0.006 (73) $/$  &    ...    & eclipsing \\[0.05cm]
         &  XMMU J050755.4-682505   &                 &                                 &                              &                               &  24.329$\pm$0.008 (73)           &                &  Be-XRB  (73) \\[0.05cm]
\hline
 10$\star$  & RX J0512.6-6717      & 05  12  41.8   &  -67  17  23.0   & 4.0 (49)   &  98   			&    ...    $/$    ...           &    ...    		         & HMXB? (38)\\[0.05cm]
\hline
 11  & Swift J0513.4-6547 (12), LXP27.3		 & 05  13  28.05 &  -65  47  20.0   & 1.9 (12)   &    ...    		& 27.28 (9) $/$ {\color{black} 27.405  (9)} 	&    ...    	                   & HMXB (12), Be-XRB {\color{black} (9)} \\[0.05cm]
\hline
12 & RX J0516.0-6916             				& 05  16  00.1   &  -69  16  09.0   & 7.3 (2)      & 99    			&    ...   	$/$    ...   	& $\sim$B1 V (25) 	& HMXB? (25) \\[0.05cm]
\hline
13 & XMMU J052016.0-692505 	& 05  20  16.0   &  -69  25  05.8  &  3 (8)  &    ...    	&    ...   	$/$    ...           & B0-3 IIIe (8) 	         & likely               \\[0.05cm]
      & 								                   &                         &			 &            &                   & 			    	          &				&  WD/Be-XRB (8)\\[0.05cm]
\hline
14 & RX J0520.5-6932, LXP8.04             		& 05 20 29.8     &  -69 31 55        & 2.9 (74)     &  100  	          & 8.03533$\pm$0.00003	(75) $/$ 24.4 (4)                                 & O8e (62), O9 Ve (4)     & Be-XRB (4)\\[0.05cm]
     &                                                                            &                         &                            &                  &              		& {\color{black} 8.032375$\pm$0.000005 (14)} $/$ 24.4302$\pm$0.0026 (59)  & 				   &                     \\[0.05cm]
     &                                                                            &                         &                            &                  &              		& {\color{black} 8.032932$\pm$0.000005 (14)} $/$ ...  & 				   &                     \\[0.05cm]
\hline
15$\star$ & RX J0523.2-7004     & 05  23  14.9   &  -70  04  12.0   & 9.9 (38)     & 101  			&    ...   	$/$    ...           &    ...    		         & HMXB (38)\\[0.05cm]
\hline
16$\star$ & RX J0524.2-6620     & 05  24  12.7   &  -66  20  50.0   & 7.0 (38)     & 102			&    ...    	$/$    ...           &    ...    			& HMXB (38) \\[0.05cm]
\hline
17$\star,\bullet$ & RX J0527.1-7005     & 05  27  07.9   &  -70  05  00.0   & 11.9 (38)   & 103  		&    ...   	$/$    ...           &    ...    			& HMXB (38)\\[0.05cm]
\hline
18$\star$ & RX J0527.3-6552     & 05  27  23.7   &  -65  52  35.0   & 15.9 (38)   & 104 			&    ...   	$/$    ...           &    ...    			& HMXB? (37,38) \\[0.05cm]
\hline
19$\star$ & RX J0529.4-6952     & 05  29  25.9   &  -69  52  11.0   & 9.8 (38)     & 105  			&    ...   	$/$    ...           &    ...    			& HMXB (37,38) \\[0.05cm]
\hline
20 &   XMMU J052947.4-655639 (6), LXP69.2,   & 05  29  47.4   &  -65  56  39       & 3--4 (6)     & 106 			&  69.232$\pm$0.002 (6) $/$    ...    & B0.5 Ve (1), B2e (5) 	     & Be-XRB (5) \\[0.05cm]
      &  RX J0529.8-6556 (5), RX J0529.7-6556 &                         &                            &                    &                            &                                             &                                   &                           \\[0.05cm] 
\hline
\end{tabular}
\end{sidewaystable*}

\begin{sidewaystable*}\scriptsize
\hspace{-8.5cm}{\bf Table 1}{\it \,\,--  continued}\\
\centering
  \begin{tabular}{@{}llccccccc@{}}
\hline
\multicolumn{2}{c}{X-ray Source}  &  R.A.                   & Dec.              & Pos.Unc.  & Liu's & ${\rm P_{pulse}/P_{orb}}$ & Spectral                 & XRB$\dag$\\
  ID                     &                   Name  &  \multicolumn{2}{c}{(J2000.0)} & (\arcsec)    & ID     &                                                & \multicolumn{2}{c}{Type}\\
   $[1]$                    &                   $[2]$        &       $[3]$                  &  $[4]$                 &  $[5]$             &  $[6]$    &                        $[7]$                    &    $[8]$             &      $[9]$\\  
\hline
21 & 	XMMU J053011.2-655122,  LXP272,            & 05  30  10.6   &  -65  51  27.0   & 2.8 (38)      & 107 			& 271.97$\pm$0.05 (6) $/$    ...    &    ...    		 & Be-XRB? (6)   \\[0.05cm] 
     & RX J0530.1-6551                                                &                         &                           &                     &                            &                         	       	          &                                   &                           \\[0.05cm]
\hline
22 & Swift J053041.9-665426, LXP28.8        &  {\color{black} 05 30 42.17}    & {\color{black} -66 54 31.0}         &  {\color{black} 0.5 (26)}      &    ...                &  $28.77521\pm0.00010$ (26) $/$    ...           &  B0-1.5Ve (26)                   & Be-XRB (26)\\[0.05cm]
\hline
23$\star$ & RX J0530.7-6606     & 05  30  47.4   &  -66  06  15.0   & 11.4 (38)    & 108 	          &    ...   	$/$    ...            &    ...    		 & HMXB (38) \\[0.05cm]
\hline
24 & RX J0531.2-6607,  EXO 0531.1-6609,       & 05  31  13.8   &  -66  07  03.0   & 2.2 (38)      & 109           	& 13.7 (7) / 25.4 (7)              & B0.7 Ve (1)             & Be-XRB (47)\\[0.05cm]
      & LXP13.7, XMMU J053113.3-660705,                          &                          &                           &                     &                            &                                              &                                 &                         \\[0.05cm]
      & EXO 053109-6609.2 (46),                            &                          &                           &                     &                            &                                              &                                 &                         \\[0.05cm]
      & XMMU J053113.1-660707,                           &                          &                           &                     &                            &                                              &                                 &                         \\[0.05cm]
      & 1RXS J053111.0-660657 (19),                    &                          &                           &                     &                            &                                              &                                 &                         \\[0.05cm]
      & IGR J05305-6559 (69)                                 &                          &                           &                     &                            &                                              &                                 &                         \\[0.05cm]
\hline
25 & XMMU J053115.4-705350 			& 05  31  15.4   &  -70  53  50.0   & $\sim$4 (48) & 110 	          &    ...   	$/$    ...             & B0 Ie (71)      	          & HMXB (48, 71)\\[0.05cm]
\hline
26$\star$ & XMMU J053118.2-660730 			& 05  31  18.2   &  -66  07  30      & $\sim$4 (48) &    ...    	&    ...   	$/$    ...             &    ...    		 & HMXB? (48)\\[0.05cm]
\hline
27 & RX J0531.5-6518,            		                   & 05  31  36.1   &  -65  18  16.0    & 18.8 (49)   & 111 		&    ...   	$/$    ...             & B2 V(e?) (1)            & HMXB (49), \\[0.05cm] 
      & 1RXS J053137.1-651759 (19)    		&                         &                             &                     &                           &                                              &                                  & BeXRB? (1)\\[0.05cm]
\hline
28$\star$ & RX J0532.3-7107, CAL 50 (19),                   & 05  32  22.7   &  -71  07  32.0    & 24.3 (38)   & 112 		&    ...   	$/$    ...             &    ...      		  & HMXB? (37,38)\\[0.05cm]
      & 1RXS J053219.5-710806 (19)                    &      	                  &                             &                     &                           &                                              &                                  &                         \\[0.05cm]
\hline
29$\star,\bullet$ & RX J0532.4-6535,                           		& 05  32  25.3   &  -65  35  09.0    & 17.4 (49)   & 113 	          &    ...   	$/$    ...             &    ...    		  & HMXB? (49) \\[0.05cm]
      & 1RXS J053226.0-653505 (19)                    &      	                  &                             &                     &                           &                                              &                                  &                        \\[0.05cm]
\hline
30 & RX J0532.5-6551, XMMUJ053232.4-655139,  & 05  32  32.0   &  -65  51  41 &  8  (51)      & 114 		&    ...    $/$    ...             & B0.5 II (50), 		 & SG-XRB (51)\\[0.05cm]
      &  1RXS J053224.1-655112 (19)                            &                         &                       &                    &                          &                                             & B0 II (1)                    &                        \\[0.05cm]
\hline
31 & 2A 0532-664, LMC X-4, LXP13.5,                &  05 32 49.79  & -66 22 13.8       & 1.02 (68)    & 115                  &  13.5 (30) /                         & O8 III (1),                   & HMXB (58)\\[0.05cm]
      &  RX J0532.8-6622 (19), CAL 49,                  & 05 32 49.4    & -66 22 13           & 0.6 (49)      &                          &  1.408$\pm$0.002 (31)   & O7 III-V (32)             &                       \\[0.05cm]
      & 1RXS J053246.1-662203 (38),    		 &                        &                             &                      &                          &                                   	           &                                  &                       \\[0.05cm] 
      &  4U 0532-66 (57), RASS 232 (49)              &                         &                             &                     &                          &                                   	           &                                  &                       \\[0.05cm]
\hline
32 & Swift J053321.3-684121 (66)      & 05 33 21.3    &  -68 41 21           &  {\color{black} 5.5 (66)}      &    ...                   &    ...    /    ...                   & B3 I (65)                &   HMXB? (66)\\[0.05cm]
\hline
33 & RX J0535.0-6700             				& 05  35  05.9   &  -67  00  15.0    &  4 (49)          & 116 	          &    ...   	/ 241.0 (34)	& B0 Ve (1) 		& Be-XRB (1) \\[0.05cm]
\hline
34 & 1A 0535-668, RX J0535.6-6651, LXP0.07,   & 05  35  41.2   &  -66  51  52.0    & 2 (63)           & 117$\equiv$118 & 0.069 (28) / 	          & B2 IIIe (27),             & Be-XRB (52) \\[0.05cm]
      & CAL G (19),  1RXS J053539.0-665158 (19)     		&                         &                             &                      &                         & 16.668$\pm$0.002 (29) &  B0.5-1 II-III (2)       &                         \\[0.05cm]  
\hline
35$\star,\bullet$ & RX J0535.8-6530,     & 05  35  53.8   &  -65  30  34.0  &  13 (49)  & 119          &    ...    $/$    ...           &    ...    	                 & HMXB? (49)\\[0.05cm]
      & 1RXS J053555.0-653039 (19)  		&                         &                             &                      &                         &                                            &                                 &                       \\[0.05cm]
\hline
\end{tabular}
\end{sidewaystable*}

\begin{sidewaystable*}\scriptsize
\hspace{-8.5cm}{\bf Table 1}{\it \,\,--  continued}\\
\centering
  \begin{tabular}{@{}llccccccc@{}}
\hline
\multicolumn{2}{c}{X-ray Source}  &  R.A.                   & Dec.              & Pos.Unc.  & Liu's & ${\rm P_{pulse}/P_{orb}}$ & Spectral                 & XRB$\dag$\\
  ID                     &                   Name  &  \multicolumn{2}{c}{(J2000.0)} & (\arcsec)    & ID     &                                                & \multicolumn{2}{c}{Type}\\
   $[1]$                    &                   $[2]$        &       $[3]$                  &  $[4]$                 &  $[5]$             &  $[6]$    &                        $[7]$                    &    $[8]$             &      $[9]$\\  
\hline
36 & 1H 0538-641, LMC X-3,                                  & 05 38 56.63  &  -64 05 03.29    & 0.05, 0.08$\rhd$ (67)      & 120         &    ...   	$/$ 1.705 (18)                     & $\sim$B2.5 V pec. (1),     & BH-HMXB (18)  \\[0.05cm]
     & XMMU J053856.7-640503 (19),   		 &                         &                             &                                                                      &                 &                                                                     & B3 V (18)          	&  \\[0.05cm]
      & 1RXS J053855.6-640457 (19),	  		 &                         &                            &                      &                          &                                           &                                  & \\[0.05cm]
      & 1RXS J053855.5-640457 (19),	  		 &                         &                            &                      &                          &                                           &                                  & \\[0.05cm]
      & RX J0538.9-6405, CAL 70 (19),  		 &               		 &                            &                      &                          &                                           &                                  & \\[0.05cm]
      & 3A 0539-641, 4U 0538-64 (19),  		 &               		 &                            &                      &                          &                                           &                                  & \\[0.05cm]
\hline
37 & 3A 0540-697, LMC X-1,                           	& 05 39 38.85  &  -69 44 35.71    & 0.02, 0.02$\rhd$ (67)    & 121  	&    ...   	$/$  			& O8 III-V (1),   	        & BH-HMXB (53,54) \\[0.05cm]
      & 1RXS J053938.8-694515 (19)    		&                        &                             &                                                                     &          & 4.2288$\pm$0.0006 (33) & O8 (f)p (16)          & \\[0.05cm]
\hline
38 & IGR J05414-6858 (13), LXP4.42		&   {\color{black}05 41 26.62}    & {\color{black}-69 01 23.0}      & {\color{black}0.52 (61)}    &    ...              & 4.4208 (61) $/$ 19.9 (61)             & 	B0-1 IIIe (61)	& Be-XRB (61)\\[0.05cm]
\hline
39 & RX J0541.4-6936             				& 05  41  22.2   &  -69  36  29.0    &  6.6 (2)   &  122 			&    ...   	$/$    ...      & B2 SG (2)		& SG-XRB? (2) \\[0.05cm]
      &                                                                           &                          &                            &                 &                               &                                       & B4 III-I (16)          & \\[0.05cm]
\hline
40 & XMMU J054134.7-682550, 			& 05  41  34.7   &  -68  25  50.0    &  4 (48)    & 123 			& $60.77\pm (2.59\times10^{-5})~(60)$   &    ...    		& HMXB? (48), \\[0.05cm]
      & LXP60.8                                                           &                          &                            &                 &                               &   $/$    ...                                                         &                             & Be-XRB? (60) \\[0.05cm]
\hline
41 & RX J0541.5-6833,                            		& 05  41  37.1   &  -68  32  32.0    & 4.5 (64)  & 124 			&    ...   	$/$    ...      & B0 III (17)	   	    & HMXB? (2)\\[0.05cm]
      & RX J0541.6-6832 (40)				&                         &                             &                 &                              &                                        &                                 &                     \\[0.05cm]
\hline
42$\star,\bullet$ & RX J0543.9-6539 (38)  & 05  43  58.0   &  -65  39  52.0  & 8.6 (38)  & 125 	&    ...   	$/$    ...      &    ...    	             & HMXB? (37)\\[0.05cm]
\hline
43 & 1SAX J0544.1-7100,  LXP96.1,      		& 05  44  06.3   &  -71  00  50.0    & 3.3 (49)  & 126 			& 96.08$\pm$0.06 (3) $/$   & B0 Ve (4) 	   & Be-XRB (4)\\[0.05cm]
      & RX J0544.1-7100 (38),                                 &                         &                             &                 &                              &  286.0 (4)                           &                           &                     \\[0.05cm]
      & AX J0548-704, AX J0544.1-7100 (40),     &                         &                             &                 &                              &                                              &                          &                      \\[0.05cm]
      & 1WGA J0544.1-7100                                   &                         &                             &                 &                              &                                              &                          &                      \\[0.05cm] 
\hline
44 & H 0544-665$\ddag$ (41)  & 05 44 15.5$\spadesuit$   &  -66 33 50.0$\spadesuit$ & 30 (56)  & 127 &    ...    $/$    ...    & B0 Ve (55,1)  & Be-XRB (55,56)\\[0.05cm]
\hline
45$\star$ & RX J0546.8-6851, RX J0547.0-6852 (38),       & 05  46  48.3   &  -68  51  47.0  & 48.7 (38)  & 128         &    ...   	$/$    ...     &    ...      		  & HMXB? (38)\\[0.05cm]
                  & 2E 0547.2-6852, 1RXS J054655.3-685142     &                         &                           &                   &                 &                                    &                             &  \\
\hline
\hline
\multicolumn{9}{c}{{\color{black} Additional sources reported in the literature after the completion of this work, thus not taken into account in the present analysis (see Section \ref{Intro})}}\\
\hline
{\color{black} A}               & {\color{black} XMMU J053833.9-691157 (48), CXOU J053833.4-691158 (45)} & {\color{black} 05 38 33.46} &  {\color{black} -69 11 58.7} & {\color{black} 0.4$^{\sq}$  (45)} & ... &    ...   	$/$    ...      & {\color{black} O9 IIIne (23)} 	  & {\color{black} HMXB (23)}\\[0.05cm]
\hline
{\color{black} B}               & {\color{black} Swift J0549.7-6812 (15), LXP6.2 (21)} & {\color{black} 05 50 06.47} & {\color{black} -68 14 55.7} & {\color{black} 1.4 (21)} & ... &  {\color{black} 6.2 (21)} $/$    ...     &    ...      		  & {\color{black} HMXB (22)}\\[0.05cm] 	
\end{tabular}

\begin{minipage}{210mm}
\footnotetext{

\noindent
Notes -- The Right Ascension is given in h m s, the Declination in \degr\, \arcmin\, \arcsec, the pulse period ${\rm P_{pulse}}$ in seconds, and the orbital period ${\rm P_{orb}}$ in days. References are given in parenthesis and are listed below.\\
$\triangle$ Based on the literature as of Dec. 2014.\\
$\dag$ This is the classification presented in the literature. In this work, we re-classify some of these sources and the final classification is given in Table \typeout{\ref{finalclass}}5.\\
$\star$ New counterparts for these X-ray sources are identified in this study.\\
$\lhd$ This pulse period is not taken into account in the present work; see comments on last paragraph of Introduction and footnote \#4.\\
$\bullet$ These sources have been re-classified as non-HMXB systems.\\
$\rhd$ Positional uncertainty in  Right Ascension, followed by that in Declination.\\
$\ddag$ We believe this source was erroneously related to 1RXS J054450.0-663445 \citep{Ebisawa et al. 2003}. The same correlation appears in the Simbad database too. In particular, this ROSAT All Sky Survey Faint Source is approximately located at 4.1\arcmin\, and 3.6\arcmin\, from the X-ray ((R.A., Decl.)${\rm _{B1950.0}}$=(05:44:11.5, -66:35:24.0) with 60\arcsec\, positional uncertainty) and optical ((R.A., Decl.)${\rm _{B1950.0}}$=(05:44:15.6, -66:34:59.0) with a 3\arcsec\, positional uncertainty of the optical match) positions, respectively, thus we propose  that these are 2 different sources. \citet{1999MNRAS.309..421S} give this source as (R.A., Decl.)${\rm _{J2000.0}}$=(05:44:15.5, -66:33:50.0) with a 30\arcsec\, positional uncertainty (at a 90\% confidence level).\\
$\spadesuit$ Transformed from (R.A., Decl.)${\rm_{B1950.0}}$=(05:44:15.6, -66:34:59.0) using NASA/IPAC Extragalactic Database (NED; https://ned.ipac.caltech.edu/forms/calculator.html). We note that in \citet{2005A&A...442.1135L} the B1950.0 coordinates were erroneously listed as J2000.0.\\
$^{\sq}$ Formal 1$\sigma$ radial positional error.

    }
    
\end{minipage}

\end{sidewaystable*}

\begin{sidewaystable*}\scriptsize
\hspace{-8.5cm}{\bf Table 1}{\it \,\,--  continued}\\

\begin{minipage}{170mm}
\footnotetext{

\noindent
{\bf References}: (1)
\citet{2002A&A...385..517N}; (2) \citet{2000A&AS..143..391S}; (3)
\citet{1998A&A...337..772C}; (4) \citet{2001MNRAS.324..623C}; (5)
\citet{1997A&A...318..490H}; (6) \citet{2003A&A...406..471H}; (7)
\citet{1995IAUC.6184....2D}; (8) \citet{2006A&A...458..285K}; (9)
\cite{2015MNRAS.447.1630C}; (10) \citet{2009ATel.1901....1B}; (11) \citet{2011A&A...529A..30D}; (12)
\citet{2009ATel.2011....1K}; (13) \citet{2010ATel.2695....1G}; (14) \citet{2014ApJ...795..154T}; (15) \citealt{2013ATel.5286....1K}; (16)
\citet{2009AJ....138..510F}; (17) \citet{2002ApJS..141...81M}; (18)
\citet{Cowley et al. 1983}; (19) SIMBAD: Identifier query; (20)
\citet{2005A&A...444L..37S}; (21) \citet{2013ATel.5309....1K}; (22) \citet{2013ATel.5293....1K}; (23) \citet{2015A&A...579A.131C}; (24) \citet{2010ATel.2601....1S}; (25)
\citet{Cowley et al. 1997}; (26) \citet{2013A&A...558A..74V}; (27)
\citet{1985MNRAS.212..565C}; (28) \citet{1982Natur.297..568S}; (29)
\citet{1980Natur.288..141S}; (30) \citet{1983ApJ...264..568K}; (31)
\citet{1977A&A....59L...9C}; (32) \citet{1978ApJ...225..548H}; (33)
\citet{1987AJ.....94..340H}; (34) \citet{1988MNRAS.232...53R}; (35)
\citet{1995PASP..107..450S}; (36) \citet{2000AJ....119.2242C}; (37)
\citet{2002A&A...388..100K}; (38) \citet{1999A&AS..139..277H}; (39)
\citet{2006A&A...459...21M}; (40) \citet{Ebisawa et al. 2003}; (41)
\citet{Johnston et al. 1979}; (42) \citet{Cowley et al. 1984}; (43)
\citet{1985AJ.....90...43C}; (44) Vasilopoulos \etal (2016, to be
subm.; private communication); (45) \citet{2006AJ....131.2164T}; (46) \citet{2007A&A...467..585B}; (47) \citet{1995A&A...302L...1H}; (48) \citet{2005A&A...431..597S}; (49) \citet{1999A&A...344..521H}; (50) \citet{2001PASP..113.1130J}; (51) \citet{1995A&A...303L..49H}; (52) \citet{1983MNRAS.202..657C}; (53) \citet{1983ApJ...275L..43H}; (54) \citet{1984ApJ...281..354W}; (55) \citet{1983MNRAS.203..279V}; (56) \citet{1999MNRAS.309..421S}; (57) \citet{2010ApJ...715..947T}; (58) \citet{1984ARA&A..22..537J}; (59) \citet{2014A&A...567A.129V}; (60) \citet{2009MNRAS.395.1662I}; (61) \citet{2012A&A...542A.109S}; (62) \citet{1994PASP..106..843S}; (63) \citet{Johnston et al. 1980}; (64) \citet{1975A&AS...21..109B}; (65) \citet{1978A&AS...31..243R}; (66) \citet{2012ATel.3993....1S}; (67) \citet{2002ApJ...576..357C}; (68) \citet{2003A&A...403..247F}; (69) \citet{2013MNRAS.428...50G}; (70) \citet{2013MNRAS.428.3607K}; (71) \citet{2012MNRAS.425..355R}; (72) \citet{2013ATel.5540....1V}; (73) \citet{2013A&A...554A...1M}; (74) \citet{2013ATel.4748....1V}; (75) \citet{2013ATel.5673....1V}.

 }
 \end{minipage}

\end{sidewaystable*}

\clearpage
\begin{sidewaystable*}
\hspace{-8.5cm}{\bf Table 2.} Optical properties of HMXBs\label{tableOpt}\\
\centering
\begin{tabular}{@{}lcccccccccccccc@{}}
\hline
Src ID & R.A. & Decl. & U & errU & B & errB & V & errV & I & errI & B-V & err(B-V) & Offset & C/part\\
            & (h\, m\, s) & (\degr\, \arcmin\, \arcsec)  &  \multicolumn{10}{c}{(mag)} & (\arcsec)& \\[0.2cm]
$[1]$ & $[2]$ & $[3]$ & $[4]$ & $[5]$ & $[6]$ & $[7]$ & $[8]$ & $[9]$ & $[10]$ & $[11]$ & $[12]$ & $[13]$ & $[14]$ & $[15]$\\
\hline
1   &  04 51 06.10 & -69 48 01.0   &  20.676  &  0.262  &   21.408  &  0.143  &    20.767  &  0.116  &       ...    &    ...    &  0.641 & 0.184  & 4.23 &    ...    \\[0.05cm]
     &  04 51 06.30 & -69 48 04.9   &  19.932  &  0.192  &   20.368  &  0.082  &    19.084  &  0.048  &    18.104  &  0.052    &  1.284 & 0.095  & 3.10 &    ...    \\[0.05cm]
     &  \textbf{04 51 06.89} & \textbf{-69 48 02.4}   &  \textbf{13.736}  &  \textbf{0.038}  &   \textbf{14.702}  &  \textbf{0.024}  &    \textbf{14.600}       &  \textbf{0.023}  &    \textbf{14.512}  &  \textbf{0.036}    &  \textbf{0.102} & \textbf{0.033}  & \textbf{0.91} & k \\[0.05cm] 
\hline
\textbf{2} & \textbf{04 55 58.89} & \textbf{-70 19 59.8} & \textbf{13.179}   &   \textbf{0.044} &  \textbf{14.104} &  \textbf{0.108}   &  \textbf{14.153}   & \textbf{0.042}  &  \textbf{14.072}  &  \textbf{0.051} & \textbf{-0.049}  &  \textbf{0.116}  &  \textbf{1.19}	&  k \\[0.05cm]
 \hline
 3  & 04 56 53.88 & -68 24 33.0   &    ...     &    ...    &  20.376  &  0.102  &  20.604   &  0.119   &  18.839   &  0.083     & -0.228 & 0.157 & 2.32 &    ...    \\[0.05cm]     
     & 04 56 54.14 & -68 24 38.9    &    ...     &    ...    &  22.354  &  0.390  &  21.147   &  0.143   &  21.553   &  0.438     &  1.207 & 0.415 & 3.89 &    ...    \\[0.05cm]   
     & 04 56 54.72 & -68 24 37.5   &    ...     &    ...    &  21.341  &  0.141  &  20.904   &  0.220   &     ...     &    ...      &  0.437 & 0.261 & 4.28  &    ...    \\[0.05cm] 
     & 04 56 54.75 & -68 24 32.8    &    ...     &    ...    &  21.659  &  0.200  &  20.826   &  0.119   &  20.600   &  0.189     &  0.833 & 0.233  & 4.23 &    ...    \\[0.05cm]
 \hline
 4  & 04 57 12.89 & -66 12 10.0    &    ...     &    ...    &  22.221  & 0.203    &  21.391   &  0.133   &  19.541   &  0.070    &  0.830  &  0.243  &  2.95 &    ...    \\[0.05cm]
 \hline
 5  & \textbf{05 00 46.08} & \textbf{-70 44 35.8}    & \textbf{16.676}  & \textbf{0.091}     &  \textbf{14.672}  & \textbf{0.028}     &  \textbf{14.727}  & \textbf{0.021}    &   \textbf{14.622}  & \textbf{0.030}   &  \textbf{-0.055}   & \textbf{0.035}  & \textbf{0.33}  & k  \\[0.05cm]
 \hline
 6   & 05 01 23.73 & -70 33 29.4    & 20.351  &  0.176    &  20.088  &  0.094   &  18.187  &  0.143  &  17.587  &  0.102    &   1.901  &   0.171  &  3.66 &    ...    \\[0.05cm]   
      & \textbf{05 01 23.85} & \textbf{-70 33 33.6}    & \textbf{13.839}  &  \textbf{0.038}    &  \textbf{14.806}  &  \textbf{0.042}   &  \textbf{14.982}  &  \textbf{0.026}  &  \textbf{15.209}  &  \textbf{0.091}    &  \textbf{-0.176}  &   \textbf{0.049}  &  \textbf{0.60} & k \\[0.05cm]  
      & 05 01 24.23 & -70 33 34.3   &    ...     &    ...    &  17.913  &  0.053   &  16.528  &  0.040   &  15.062  &  0.042    &   1.385   &  0.066   &  2.09 &    ...    \\[0.05cm]           
 \hline
 7  &  \textbf{05 02 51.82} & \textbf{-66 26 26.8}   & \textbf{13.051}   &  \textbf{0.012}   &  \textbf{14.171}  &  \textbf{0.026}   &  \textbf{14.364}   &  \textbf{0.193}  & \textbf{13.988}    &  \textbf{0.035}  &  \textbf{-0.193}  &  \textbf{0.195}  & \textbf{2.22}  & k \\[0.05cm] 
     &  05 02 51.96 & -66 26 20.8   & 19.198   &  0.151   &  20.637  &  0.103   &  19.983   &  0.088   &  19.454  &  0.112  &    0.654  &  0.135  & 4.77 &    ...    \\[0.05cm] 
 \hline
 8  &  05 07 37.68 & -68 47 51.9   &  17.636  &  0.070   &  17.008  &  0.098   &   17.577  &  0.052   &   17.706  &  0.062  & -0.569   &   0.111  &  3.10 &    ...    \\[0.05cm] 
     &  05 07 37.82 & -68 47 48.6   &  18.545  &  0.093   &  17.176  &  0.179   &   17.139  &  0.125   &   16.584  &  0.054  &    0.037  &  0.218  &  0.62  &    ...    \\[0.05cm]        
     &  \textbf{05 07 38.18} & \textbf{-68 47 47.8}   &  \textbf{15.869}  &  \textbf{0.050}    &  \textbf{16.180}     &  \textbf{0.110}     &   \textbf{16.260}    &  \textbf{0.060}      &   \textbf{16.448}  &  \textbf{0.099}  &  \textbf{-0.080}  &  \textbf{0.125}   &  \textbf{1.93}  & n \\[0.05cm] 
     &  05 07 38.23 & -68 47 45.3   &  16.065  &  0.052   &  16.323  &  0.114   &   16.427  &  0.067   &   16.601  &  0.046  &  -0.104  &   0.132  &  4.12 &    ...    \\[0.05cm] 
     &  05 07 38.25 & -68 47 52.3   &  17.598  &  0.080   &  18.084  &  0.050   &  18.046   & 0.044    &   18.248  &  0.053  &  0.038   &   0.067  &  3.85  &    ...    \\[0.05cm]  
 \hline
9   &   \textbf{05 07 55.49} & \textbf{-68 25 04.8} & \textbf{14.018} & \textbf{0.080} & \textbf{15.016} & \textbf{0.022} & \textbf{14.945} & \textbf{0.026}  & \textbf{14.894} & \textbf{0.038} & \textbf{0.071} & \textbf{0.034} &  \textbf{0.64}  & k \\[0.05cm] 
           &  05 07 55.80 & -68 25 08.0 & 20.317 & 0.175 & 18.513 & 0.061 & 17.046  & 0.035 & 15.672 & 0.186 & 1.467 &  0.070 & 4.08  &    ...    \\[0.05cm]  
\hline
 10  & 05 12 41.26 & -67 17 26.8    &  20.222  &  0.181   &   18.707   &  0.038   &  17.327   &  0.028   &   15.764  &  0.025  &   1.380  &  0.047   &  4.89 &    ...    \\[0.05cm]
     &  \textbf{05 12 41.33} & \textbf{-67 17 23.4}   &  \textbf{15.369}  &  \textbf{0.024}   &   \textbf{16.090}      &  \textbf{0.089}   &  \textbf{16.187}   &  \textbf{0.024}   &  \textbf{16.394}   &  \textbf{0.054}  & \textbf{-0.097}  &  \textbf{0.092}   &  \textbf{2.74}  & n \\[0.05cm]
     &  05 12 41.49 & -67 17 18.4   &  18.959  &  0.100   &   19.425   &  0.064   &  19.409   &  0.052   &   19.213  &  0.055  &   0.016  &  0.082   &  4.93 &    ...    \\[0.05cm]
     &  05 12 42.31 & -67 17 23.1   &  16.618  &  0.042   &   17.205    &  0.03     &  17.254   &  0.027   &  17.407   &  0.045  & -0.049  &  0.040   &  2.94 &    ...    \\[0.05cm]
 \hline
 11  & 05 13 27.60 & -65 47 16.2    &    ...     &    ...    &  21.210   &  0.116   &  20.537  &  0.111   &     ...    &    ...    &  0.673    &  0.161   &  4.72  &    ...    \\[0.05cm]   
     & 05 13 28.01 & -65 47 24.1    &  20.405  &  0.169    &  20.389  &  0.070    &  20.039  &  0.074  &  20.112  &  0.119    &  0.350    &  0.102   &  4.08  &    ...    \\[0.05cm]   
\hline
\end{tabular}
\end{sidewaystable*}

\begin{sidewaystable*}
\hspace{-8.5cm}{\bf Table 2}{\it \,\,--  continued}\\
\centering
\begin{tabular}{@{}lcccccccccccccc@{}}
\hline
Src ID & R.A. & Decl. & U & errU & B & errB & V & errV & I & errI & B-V & err(B-V) & Offset & C/part\\
            & (h\, m\, s) & (\degr\, \arcmin\, \arcsec)  &  \multicolumn{10}{c}{(mag)} & (\arcsec)& \\[0.2cm]
$[1]$ & $[2]$ & $[3]$ & $[4]$ & $[5]$ & $[6]$ & $[7]$ & $[8]$ & $[9]$ & $[10]$ & $[11]$ & $[12]$ & $[13]$ & $[14]$ & $[15]$\\
\hline
    & \textbf{05 13 28.26} & \textbf{-65 47 18.4}    &  \textbf{14.122}  &  \textbf{0.036}    &  \textbf{15.067}  &  \textbf{0.088}   &  \textbf{15.098}  &  \textbf{0.027}   &  \textbf{15.328}  &  \textbf{0.039}    & \textbf{ -0.031}   &  \textbf{0.092}   &  \textbf{2.06} & k \\[0.05cm]   
\hline
12  &  05 15 59.58 & -69 16 05.7    & 20.489  &  0.316    &  20.013  &  0.189   &  19.186  & 0.269   &  18.485  &  0.284   &   0.827   &   0.329    &   4.32  &    ...    \\[0.05cm]
      &  05 15 59.67 & -69 16 11.0   &    ...    &    ...    &  19.056  &  0.069   &  17.603   &  0.100  &  16.533  &  0.089   &   1.453   &   0.121    &   3.01  &    ...    \\[0.05cm]
      & \textbf{05 15 59.92} & \textbf{-69 16 07.6}    & \textbf{14.228}  &  \textbf{0.031}   &  \textbf{15.011}   &  \textbf{0.029}   &  \textbf{15.126}  &  \textbf{0.041}  &  \textbf{14.99}    &  \textbf{0.049}   &  \textbf{-0.115}   &   \textbf{0.050}    &   \textbf{1.66}  & k,A \\[0.05cm]
      &  05 16 00.39 & -69 16 06.0   & 18.645  &  0.088   &  18.954   &  0.071   &  18.928  &  0.152  &  17.93    &  0.092   &   0.026   &   0.168    &  3.42 &    ...    \\[0.05cm]
      &  05 16 00.79 & -69 16 09.6   &    ...    &    ...    & 18.254   &  0.062    &  16.783  &  0.067  &  15.449  &  0.051   &  1.471    &   0.091   &   3.70 &    ...    \\[0.05cm]
 \hline
13  &  05 20 15.33 & -69 25 02.5   & 19.563  & 0.147    &   18.531  & 0.071  &    17.202  & 0.074    &   16.377  & 0.043  &   1.329   &   0.103  &   4.33  &    ...    \\[0.05cm]
      &  \textbf{05 20 16.15} & \textbf{-69 25 05.3}   & \textbf{13.905}   & \textbf{0.031}    &  \textbf{14.785}  & \textbf{0.031}   &    \textbf{14.944}  & \textbf{0.065}    &   \textbf{14.665}  & \textbf{0.073}  &  \textbf{-0.159}   &  \textbf{0.072}   &  \textbf{0.86} & k \\[0.05cm]
      &  05 20 16.83 & -69 25 07.0    &    ...     &    ...    &  20.297  & 0.129   &   18.587  & 0.136    &   19.366  & 0.148  &    1.710   &  0.187   &  4.80 &    ...    \\[0.05cm]
 \hline
14$\spadesuit$     &  \textbf{05 20 29.70}  &  \textbf{-69 31 55.24}  &     ...     &    ...     &  \textbf{14.312}  &   \textbf{0.087} &  \textbf{16.004} &   \textbf{0.471} &     ...     &    ...     &  \textbf{-1.692} &  \textbf{0.479}  &  \textbf{0.60}   & k,A \\[0.05cm]
         &  05 20 29.91  &  -69 31 51.35  &   18.306  &  0.095  &  18.442  &   0.114  & 17.612  &  0.153   & 18.429    &  0.137     &  0.830  &  0.191  &  3.70   &  \\[0.05cm]
 \hline
15  & 05 23 14.02 & -70 04 13.9   &    ...     &    ...    &  19.003  &  0.170   &  18.795  &  0.120  &  19.437  &  0.119    &   0.208  &   0.208  &  4.85 &    ...    \\[0.05cm]
      & 05 23 14.29 & -70 04 13.2  &    ...     &    ...    &  19.023  &  0.125  &  18.599  &  0.077  &    ...     &    ...     &   0.424  &   0.147  &  3.32  &    ...    \\[0.05cm]
      & 05 23 14.62 & -70 04 07.4  &    ...     &    ...    &  20.876  &  0.260  &   20.111  &  0.191  &  20.250  &  0.224    &   0.765  &   0.323  &  4.82  &    ...    \\[0.05cm]
      & 05 23 14.71 & -70 04 15.0  & 18.631   &  0.085   &  18.729  &  0.074  &   18.095  &  0.158  &  18.345 &   0.089    &  0.634   &   0.174  &  3.14   &    ...    \\[0.05cm]
      & 05 23 15.05 & -70 04 11.7  & 19.594   &  0.137   &  19.745  &  0.153  &   18.555  &  0.120   &  18.088  &  0.084    &   1.190  &   0.194  &  0.85 &    ...    \\[0.05cm]
      & \textbf{05 23 15.58} & \textbf{-70 04 11.1}  & \textbf{16.573}   &  \textbf{0.052}   &  \textbf{15.991}  &  \textbf{0.080}  &   \textbf{15.129}  &  \textbf{0.063}  &  \textbf{14.094}  &  \textbf{0.073}    &   \textbf{0.862}  &   \textbf{0.102}  &  \textbf{3.62}  & n,A \\[0.05cm]
 \hline
16 & 05 24 12.46 & -66 20 50.4  & 19.032  &  0.082   &    19.093  &  0.035   &   19.190  &   0.073   &  19.120   &  0.056    &  -0.097  &   0.081  & 1.49  &    ...    \\[0.05cm] 
 \hline
17 & \textbf{05 27 06.99} & \textbf{-70 04 58.4}    & \textbf{17.373} &  \textbf{0.047}   &   \textbf{16.516}  &  \textbf{0.043}    &   \textbf{15.308}  &  \textbf{0.050}   &  \textbf{13.713}   &  \textbf{0.051}    &   \textbf{1.208}  &   \textbf{0.066}  &   \textbf{4.91}  & n,A \\[0.05cm]
      & 05 27 08.00 & -70 05 03.6    & 21.121 &  0.427   &   21.455  &  0.263    &    20.790  &  0.285   &    ...     &    ...     &    0.665 &   0.388  &  3.59  &    ...    \\[0.05cm] 
      & \textbf{05 27 08.17} & \textbf{-70 05 00.3}   & \textbf{20.754} &  \textbf{0.272}   &  \textbf{19.899}   &  \textbf{0.095}    &   \textbf{19.83}    &  \textbf{0.31}     &     ...     &    ...     &  \textbf{ 0.069}  &  \textbf{0.324}  &   \textbf{1.40}  & n,A \\[0.05cm]
      &  05 27 08.25 & -70 04 58.1   & 20.998 &  0.360   &   20.380   &  0.132    &   19.994  &  0.191  &     ...     &    ...     &   0.386  &  0.232  &   2.59 &    ...    \\[0.05cm]
 \hline
18  & 05 27 23.14 & -65 52 38.0   & 20.643  &  0.152    &  20.603 &  0.066   &  20.407  &  0.051   &   20.198  &  0.126    &  0.196    &    0.083  &  4.53  &    ...    \\[0.05cm]
      & 05 27 23.26 & -65 52 35.8    &    ...    &    ...    &  22.513 &  0.304   &  21.787  &  0.150    &     ...     &    ...     &  0.726   &    0.339  &  2.92 &    ...    \\[0.05cm]  
      & 05 27 23.54 & -65 52 32.0    & 21.723  &  0.335   &  21.671 &  0.128   &  21.192  &  0.126    &  20.662   & 0.166     &  0.479    &   0.180  &  3.15  &    ...    \\[0.05cm] 
      & 05 27 24.18 & -65 52 31.0    & 21.467  &  0.339   &  22.292 &  0.211   &  21.146  &  0.126    &     ...     &    ...     &  1.146    &   0.246  &  4.98  &    ...    \\[0.05cm] 
      & 05 27 24.21 & -65 52 36.6   &    ...    &    ...    &  22.361 &  0.192   &  22.029  &  0.216   &     ...     &    ...      &  0.332    &   0.289  &  3.54   &    ...    \\[0.05cm] 
\hline
19 &  05 29 25.47 & -69 52 06.7   &  21.021 &  0.266   &  20.185  &  0.115    &  19.542  &  0.082   &   18.668  &   0.093    &  0.643   &  0.141  &  4.84   &    ...    \\[0.05cm] 
      &  05 29 26.10 & -69 52 13.3  &  21.217 &  0.512   &  20.531  &  0.177   &   19.804  &  0.137   &  18.779   &  0.172     &  0.727   &  0.224  &  2.47    &    ...    \\[0.05cm] 
\hline
\end{tabular}
\end{sidewaystable*}

\begin{sidewaystable*}
\hspace{-8.5cm}{\bf Table 2}{\it \,\,--  continued}\\
\centering
  \begin{tabular}{@{}lcccccccccccccc@{}}
\hline
Src ID & R.A. & Decl. & U & errU & B & errB & V & errV & I & errI & B-V & err(B-V) & Offset & C/part\\
            & (h\, m\, s) & (\degr\, \arcmin\, \arcsec)  &  \multicolumn{10}{c}{(mag)} & (\arcsec)& \\[0.2cm]
$[1]$ & $[2]$ & $[3]$ & $[4]$ & $[5]$ & $[6]$ & $[7]$ & $[8]$ & $[9]$ & $[10]$ & $[11]$ & $[12]$ & $[13]$ & $[14]$ & $[15]$\\
\hline
   &  05 29 26.22 & -69 52 10.3   &    ...    &    ...    &  19.757  &  0.126   &   20.085  &  0.145   &     ...     &    ...     &  -0.328  &  0.192  &  1.78   &    ...    \\[0.05cm] 
      &  05 29 26.48 & -69 52 08.0   &    ...    &    ...    &  20.385  &  0.166   &   20.142  &  0.170   &     ...     &    ...     &   0.243   &  0.238  &  4.22  &    ...    \\[0.05cm] 
\hline
20  & 05 29 48.10 & -65 56 55.2    & 20.949   & 0.213     &  21.333  &  0.091  &   20.951  & 0.097   &   20.734  &  0.170     &   0.382   &  0.133  &  4.60  &    ...    \\[0.05cm]
       & 05 29 48.45 & -65 56 50.5    &    ...     &    ...    &  20.985  &  0.082  &   20.772  & 0.106   &   20.615  &  0.178     &   0.213   &  0.134  &  0.58  &    ...    \\[0.05cm]
       & 05 29 48.66 & -65 56 46.5    &    ...     &    ...    &  22.094  &  0.172  &   20.711  & 0.125   &    ...      &    ...      &   1.383   &  0.213  &  4.73  &    ...    \\[0.05cm]  
 \hline
21 &      ...        &     ...            &     ...     &    ...    &    ...     &    ...    &     ...     &    ...    &    ...      &    ...      &    ...     &    ...      &    ...     &    ...     \\[0.05cm]
\hline
22  & 05 30 41.39 & -66 54 26.7  & 21.529  &  0.470   & 21.665   &  0.179   &  22.032   &  0.240    & 20.662    &  0.235     &  -0.367   &  0.299      &  3.80  &    ...     \\[0.05cm]
       & \textbf{05 30 42.13} & \textbf{-66 54 30.1}  & \textbf{14.312}  & \textbf{0.022}    & \textbf{15.007}   &  \textbf{0.033}   &  \textbf{15.321}   &  \textbf{0.018}    & \textbf{15.586}    & \textbf{0.023}      &  \textbf{-0.314}   &  \textbf{0.038}      &  \textbf{1.71}  & k \\[0.05cm]
 \hline
23 & 05 30 47.25 & -66 06 15.0  & 21.734  & 0.283  & 20.852 &  0.084 &  20.684  &  0.108   &    ...      &    ...      &   0.168 &  0.137  &   0.93  &    ...    \\[0.05cm]   
      & 05 30 47.84 & -66 06 17.6   &  21.174  &  0.248   &  21.485 &  0.121  &  21.576   &  0.200     &    ...      &    ...      &  -0.091  &  0.234  &  3.74 &    ...    \\[0.05cm]  
      & 05 30 47.92 & -66 06 16.3   &    ...     &    ...    &  22.493 &  0.217  &  22.127  &  0.345   &    ...      &    ...      &   0.366  &  0.408  &  3.42   &    ...    \\[0.05cm]   
 \hline
24 & 05 31 13.61 & -66 07 04.5    & 20.479   & 0.155    &    20.070  &  0.103   &   20.288   &  0.110   &    ...     &    ...     &  -0.218 &  0.151 &   1.89   &    ...    \\[0.05cm]    
 \hline
25 &  \textbf{05 31 15.53} & \textbf{-70 53 48.5}    & \textbf{13.158}   & \textbf{0.038}    &   \textbf{14.009}  &  \textbf{0.035}    &  \textbf{14.051}   &  \textbf{0.025}  &  \textbf{13.799}  &  \textbf{0.028}     &  \textbf{-0.042} &  \textbf{0.043}  &   \textbf{1.64}   & k,A \\[0.05cm]
 \hline
26   &  05 31 18.10 & -66 07 31.6   &  20.965  &  0.210  &   20.390   &  0.065    &  20.097   &  0.145  &  19.769   &  0.294    &   0.293  &   0.159  & 1.69  &    ...    \\[0.05cm] 
        & 05 31 18.12 & -66 07 28.9    & 18.078  &  0.046   &   18.160   &  0.075    &  18.157   &  0.035  &  18.117   &  0.073    &   0.003  &   0.083  &  1.18 &    ...     \\[0.05cm] 
        & \textbf{05 31 18.66} & \textbf{-66 07 31.6}    & \textbf{16.471}  &  \textbf{0.034}   &   \textbf{16.804}  &  \textbf{0.021}    &  \textbf{16.873}   &  \textbf{0.023}  &    ...     &    ...     &  \textbf{-0.069}  &   \textbf{0.031}  &  \textbf{3.21}  & n,A \\[0.05cm] 
 \hline
27 & 05 31 36.20 & -65 18 17.5  &    ...     &    ...    &  21.921  &  0.163   &  21.731  &  0.239   &    ...     &    ...     &   0.190  &   0.289  &  1.61  &    ...     \\[0.05cm] 
      &  \textbf{05 31 36.81} & \textbf{-65 18 16.1}   & \textbf{15.401}   & \textbf{0.031}   &  \textbf{15.875}  &  \textbf{0.029}   &  \textbf{16.007}  &  \textbf{0.021}   &  \textbf{16.226}  &  \textbf{0.153}    &  \textbf{-0.132}  &  \textbf{0.036}  &  \textbf{4.48} & k \\[0.05cm] 
 \hline
28 &  05 32 21.72 & -71 07 31.4  &  20.456  &  0.202    &  20.580   &  0.091   &  20.179  &  0.061   &   19.879  &  0.105    &    0.401 &   0.110  &  4.77 &    ...    \\[0.05cm]   
      &  05 32 22.92 & -71 07 30.7  &    ...     &    ...    &  21.411  &  0.150   &  21.024  &  0.116   &     ...     &    ...     &    0.387  &   0.190  &  1.71 &    ...    \\[0.05cm]    
      &  05 32 23.39 & -71 07 33.6  &    ...     &    ...    &  20.855  &  0.111   &  21.053  &  0.133   &     ...     &    ...     &  -0.198  &   0.173  &  3.73  &    ...    \\[0.05cm]  
      &  05 32 23.45 & -71 07 31.5  & 20.348  &  0.275   &  20.769  &  0.129   & 19.437   &  0.067   &   18.471  &  0.052    &    1.332  &   0.145  &   3.69  &    ...    \\[0.05cm]  
 \hline
29  & 05 32 24.93 & -65 35 05.5    &    ...     &    ...    &  22.509  &  0.217   &  22.432  &  0.298   &  20.852   &  0.28     &   0.077   &   0.369   &  4.20 &    ...    \\[0.05cm]  
      & 05 32 25.44 & -65 35 07.9    &  19.613  &  0.112   &  20.525  &  0.054   &  19.983  &  0.059   &  19.008   &  0.066   &   0.542   &   0.080   &  1.38 &    ...    \\[0.05cm]
 \hline
30 &  \textbf{05 32 32.62} & \textbf{-65 51 40.2}   & \textbf{12.212}  &  \textbf{0.038}    &  \textbf{12.474}  &  \textbf{0.138}    &  \textbf{13.048}  &  \textbf{0.076}   &  \textbf{13.162}  &  \textbf{0.146}   &  \textbf{-0.574}  & \textbf{0.158}  &  \textbf{3.85} & k,A \\[0.05cm]
 \hline
31 & \textbf{05 32 49.54} &  \textbf{-66 22 12.9} & \textbf{15.449} &  \textbf{0.070} &   \textbf{13.940} &   \textbf{0.029} &  \textbf{14.167} &  \textbf{0.029} &    ...     &    ...     & \textbf{-0.227} &  \textbf{0.041} & \textbf{1.72} &  k,A \\[0.05cm]
 \hline
32  & \textbf{05 33 20.80} & \textbf{-68 41 23.2}  & \textbf{11.673}  &    ...    & \textbf{12.684} &    ...    & \textbf{12.776} &    ...    &  \textbf{12.239} & \textbf{0.142}    & \textbf{-0.092} &    ...    &  \textbf{3.50} & k \\[0.05cm]      
                      &   05 33 21.65 & -68 41 19.3  &    ...    &    ...      & 20.801 & 0.422   & 19.571 & 0.304    &    ...    &    ...    & 1.230  & 0.520    & 2.55 &    ...    \\[0.05cm]
\hline
33   & \textbf{05 35 05.99} & \textbf{-67 00 15.4}    & \textbf{13.941}   & \textbf{0.026}    &  \textbf{14.728}  &  \textbf{0.029}  &   \textbf{14.873}  &  \textbf{0.023}  &  \textbf{14.910}     & \textbf{0.032}   &  \textbf{-0.145}  &  \textbf{0.037}  &  \textbf{0.65}  & k  \\[0.05cm]
\hline
\end{tabular}
\end{sidewaystable*}

\begin{sidewaystable*}
\hspace{-8.5cm}{\bf Table 2}{\it \,\,--  continued}\\
\centering
  \begin{tabular}{@{}lcccccccccccccc@{}}
\hline
Src ID & R.A. & Decl. & U & errU & B & errB & V & errV & I & errI & B-V & err(B-V) & Offset & C/part\\
            & (h\, m\, s) & (\degr\, \arcmin\, \arcsec)  &  \multicolumn{10}{c}{(mag)} & (\arcsec)& \\[0.2cm]
$[1]$ & $[2]$ & $[3]$ & $[4]$ & $[5]$ & $[6]$ & $[7]$ & $[8]$ & $[9]$ & $[10]$ & $[11]$ & $[12]$ & $[13]$ & $[14]$ & $[15]$\\
\hline
      & 05 35 06.07 & -67 00 13.6    &    ...     &    ...    &  18.726  &  0.126  &  18.842  &  0.167   &  18.905  & 0.159   &  -0.116  &  0.209  &  1.74 &    ...    \\[0.05cm]
 \hline
34 & \textbf{05 35 41.01} & \textbf{-66 51 53.1}    & \textbf{13.828}  &  \textbf{0.025}   &  \textbf{14.584}  &  \textbf{0.029}  &  \textbf{14.824}  &  \textbf{0.022}   &  \textbf{15.281}   & \textbf{0.358}    &  \textbf{-0.240}  &   \textbf{0.036}  &   \textbf{1.58} & k \\[0.05cm]
      &  05 35 41.20 & -66 51 51.9   & 17.939  &  0.230   &  18.412  &  0.161  &  18.311  &  0.087   &   17.745  &  0.134   &   0.101   &  0.183  &   0.06 &    ...    \\[0.05cm] 
      &  05 35 41.66 & -66 51 48.6   &    ...    &    ...    &  17.606  &  0.041 &  18.363  &  0.137    &  17.926  &  0.045    & -0.757   &  0.143   &  4.38 &    ...     \\[0.05cm] 
 \hline
35 & 05 35 53.40 & -65 30 31.9     & 19.892  &   0.166   &  19.948  &  0.128  &   19.031   &  0.037   &   18.207  &   0.055    &   0.917   &  0.133  &  3.21  &    ...    \\[0.05cm]
      & 05 35 53.48 & -65 30 34.8    &    ...    &    ...    &  22.330  &  0.226   &   21.327   &  0.155   &     ...     &    ...      &  1.003   &  0.274  &  2.16 &    ...    \\[0.05cm]
      & 05 35 53.62 & -65 30 38.3    & 20.031  &  0.167   &  19.775  &  0.064  &   18.847   &  0.055   &  17.868   &   0.067    &   0.928  &  0.084  &  4.44  &    ...    \\[0.05cm]
      & 05 35 54.12 & -65 30 29.8    &    ...    &    ...    &  21.152  &  0.113  &   20.406   &  0.084   &  19.668   &   0.093    &   0.746  &  0.141  &  4.64  &    ...    \\[0.05cm]  
 \hline
36$\star$ &      ...        &     ...            &     ...     &    ...    &    ...     &    ...    &     ...     &    ...    &    ...      &    ...      &    ...     &    ...      &    ...      &    ...    \\[0.05cm] 
 \hline
37   &  05 39 38.73 & -69 44 40.1   & 16.945 &  0.256 & 18.094 & 0.257 & 17.234 &  0.204 &    ...    &    ...    & 0.860 & 0.328  &    4.46  &    ...    \\[0.05cm]         
       &  \textbf{05 39 38.84} & \textbf{-69 44 35.6}   &  \textbf{13.819}   &   \textbf{0.087}    &  \textbf{14.535}  &   \textbf{0.050}     &    \textbf{14.612}  &    \textbf{0.171}  &   \textbf{13.886}  &    \textbf{0.093}   &   \textbf{-0.077}  &   \textbf{0.178}  &  \textbf{0.10} & k \\[0.05cm]
       &  05 39 39.39 & -69 44 38.0   &    ...     &    ...     &  17.928  &  0.136   &   16.596  &   0.151  &    ...     &    ...     &   1.332  &  0.203  & 3.67 &    ...    \\[0.05cm] 
 \hline
38  & 05 41 26.64 & -69 01 21.3  & 19.183  & 0.238  & 19.042  & 0.103 & 17.997 & 0.057 & 16.731 & 0.477 & 1.045   & 0.118 & 1.75  &    ...    \\[0.05cm] 
       & \textbf{05 41 26.67} & \textbf{-69 01 23.3}  & \textbf{14.463}  & \textbf{0.024}  & \textbf{15.296}  & \textbf{0.022} & \textbf{15.329} & \textbf{0.024} & \textbf{15.527} & \textbf{0.025} & \textbf{-0.033} & \textbf{0.033} & \textbf{0.43} & k \\[0.05cm] 
       & 05 41 27.29 & -69 01 23.0  & 21.772  & 0.479  & 19.755  & 0.110  & 20.087 & 0.110 & 19.516 & 0.133 & -0.332 & 0.156 & 3.60 &    ...    \\[0.05cm] 
\hline
39  & 05 41 21.84 & -69 36 30.1    &    ...     &    ...     &   20.999   &  0.315   &   20.644  &  0.197   &  20.183   &  0.128  &   0.355 &  0.372   &  2.18 &    ...    \\[0.05cm]
      & 05 41 21.89 & -69 36 28.4    & 19.565   &  0.132    &   19.271   &  0.130   &   19.067  &  0.095   &  18.999   &  0.082   &   0.204 &   0.161  &  1.75 &    ...    \\[0.05cm]
      & 05 41 22.19 & -69 36 25.3    & 19.991   &  0.181    &   19.964   &  0.096   &  19.085   &  0.099   &  18.342   &  0.08     &   0.879 &   0.138  &  3.66 &    ...    \\[0.05cm]
      & 05 41 22.33 & -69 36 31.0    & 19.898   &  0.114    &   19.509   &  0.058   &  18.945   &  0.051   &  18.151   &  0.076   &   0.564 &   0.077  &  2.14 &    ...    \\[0.05cm]
      & 05 41 22.60 & -69 36 25.3   & 19.413   &  0.154    &   19.406   &  0.055   &  18.414   &  0.064   &   17.437   &  0.059   &   0.992 &   0.084  &  4.23  &    ...    \\[0.05cm]
      & \textbf{05 41 22.98}  &  \textbf{-69 36 30.0}  &  \textbf{18.711}  &   \textbf{0.081}   &  \textbf{18.714}   &  \textbf{0.074}   &  \textbf{18.650}   &   \textbf{0.054}  &   \textbf{18.539}   &  \textbf{0.136}   &   \textbf{0.064} &    \textbf{0.092}  &   \textbf{4.21} & A \\[0.05cm]
 \hline
40 &  \textbf{05 41 34.34} & \textbf{-68 25 48.3}   &  \textbf{13.138}  &  \textbf{0.031}     & \textbf{14.072}    &  \textbf{0.030}     &  \textbf{14.043}  &  \textbf{0.019}    &  \textbf{13.911}  &  \textbf{0.040}     &    \textbf{0.029}  &   \textbf{0.036}  &  \textbf{2.59} & k \\[0.05cm]
      &  05 41 35.30 & -68 25 49.4   &    ...     &    ...     &  20.152   &  0.213   &  20.903  &  0.160    &    ...     &    ...    &  -0.751   &  0.266  &  3.38 &    ...     \\[0.05cm]
      &  05 41 35.30 & -68 25 51.8   &  19.499  &  0.166     &  20.177  &  0.127   &  19.112  &  0.074    &  18.097  &  0.053   &   1.065   &   0.147  &  3.78 &    ...     \\[0.05cm]  
 \hline
41     &   05 41 36.74 & -68 32 36.4   &    ...     &    ...     &  21.817  &  0.138    &   20.915  &  0.125   &   20.900  &  0.245  &   0.902  &  0.186   &   4.81 &     ...    \\[0.05cm]
         &  \textbf{05 41 37.54} & \textbf{-68 32 33.3}    & \textbf{13.121}   & \textbf{0.030}       &  \textbf{14.136}  &  \textbf{0.026}    &   \textbf{14.240}     &  \textbf{0.044}   &  \textbf{14.356}   &  \textbf{0.022}   &  \textbf{-0.104}  &  \textbf{0.051}  &   \textbf{2.73} & k,A \\[0.05cm]
 \hline
42 & 05 43 57.76 & -65 39 52.2    &    ...     &    ...     &  21.801 &  0.113   &  21.438 &  0.109   &  20.644 &  0.165     &   0.363  &   0.157  &   1.52 &     ...    \\[0.05cm]
      & \textbf{05 43 58.43} & \textbf{-65 39 53.6} &  \textbf{20.138} &  \textbf{0.165}    &  \textbf{20.436} &  \textbf{0.052}   &  \textbf{20.063} &  \textbf{0.049}   &  \textbf{19.018} &  \textbf{0.065}     &   \textbf{0.373}  &   \textbf{0.071}  &   \textbf{3.12} & n,A \\[0.05cm]
      & 05 43 58.43 & -65 39 55.8    &    ...     &    ...     &  23.239 &  0.493   &  23.409 &  0.524   &    ...    &    ...     &  -0.170  &   0.719  &   4.68 &    ...     \\[0.05cm]
\hline
\end{tabular}
\end{sidewaystable*}

\begin{sidewaystable*}
\hspace{-8.5cm}{\bf Table 2}{\it \,\,--  continued}\\
\centering
  \begin{tabular}{@{}lcccccccccccccc@{}}
\hline
Src ID & R.A. & Decl. & U & errU & B & errB & V & errV & I & errI & B-V & err(B-V) & Offset & C/part\\
            & (h\, m\, s) & (\degr\, \arcmin\, \arcsec)  &  \multicolumn{10}{c}{(mag)} & (\arcsec)& \\[0.2cm]
$[1]$ & $[2]$ & $[3]$ & $[4]$ & $[5]$ & $[6]$ & $[7]$ & $[8]$ & $[9]$ & $[10]$ & $[11]$ & $[12]$ & $[13]$ & $[14]$ & $[15]$\\
\hline
43   &  05 44 06.34 & -71 00 45.5   &    ...    &    ...    &  19.802  &  0.134  & 18.426  &  0.028   &   17.356  &  0.131  &   1.376  &  0.137  &  4.54  &     ...    \\[0.05cm] 
       &  05 44 06.66 & -71 00 49.1    & 20.317  &  0.220  &  20.552 &  0.120 &  20.009  &  0.076   &   19.946  &  0.142  &   0.543  &  0.142  &  1.95   &    ...     \\[0.05cm]  
       &  05 44 06.80 & -71 00 51.2   &    ...    &    ...    &  20.560  &  0.140  &  19.469  &  0.061   &   18.336  &  0.055  &  1.091  &   0.153  &  2.70 &    ...    \\[0.05cm] 
 \hline
44 &  \textbf{05 44 15.77} &   \textbf{-66 33 47.1} &    \textbf{14.127} &   \textbf{0.036} &   \textbf{15.075} &  \textbf{0.018} &  \textbf{15.201} &  \textbf{0.021} &   \textbf{15.202} &   \textbf{0.078} &   \textbf{-0.126} &  \textbf{0.028} &   \textbf{3.33} & k\\[0.05cm]             
 \hline
45 &  05 46 48.87 & -68 51 50.2   &  20.362  &  0.225    &  20.537  &  0.091  &  19.803  &  0.061  & 19.042  &  0.098  &   0.734  &   0.110   &  4.46 &    ... \\
\hline

\end{tabular}

 \begin{minipage}{195mm}
\footnotetext{

\noindent
Notes -- Three dots indicate that an entry was not available in the MCPS catalog. The most likely counterpart of each X-ray source is indicated in bold. For sources without a match shown in bold in this table, its most likely counterpart is presented either in the 1$\sigma<$ search radius $<2\sigma$ area or even at larger than 10\arcsec\, distances (Tables \typeout{\ref{tableBright}}3 and \typeout{\ref{tableMatchesOut}}4, respectively). An input "n"  in Column 15 indicates a newly identified counterpart from this work, a "k" a known in the literature counterpart, and an "A" those counterparts that are discussed in Appendix \ref{appendixA}.\\\\ Additional identifications and notes for the sources follow below, which however by no means consist a complete list.\\

\noindent
\textbf{Src 1:} [Massey2002]9775 (blue star) at 0.85\arcsec\, from X-ray source; 2MASS J04510685-6948032 at 0.32\arcsec\, from X-ray source; [MACHO]44.1741.17\\
\textbf{Src 5:} [MACHO]23.3300.10 (24); [Massey2002]48644 (blue star) at 0.14\arcsec\, from X-ray source; 2MASS J05004604-7044360$\equiv$USNO-B1.0 0192-00057570 (19)\\
\textbf{Src 6:} [MACHO]23.3424.11 blue variable star (\citealt{2002AJ....124.2039K}) and [Massey2002]51265 source at 1.17\arcsec\, and 1.04\arcsec\, from X-ray position, respectively\\
\textbf{Src 7:} [MACHO]53.3727.19 blue variable star
(\citealt{2002AJ....124.2039K}) at 3.18\arcsec\, from X-ray position\\
\textbf{Src 8:} 1.1\arcsec\, from the MCPS star cluster LMC0540 ((R.A., Decl.)${\rm _{J2000.0}}$=(05:07:38, -68:47:50), log(age)$\sim$8, and V$=$13.45 mag; \citealt{2010A&A...517A..50G}); 
5.9\arcsec\, from the cluster OGLE-CL LMC 129 ((R.A., Decl.)${\rm _{J2000.0}}$=(05:07:38.63, -68:47:45.9), apparent radius $\sim$20\arcsec, and log(age)$\sim$7.9; \citealt{2000AcA....50..337P});
 [Massey2002]73045 and 2MASS J05073783-6847492 at 4.29\arcsec\, and 0.44\arcsec\, from the X-ray source, respectively\\
\textbf{Src 11:} 2MASS J05132826-6547187 source at 1.82\arcsec\, from X-ray source (12); [MACHO]59.5431.442 with ${\rm <R>}\sim15.6$ and ${\rm V-R}\sim0.05$ (\citealt{2009ATel.2012....1S})\\  
\textbf{Src 12:} [Massey2002]101869 at 4.83\arcsec\, from X-ray source\\ 
\textbf{Src 13:} [Massey2002]118611 (blue star) at 1.33\arcsec\, from X-ray source; [OGLE-II]05201625-6925053 with (R.A., Decl.)${\rm _{J2000.0}}$=(05:20:16.25,-69:25:05.3) is a Type-4 object (i.e. light curve similar to Galactic Be stars; Sabogal \etal 2005) at 1.35\arcsec\, from X-ray source\\
\textbf{Src 15:} [MACHO]6.6940.6123 at 1.25\arcsec\, from X-ray position\\
\textbf{Src 16:} 2MASS J05241180-6620512 at 5.05\arcsec\, from X-ray position\\
\textbf{Src 18:} 2MASS J05272496-6552385 at 8.53\arcsec\, from X-ray position\\
\textbf{Src 20:} 2MASS J05294785-6556437 (at 8\arcsec) $\equiv$ UCAC2 2849264 (19)\\
\textbf{Src 21:} 2MASS J05301136-6551240 at 5.54\arcsec\, from the X-ray source (19)\\
\textbf{Src 24:} 2MASS J05311318-6607092 at 7.27\arcsec\, from the X-ray source\\
\textbf{Src 25:} 2MASS J05311556-7053485 at 1.66\arcsec\, from the X-ray source (19) and LMC V3134 variable source with ((R.A., Decl.)${\rm _{J2000.0}}$ = (05:31:15.66,-70:53:49.9) at 1.29 \arcsec\\ 
\textbf{Src 27:} 2MASS J05313676-6518162 at 4.18\arcsec\, from the X-ray source\\
\textbf{Src 28:} 2MASS J05322417-7107336 at 7.35\arcsec\, from the X-ray source and [MACHO]14.8376.548 (38) match\\
\textbf{Src 29:} 2MASS J05322587-6535012 at 8.54\arcsec\, from the X-ray source\\
\textbf{Src 30:} USNO-A2.0 0225-02109716 (19)\\

    }
    
 \end{minipage}

\end{sidewaystable*}

\begin{sidewaystable*}
\hspace{-8.5cm}{\bf Table 2}{\it \,\,--  continued}\\

 \begin{minipage}{195mm}
\footnotetext{

\noindent
\textbf{Src 31:} Source 316 in \cite{1999A&A...344..521H} is 2MASS
J05324953-6622132 with $J=14.586\pm0.037$ mag, $H=14.780\pm0.075$ mag,
and $K=14.750\pm0.115$ mag; Variable star LMC V3213 (from the Extragalactic Variable Stars Catalogue Vol. V of \citealt{2009yCat....102025S}) at 2.18\arcsec\, from X-ray position, classified as an X-ray pulsar\\
\textbf{Src 33:} 2MASS J05350598-6700157 at 0.56\arcsec\, from the
X-ray source; Mira (Omicron) Ceti-type variable star LMC V3394 at
0.8\arcsec\, from the X-ray source (from the Extragalactic Variable
Stars Catalogue Vol. V of \citealt{2009yCat....102025S}). These are
long-period variable giants with characteristic late-type emission
spectra (Me, Ce, Se) and light amplitudes from 2.5 to 11 mag in
V. Their periodicity is well pronounced, and the periods lie in the
range between 80 and 1000 days. Infrared amplitudes are usually less
than in the visible and may be $<$2.5 mag. For example, in the K band
they usually do not exceed 0.9 mag. \cite{1999A&A...344..521H}
classified this system as a Be-XRB (instead of a Mira variable).\\
\textbf{Src 34:} 2MASS J05354101-6651535 at 1.88\arcsec\, from X-ray position\\
\textbf{Src 35:} 2MASS J05355458-6530382 at 6.45\arcsec\, from X-ray position, and [2MASX]05355466-6530384 (from the 2MASS Extended Source Catalog of ) at 6.97\arcsec\\
\textbf{Src 36:} Variable star LMC V3758 (from the Extragalactic Variable Stars Catalogue Vol. V of \citealt{2009yCat....102025S}) at 1.83\arcsec\, from X-ray position, classified as a rotating ellipsoidal variable and a close binary system that is a source of strong, variable X-ray emission.\\
\textbf{Src 37:} extended objects SNR0540-69.7 (Supernova remnant) and
LMC-N159F (Nebula w/probable embedded cluster) at 4.0\arcsec\, and
4.8\arcsec, respectively \citep{2008MNRAS.389..678B}; 2MASS objects
J05393883-6944356, J05394007-6944337, and J05393990-6944413 at
3.69\arcsec, 4.42\arcsec, and 4.86\arcsec, respectively from X-ray position\\
\textbf{Src 38:} [Massey2002]176277 and 2MASS J05412663-6901224 at 1.58\arcsec\, and 2.54\arcsec\, from the X-ray position (with 3.0\arcsec\, positional accuracy), respectively\\
\textbf{Src 39:} 2MASS J05412014-6936229 (9.12\arcsec\, from X-ray source)$\equiv$SSTISAGEMC J054120.18-693622.9 (19)\\
\textbf{Src 40:} [Massey2002]176466 (blue star) and 2MASS J05413431-6825484 at 2.30\arcsec\, and 2.64\arcsec\, from the X-ray source, respectively\\
 \textbf{Src 41:} [Massey2002]176538 (blue star) and 2MASS J05413431-6825484 at 2.84\arcsec\, and 2.64\arcsec\, from the X-ray source, respectively\\
\textbf{Src 43:} 2MASS J05440640-7100454 at 4.55\arcsec\, from X-ray source (19)\\

\noindent
$\spadesuit$ None of the matches listed within 5\arcsec\, from the X-ray source position has been presented anywhere in the literature. Although the coordinates of the sources are somehow compatible, their magnitudes are totally different (especially the $B-V$ color). We believe that these works blend together multiple point sources, resulting in erroneous $B$ and/or $V$ magnitudes. For more details, see notes in Appendix \ref{appendixTOOblue}.\\
$\star$ There is no coverage by either the OGLE-II (\citealt{2000AcA....50..307U}), OGLE-III (\cite{2008AcA....58...89U}) or MCPS (\citealt{2004AJ....128.1606Z}) surveys. Negueruela \& Coe (2002) list the following magnitudes: $B=16.85$, $V=16.74$, $R=16.89$ with 0.05 errors, and they mention that the optical counterpart of LMC X-3 is very variable (varying between $V=16.5$ mag and $V=17.3$ mag).

   }
 
 \end{minipage}

\end{sidewaystable*}

\clearpage
\begin{sidewaystable*}
\hspace{-6.5cm}{\bf Table 3.} Bright optical counterparts of HMXBs within 1$\sigma<$ search radius $<2\sigma$\label{tableBright}\\
\centering
  \begin{tabular}{@{}lcccccccccccccc@{}}
\hline
Src ID & R.A. & Decl. & U & errU & B & errB & V & errV & I & errI & B-V & err(B-V) & Offset & C/part\\
            & (h\, m\, s) & (\degr\, \arcmin\, \arcsec)  &  \multicolumn{10}{c}{(mag)} & (\arcsec)& \\[0.2cm]
$[1]$ & $[2]$ & $[3]$ & $[4]$ & $[5]$ & $[6]$ & $[7]$ & $[8]$ & $[9]$ & $[10]$ & $[11]$ & $[12]$ & $[13]$ & $[14]$ & $[15]$\\
\hline
2        &    04   55  57.31   &   -70  19   56.8   &     ...          &     ...        &   19.927    &  0.073    &  18.896   &  0.084   &  18.156  &  0.054     &  1.031    &   0.111    &   	      9.05 &    ...    \\[0.05cm]  
          &     04   55  57.69   &   -70  19   57.0   &  19.977   &  0.188    &  20.962     &  0.133   &  20.537   &   0.127   &     ...     &     ...     &  0.425     &   0.184     &           7.30  &    ...    \\[0.05cm]   
          &     04   55  57.97   &   -70  19   56.5   &     ...          &     ...        &  20.788     &  0.098    &  20.112   &   0.142   & 19.986    &  0.190     &  0.676     &   0.173     &           6.52  &    ...    \\[0.05cm]   
          &     04   55  58.88   &   -70  20   08.4   &     ...          &     ...        &   21.542    &  0.119    &  21.055   &   0.144   &  20.617   &  0.202     &   0.487    &    0.187    &            7.45 &    ...    \\[0.05cm]   
          &     04   55  59.49   &   -70  19   52.0   &     ...          &     ...        &  21.280    &  0.122    &  20.611   &   0.145   &     ...     &     ...     &   0.669     &   0.189    &            9.45 &    ...    \\[0.05cm]   
          &     04   55  59.72   &   -70  20   04.7   &  20.952   &  0.306    &  21.841     &  0.183    &  20.772   &   0.174   &     ...     &     ...     &  1.069     &   0.253    &   	     5.54  &    ...    \\[0.05cm]   
          &     04   56  00.41   &   -70  20   03.9   &  20.909   &  0.242    &   21.589    &  0.108    &  20.993   &   0.112    &  20.170    &  0.133     &   0.596    &    0.156     &           8.16 &    ...    \\[0.05cm]   
          &     04   56  00.66   &   -70  19   58.5   &  19.918   &  0.146    &  19.425    &  0.084    &  18.481   &   0.103   &   17.787  &  0.055      &  0.944     &   0.133     &           9.22  &    ...    \\[0.05cm]  
          &     04   56  00.85   &   -70  20   01.3   &  19.438   &  0.083    &  19.530    &   0.058   &  19.470   &    0.077  &  19.433   &  0.078      &  0.060     &   0.096     &           9.85  &    ...    \\[0.05cm]
\hline
3     & 04 56 53.35 & -68 24 40.4    &  19.991  & 0.179    &  19.845  & 0.050  &  19.895  & 0.075    &  19.675   & 0.130      & -0.050  & 0.090  & 6.84 &    ...    \\[0.05cm]
       & 04 56 53.65 & -68 24 26.6    &  18.795  & 0.124    &  19.777  & 0.053  &  19.580   &  0.043   &  19.672   & 0.095      & 0.197   & 0.068  & 8.76 &    ...     \\[0.05cm]   
       & \textbf{04 56 55.07} & \textbf{-68 24 27.6}    &  \textbf{18.092}  & \textbf{0.078}    &  \textbf{18.378}  & \textbf{0.035}   &  \textbf{18.283}   &  \textbf{0.051}   &  \textbf{18.287}   & \textbf{0.063}     & \textbf{0.095}    & \textbf{0.062}  & \textbf{9.10} & n \\[0.05cm] 
\hline       
4     &  04 57 11.55 & -66 12 08.0   &  20.412  & 0.183    &   20.090   & 0.064    &  19.956 &  0.059   &  19.745   &  0.102      &  0.134 & 0.087  & 5.55 &    ...    \\[0.05cm]    
\hline
6     &  05 01 25.12 & -70 33 32.1    &  18.590   & 0.061    &  18.931  & 0.052   &  18.851  &  0.072   &  19.074  &  0.153      &  0.080  &  0.089  &  6.14  &    ...    \\[0.05cm] 
\hline
7     &  05 02 51.49 & -66 26 19.4   &  19.548   & 0.128    &  19.271  & 0.046   &  19.200   &  0.053   &  18.933  &  0.091    &   0.071  & 0.070  & 5.68 &    ...    \\[0.05cm]   
       &  05 02 51.57 & -66 26 34.8    &  19.785   & 0.140    &  19.663  & 0.090   &  19.504 &  0.086   &  19.040   & 0.181     &   0.159  &  0.124 & 9.80 &    ...    \\[0.05cm] 
       &  05 02 52.77 & -66 26 26.9    &  20.199   & 0.219    &   20.195 & 0.096   &  19.996 & 0.071    &   20.160    &  0.210   &   0.199  &  0.119 & 7.25 &    ...    \\[0.05cm]  
\hline
8     &  05 07 36.87 & -68 47 49.9  &   17.481   &  0.057    &  17.967   &  0.047   &  17.959   &  0.045   &  17.963   &  0.046   &     0.008 &   0.065  &   5.64 &    ...    \\[0.05cm]
       &  05 07 36.97 & -68 47 45.4  &   15.752   &   0.051   &  16.172   &  0.031   &  16.024   &  0.067   &  16.269   &  0.039   &    0.148 &    0.074  &   6.20 &    ...    \\[0.05cm]
       &  05 07 37.30 & -68 47 52.8  &  17.828   &  0.065     &  17.873   &  0.110   &  17.835   &  0.042   &  17.948   &  0.050   &     0.038 &   0.118  &   5.01 &    ...    \\[0.05cm]       
       &  05 07 37.53 & -68 47 43.2  &   17.664   &  0.064    &  17.596   &  0.087   &  17.771   &  0.063   &  17.238   &  0.095   &   -0.175 &   0.107  &   6.14 &    ...    \\[0.05cm]
       &  05 07 37.95 & -68 47 55.6  &    ...       &    ...     &  19.733   &  0.152   &  19.613   &  0.094   &    ...     &    ...     &     0.120 &  0.179  &   6.56 &    ...    \\[0.05cm]
       &  05 07 38.70 & -68 47 45.2  &   15.932   &  0.054    &  16.440   &  0.169   &  16.363   &  0.038   &  16.375   &  0.061   &    0.077 &    0.173  &   5.81 &    ...    \\[0.05cm]
       &  05 07 38.71 & -68 47 53.1  &    ...      &    ...      &  20.107   &  0.091   &  19.974   &  0.065   &  19.570   &  0.108     &    0.133 &  0.112  &   5.99 &    ...     \\[0.05cm]
       &  05 07 38.80 & -68 47 56.2  &   19.095   &  0.158    &  18.868   &  0.156   &  18.871   &  0.088   &    ...      &    ...    &    -0.003 &  0.179  &   8.69 &    ...    \\[0.05cm]       
       &  05 07 38.88 & -68 47 43.0  &   17.472   &  0.123    &  17.326   &  0.193   &  17.718   &  0.073   &  18.032   &  0.127   &   -0.392 &   0.206  &   8.03 &    ...    \\[0.05cm]       
       &  05 07 38.93 & -68 47 48.7  &   16.286   &  0.061    &  16.499   &  0.055   &  16.591   &  0.033   &  16.717   &  0.042   &   -0.092 &   0.064  &   5.61 &    ...    \\[0.05cm]
       &  05 07 39.04 & -68 47 45.7  &    ...       &    ...     &  18.581   &  0.325   &  18.655   &  0.121   &    ...      &    ...    &    -0.074 &  0.347  &   7.04 &    ...    \\[0.05cm]
\hline
\end{tabular}
\end{sidewaystable*}

\begin{sidewaystable*}
\hspace{-8.5cm}{\bf Table 3}{\it \,\,--  continued}\\
\centering
  \begin{tabular}{@{}lcccccccccccccc@{}}
\hline
Src ID & R.A. & Decl. & U & errU & B & errB & V & errV & I & errI & B-V & err(B-V) & Offset & C/part\\
            & (h\, m\, s) & (\degr\, \arcmin\, \arcsec)  &  \multicolumn{10}{c}{(mag)} & (\arcsec)& \\[0.2cm]
$[1]$ & $[2]$ & $[3]$ & $[4]$ & $[5]$ & $[6]$ & $[7]$ & $[8]$ & $[9]$ & $[10]$ & $[11]$ & $[12]$ & $[13]$ & $[14]$ & $[15]$\\
\hline
       &  05 07 39.32 & -68 47 48.6  &   16.608   &  0.091    &  16.940   &  0.057   &  17.025   &  0.035   &  17.259   &  0.056   &   -0.085 &   0.067  &   7.72 &    ...    \\[0.05cm]
       &  05 07 39.34 & -68 47 50.7  &    ...       &    ...     &  18.385   &  0.122   &  18.439   &  0.084   &  18.396   &  0.094   &    -0.054 &   0.148 &   7.99 &    ...    \\[0.05cm]      
\hline
10     & 05 12 40.29 & -67 17 22.9   & 19.947   &  0.190   &  19.707    &  0.088     &  19.741    &   0.072   &  19.532   &  0.276    &  -0.034   &  0.114  &  8.72 &    ...    \\[0.05cm]       
       & 05 12 40.41 & -67 17 18.1   & 19.721   &  0.099   &  19.815    &  0.056     &  19.765    &   0.092   &  19.528   &  0.102    &   0.050   &   0.108 &  9.39 &    ...    \\[0.05cm]               
\hline
12  & 05 15 59.40 & -69 16 14.9  &  17.895   &  0.051   &  18.068   &   0.041   &  18.105   &  0.185   &  17.808   &  0.118   &   -0.037    &  0.189    &   6.99  &    ...    \\[0.05cm]
       & 05 16 00.03 & -69 16 14.4   &  17.608   &  0.048   &  17.862   &  0.044    & 17.885   &  0.053   &  17.712   &  0.077    &   -0.023   &   0.069   &   5.42 &    ...    \\[0.05cm]
       & 05 16 00.79 & -69 16 04.4  & 19.959   &   0.120   &  19.803   &   0.083   &  19.613   &  0.273   &  19.554   &  0.278   &    0.190   &   0.285    &   5.89  &    ...    \\[0.05cm]
       & 05 16 01.22 & -69 16 12.4  &  17.650    &  0.047   &  18.090     &  0.058    & 18.035   &  0.082   &  18.113   &  0.112    &    0.055   &   0.100   &    6.82 &    ...    \\[0.05cm]
\hline
13  & 05 20 17.33 & -69 25 02.4   &  18.496   &  0.073   &  18.638   &  0.067   &  19.028   &  0.501   &  18.412   &  0.204  &  -0.390    &  0.505     &  7.47 &    ...    \\[0.05cm]   
\hline
14  &  05 20 29.01  &  -69 31 52.03  &   19.575  &  0.128  &  19.299  &    0.072 & 18.161  &  0.172   &     ...     &    ...      &  1.138  &  0.186   &  5.09   & ... \\[0.05cm]
       &  05 20 29.31  &  -69 31 59.84  &   19.167  &  0.130  &  19.349  &   0.094  & 17.431  &  0.124   &  17.196   &  0.110     &  1.918  &  0.156   &  5.48   &    ...    \\[0.05cm]
\hline
15  & \textbf{05 23 14.22} & \textbf{-70 04 15.8}  &  \textbf{16.533}  &  \textbf{0.049}   &  \textbf{16.997}   &  \textbf{0.062}   &  \textbf{16.965}   &  \textbf{0.065}   &  \textbf{17.151}   &  \textbf{0.097}    &   \textbf{0.032}   &   \textbf{0.090}     &   \textbf{5.18} &  n,A \\[0.05cm]
\hline
16  & \textbf{05 24 11.83} & \textbf{-66 20 50.9}  &  \textbf{13.837}  &  \textbf{0.024}  &  \textbf{14.813}   &  \textbf{0.014}  &  \textbf{14.867} &  \textbf{0.022}  &  \textbf{14.866}  &  \textbf{0.021} &  \textbf{-0.054}  &  \textbf{0.026}  &  \textbf{5.31} & n  \\[0.05cm] 
\hline
18    & \textbf{05 27 22.51} & \textbf{-65 52 40.5} & \textbf{18.266}  & \textbf{0.058}  & \textbf{18.586}  & \textbf{0.034}  & \textbf{18.494}  & \textbf{0.030}  & \textbf{18.552}  & \textbf{0.047}    & \textbf{0.092}   & \textbf{0.045}    & \textbf{9.10} & n \\[0.05cm] 
        & 05 27 23.81 & -65 52 26.1  &  20.131  &  0.113  &  20.084   &  0.045   &  19.956  &  0.061  &  19.806  &  0.107   &  0.128   &  0.076    &  8.95 &    ...    \\[0.05cm] 
\hline
19   & 05 29 24.23 & -69 52 10.3  &  20.100   &  0.149  &  19.362  &  0.079  &  19.381  &  0.073  &  19.122  &  0.083    &  -0.019     &  0.108    &  8.64 &     ...    \\[0.05cm] 
        & 05 29 24.60 & -69 52 09.9 &    ...    &    ...     & 19.703 & 0.078 & 18.804 & 0.052 & 17.641 & 0.134 & 0.899  & 0.094 &  6.79    &    ...    \\[0.05cm]
        & 05 29 27.24 & -69 52 14.6 & 20.45    & 0.205     & 19.686  & 0.098  & 18.808  & 0.075  & 17.523  & 0.096 &   0.878    &  0.123  &  7.81   &    ...    \\[0.05cm]
        & \textbf{05 29 25.17} & \textbf{-69 52 15.8}  &  \textbf{18.782}  &  \textbf{0.127}  &  \textbf{18.693}  &  \textbf{0.110}   &  \textbf{18.811}  &  \textbf{0.139}  &  \textbf{18.853} &  \textbf{0.156}  & \textbf{-0.118}   & \textbf{0.177}    &   \textbf{6.12} & n,A \\[0.05cm]
        & 05 29 26.36  & -69  52   17.8  & 20.671  & 0.233 & 20.002  & 0.118  &  19.757 &  0.145 &  18.903 & 0.136  & 0.245   & 0.187   & 7.24   &    ...    \\[0.05cm] 
        & 05 29 27.14 & -69 52 03.6  &  19.838  &  0.164  &  19.382  &  0.114  & 19.530   &  0.228  &  19.369  &  0.168    &   -0.148    &   0.255   &  9.76  &    ...    \\[0.05cm] 
        & 05 29 27.33 &	-69 52 07.6  & 21.127    & 0.343   & 19.865  & 0.149	&  19.599	&  0.175	& 18.662	& 0.143  & 0.266   & 0.230  &  8.12	  &    ...    \\[0.05cm] 
\hline
20 & \textbf{05 29 47.82} & \textbf{-65 56 43.4}     &  \textbf{13.761}  &  \textbf{0.029}   &  \textbf{14.477}  &  \textbf{0.038}  &  \textbf{14.594}  &  \textbf{0.022}  &  \textbf{14.741}  &  \textbf{0.106}     &  \textbf{-0.117}     &  \textbf{0.044}   &  \textbf{8.36} & k \\[0.05cm] 
\hline
21 & \textbf{05 30 11.37} & \textbf{-65 51 23.5}  &  \textbf{13.933}  &  \textbf{0.034}  &  \textbf{14.894}  &  \textbf{0.016}  &  \textbf{14.877}  &  \textbf{0.022}  &  \textbf{14.691} &  \textbf{0.023}  &  \textbf{0.017}  &  \textbf{0.027}   &   \textbf{5.89} & k \\[0.05cm] 
\hline
22 & 05 30 42.27 &  -66 54 38.6 & 18.238  &  0.055   & 18.360   &  0.060    &  18.556    & 0.056    & 18.680   &  0.067   & -0.196  &  0.082  &  9.80  &    ...    \\[0.05cm]
\hline
23 & 05 30 46.70 & -66 06 18.3  &  20.617  &  0.168  &  20.358  &  0.039  & 20.061  &  0.067  &  19.815  &   0.106   &  0.297   &  0.078  &   5.36 &    ...    \\[0.05cm] 
\hline
24 & \textbf{05 31 13.08} & \textbf{-66 07 06.7}  &  \textbf{13.862}  & \textbf{0.029}  & \textbf{14.749} &  \textbf{0.014} &  \textbf{14.798}  & \textbf{0.051} & \textbf{14.763}  &  \textbf{0.039}  & \textbf{-0.049}  & \textbf{0.053}  &  \textbf{5.71}  & k,A \\[0.05cm] 
      & 05 31 13.14 & -66 07 09.3 & 16.131  & 0.041 &  16.157  &  0.029  & 15.740  & 0.123  & 15.080  &  0.038 & 0.417 & 0.126  &  7.44 &    ...     \\[0.05cm]
      & 05 31 15.20 &  -66 07 05.0  &  17.496  &  0.038 &  17.714  &   0.018  &   17.741  &   0.026  &    ...    &    ...    &   -0.027  &    0.032   &   8.76 &    ...    \\[0.05cm]
\hline
\end{tabular}
\end{sidewaystable*}

\begin{sidewaystable*}
\hspace{-8.5cm}{\bf Table 3}{\it \,\,--  continued}\\
\centering
  \begin{tabular}{@{}lcccccccccccccc@{}}
\hline
Src ID & R.A. & Decl. & U & errU & B & errB & V & errV & I & errI & B-V & err(B-V) & Offset & C/part\\
            & (h\, m\, s) & (\degr\, \arcmin\, \arcsec)  &  \multicolumn{10}{c}{(mag)} & (\arcsec)& \\[0.2cm]
$[1]$ & $[2]$ & $[3]$ & $[4]$ & $[5]$ & $[6]$ & $[7]$ & $[8]$ & $[9]$ & $[10]$ & $[11]$ & $[12]$ & $[13]$ & $[14]$ & $[15]$\\
\hline
    & \textbf{05 13 28.26} & \textbf{-65 47 18.4}    &  \textbf{14.122}  &  \textbf{0.036}    &  \textbf{15.067}  &  \textbf{0.088}   &  \textbf{15.098}  &  \textbf{0.027}   &  \textbf{15.328}  &  \textbf{0.039}    & \textbf{ -0.031}   &  \textbf{0.092}   &  \textbf{2.06} & k,A \\[0.05cm]   
\hline
25 & 05 31 16.06 & -70 53 43.0  & 19.020   &  0.090   &  18.960 &  0.065  &  18.784  &  0.063  & 18.985  & 0.096  &  0.176   &  0.091   &   7.68 &    ...    \\[0.05cm]
      & 05 31 16.76 & -70 53 44.2  &  20.102  &  0.122  &  19.662  &  0.081   &  19.547  &  0.115 &  19.246 &  0.110  &  0.115   &   0.141  &    8.86 &    ...    \\[0.05cm]
\hline
26   & 05 31 17.35 & -66 07 23.7  & 19.942  &  0.133	&  20.064	&  0.089  & 20.029   &  0.045   & 20.189  &  0.100  & 0.035 & 0.100 &  8.12 &    ...     \\[0.05cm]
        & 05 31 19.21 & -66 07 26.1  & 19.700  &  0.114  &  19.758  &  0.050  &  19.680  &  0.049  &  19.669  &  0.092   &  0.078   &  0.070   &   7.26  &    ...    \\[0.05cm]
\hline
27  &  05 31 36.45 & -65 18 20.6  &  20.586  &  0.223  &  19.857  &  0.061  &  19.744  &  0.052  &  19.746  &  0.088  &   0.113   &   0.080   &   5.13 &    ...    \\[0.05cm]
       & 05 31 36.75 & -65 18 23.1  & 20.649 &  0.233	&  20.006	  &  0.274	&  20.040	 &  0.062 &  20.002  &  0.139 & -0.034   & 0.281  &  8.18 &    ...    \\[0.05cm]
\hline
29  & \textbf{05 32 24.31} & \textbf{-65 35 01.1}  &    ...    &    ...    & \textbf{19.771} &  \textbf{0.044} &  \textbf{19.576} &  \textbf{0.046}  & \textbf{19.703}   &  \textbf{0.097}   &   \textbf{0.195}   &    \textbf{0.064}    &   \textbf{9.96} & n,A \\[0.05cm]
       & \textbf{05 32 25.85} & \textbf{-65 35 01.3} & \textbf{20.727} & \textbf{0.178} & \textbf{19.536} & \textbf{0.037} & \textbf{18.336} & \textbf{0.026} & \textbf{17.173} & \textbf{0.054} & \textbf{1.200}  & \textbf{0.045}  & \textbf{8.44} & n,A \\[0.05cm]
\hline
30  & 05 32 31.92 & -65 51 46.4  &  19.150  &  0.147  &  19.360  &  0.046  &  19.244  &  0.092  &  18.981  &  0.121   &  0.116    &  0.103   &   5.39 &    ...    \\[0.05cm] 
\hline
31  & 05 32 48.81 &  -66 22 14.1 &  20.506 &  0.247  &  20.178 &  0.063 &  19.859 & 0.094  & 20.08  &  0.169 &  0.319  & 0.113  & 5.91 &    ...    \\[0.05cm]
       & 05 32 49.33 &  -66 22 19.6 & 19.343  &   0.089  &  19.204 & 0.091 &  19.062  & 0.101 &  19.667 & 0.325 & 0.142 &  0.136 & 6.42  &    ...    \\[0.05cm]
       & 05 32 51.07 &  -66 22 17.0 &  17.846 &  0.046  & 18.066  & 0.090 &  18.243 &  0.103  & 18.297 &  0.042 & -0.177 & 0.137 & 8.33 &    ...    \\[0.05cm]
\hline
33    & 05 35 04.28 & -67 00 14.9  &  18.128   &  0.050    &  18.444   &  0.066   &  18.53      &  0.114  &  18.585  &  0.062   &  -0.086  &  0.132    &  9.54  &    ...    \\[0.05cm]
         & 05 35 06.78 & -67 00 14.4  &  19.006   &  0.076   &  19.190   &  0.061    &  19.092   &  0.046  &  19.016   &  0.060    &   0.098  &   0.076   &  5.43 &    ...    \\[0.05cm]
\hline
34  & 05 35 41.20 & -66 51 51.9  &  17.939  &  0.230   &  18.412  &  0.161   &  18.311   &  0.087  &  17.745   &  0.134  &    0.101  &   0.183  &    0.06  &    ...    \\[0.05cm]
       & 05 35 41.66 & -66 51 48.6  &    ...    &    ...    &  17.606  &  0.041   &  18.363   &  0.137  & 17.926    &  0.045  &   -0.757  &   0.143  &    4.38  &    ...     \\[0.05cm]
\hline
35     & 05 35 52.34 & -65 30 35.3 &    ...    &    ...    &  21.877 & 0.147 & 21.387 & 0.115 &    ...    &    ...    & 0.490 & 0.187 & 9.16  &    ...    \\[0.05cm]       
     & 05 35 53.41 & -65 30 43.6 & 21.007 & 0.303 & 20.918 & 0.082 & 20.447 & 0.074 & 19.814 & 0.106 & 0.471 & 0.110 & 9.86  &    ...    \\[0.05cm]       
     & 05 35 54.60 & -65 30 29.4 & 20.027 & 0.189 & 20.430 & 0.085 & 19.849 & 0.048 & 18.816 & 0.064 & 0.581 &  0.098 &  6.79  &    ...    \\[0.05cm]       
     & \textbf{05 35 54.64} & \textbf{-65 30 37.7} & \textbf{17.915} & \textbf{0.054} & \textbf{18.026} & \textbf{0.105} & \textbf{16.952} & \textbf{0.101} & \textbf{15.720} & \textbf{0.102} & \textbf{1.074} &  \textbf{0.146} &  \textbf{6.40}  & n,A \\[0.05cm]     
     & 05 35 54.69 &  -65 30 31.5  &    ...     &    ...     & 21.440 & 0.165  & 21.156  & 0.146  &     ...    &    ...    & 0.284 & 0.220 & 6.11 &    ...    \\[0.05cm]
\hline
 37      &  05 39 39.90 & -69 44 41.1   & 15.873   &   0.213   &   16.287  &  0.185   &  16.023  &   0.117   &  15.547  &   0.136  &    0.264  &  0.219 &  7.64 &    ...    \\[0.05cm] 
     $\star$  & 05 39 40.03 & -69 44 33.4   & 11.745   &  0.033   &   12.295  &  0.035   &  12.080     &  0.054   &  11.493  &   0.180  &   0.215  &  0.064 &  6.54 & ... \\[0.05cm]
\hline
38       & 05 41 26.58 & -69 01 16.8 & 19.274 & 0.110  & 19.451 & 0.036 & 19.250 &  0.042 & 19.125 & 0.058 & 0.201  &  0.055 &  6.18  &    ...    \\[0.05cm] 
           & 05 41 26.80 & -69 01 28.5 & 19.913 & 0.127  & 19.745 & 0.067 & 19.610 &  0.050 & 19.454 & 0.071 & 0.135  &  0.084 & 5.57  &    ...    \\[0.05cm] 
\hline
39  &  05 41 20.55 & -69 36 31.6 &    ...    &    ...    &  20.113 & 0.256 &  19.884 & 0.292 &    ...    &    ...    & 0.229 &  0.388 &  9.01  &    ...    \\[0.05cm] 
       &  05 41 22.31 & -69 36 38.8 & 20.957  & 0.206 &  20.247 &  0.088  & 20.084 &  0.095 &  20.022 &  0.121 &  0.163 &  0.129  &  9.82  &    ...    \\[0.05cm] 
       &  \textbf{05 41 22.93} & \textbf{-69 36 33.3}  &  \textbf{17.122}  &  \textbf{0.045}  &  \textbf{17.399}  &  \textbf{0.033}  &  \textbf{17.326}  &  \textbf{0.036}  &  \textbf{17.243}  &  \textbf{0.059}   &   \textbf{0.073}  &   \textbf{0.049}   &  \textbf{5.74} & n?,A \\[0.05cm]
       &  05 41 21.78 & -69 36 20.2 & 18.487 & 0.135   & 17.813 &  0.080  &  16.767 &  0.049 &  15.642 &  0.056 & 1.046  &  0.094 &  9.11  &    ...    \\[0.05cm]   
\hline
\end{tabular}
\end{sidewaystable*}

\begin{sidewaystable*}
\hspace{-8.5cm}{\bf Table 3}{\it \,\,--  continued}\\
\centering
  \begin{tabular}{@{}lcccccccccccccc@{}}
\hline
Src ID & R.A. & Decl. & U & errU & B & errB & V & errV & I & errI & B-V & err(B-V) & Offset & C/part\\
            & (h\, m\, s) & (\degr\, \arcmin\, \arcsec)  &  \multicolumn{10}{c}{(mag)} & (\arcsec)& \\[0.2cm]
$[1]$ & $[2]$ & $[3]$ & $[4]$ & $[5]$ & $[6]$ & $[7]$ & $[8]$ & $[9]$ & $[10]$ & $[11]$ & $[12]$ & $[13]$ & $[14]$ & $[15]$\\
\hline
40 & 05 41 33.57 & -68 25 52.2  &  20.065  &  0.129  &  18.948  &  0.077  &  19.064  &  0.076  &  18.240   &  0.059   &  -0.116   &  0.108   &  6.63 &     ...    \\[0.05cm]
\hline
42  & \textbf{05 43 57.78} & \textbf{-65 39 46.33} &    ...    &    ...    &   \textbf{20.280} & \textbf{0.051} &   \textbf{19.357} &   \textbf{0.065} &   \textbf{18.260}  & \textbf{ 0.045} &   \textbf{0.923} &   \textbf{0.083}   &   \textbf{5.82} & n,A \\[0.05cm]
       & \textbf{05 43 59.57} & \textbf{-65 39 53.0}  &  \textbf{20.445}  &  \textbf{0.154}  &  \textbf{20.171}  &  \textbf{0.055}  &  \textbf{19.956}  &  \textbf{0.050}  &  \textbf{19.707}  &  \textbf{0.084}    &  \textbf{0.215}   &    \textbf{0.074}   &    \textbf{9.76} & n,A \\[0.05cm]
\hline
43  & \textbf{05 44 05.13} & \textbf{-71 00 50.8}  &  \textbf{14.598}  &  \textbf{0.031}  &  \textbf{15.292}  &  \textbf{0.029}  &  \textbf{15.235}  &  \textbf{0.022}  &  \textbf{15.220}   &  \textbf{0.029}   &    \textbf{0.057}  &     \textbf{0.036}   &    \textbf{5.75} & k \\[0.05cm]
\hline
44 & 05  44  14.01 &  -66  33  48.4 &    ...     &    ...    &  20.865 & 0.069 &   20.705 & 0.067 &   20.337 & 0.123 &   0.160  &   0.096   &      9.01   &    ...      \\[0.05cm]             
     & 05  44  14.68 &  -66  33  42.9 &    ...     &    ...    &  21.123 & 0.074 &   20.935 & 0.089 &   20.463 & 0.113 &   0.188  &   0.116   &      8.61   &    ...    \\[0.05cm]             
     & 05  44  16.32 &  -66  33  41.4 &    ...     &    ...    &  20.966 & 0.071 &   20.685 & 0.076 &    ...     &    ...    & 0.281  &   0.104   &      9.93   &    ...    \\[0.05cm]             
     & 05  44  16.34 &  -66  33  51.4 &    ...     &    ...    &  22.719 & 0.339 &   21.504 & 0.127 &   20.583 & 0.172 &   1.215  &   0.362   &      5.23   &    ...    \\[0.05cm]             
     & 05  44  16.68 &  -66  33  56.2 &   20.853 & 0.242   &  19.898 & 0.049 &   19.203 & 0.041 &   18.059 & 0.055 &   0.695  &   0.064   &      9.38   &    ...    \\[0.05cm]             
     & 05  44  16.86 &  -66  33  45.4 &   18.886 & 0.101   &  18.678 & 0.026 &   18.536 & 0.029 &   18.422 & 0.058 &   0.142  &   0.039   &      9.32   &    ...    \\[0.05cm]            
\hline
45  & \textbf{05 46 49.39} & \textbf{-68 51 46.2}  &  \textbf{18.150}   &  \textbf{0.071}  &  \textbf{18.480}   & \textbf{0.041}  &  \textbf{18.440}   &  \textbf{0.038}  &  \textbf{18.520}   &  \textbf{0.060}   &    \textbf{0.040}   &  \textbf{0.056}   &    \textbf{5.93} & n,A \\
\hline

\end{tabular}

 \begin{minipage}{195mm}
\footnotetext{

\noindent
Notes -- Similarly to Table \typeout{\ref{tableOpt}}2, the most likely counterpart of each X-ray source is indicated in bold. Three dots indicate that an entry was not available in the MCPS catalog. Bright refers to stars with $V\lesssim20$ mag and $B-V\lesssim0.2$ mag. 2$\sigma$ search radius corresponds to 10\arcsec\, from the X-ray source position.\\ X-ray sources, which do not appear in this table, do not have any bright optical match within 5\arcsec\, to 10\arcsec\, from their position. Their most likely counterpart is found either within 5\arcsec\, from the X-ray position (Table \typeout{\ref{tableOpt}}2) or at larger than 10\arcsec\, distances (especially for those X-ray sources with larger than 10\arcsec\, positional uncertainties; see Table \typeout{\ref{tableMatchesOut}}4). An input "n"  in Column 15 indicates a newly identified counterpart from this work, a "k" a known in the literature counterpart, and an "A" those counterparts that are discussed in the Appendix \ref{appendixA}.\\\\ Additional identifications and notes for these sources follow below, which however by no means consist a complete list.\\

\noindent
\textbf{Src 8:} [OGLE-CL]129 at 5.21\arcsec\, (R.A., Dec.)${\rm _{J2000.0}}$=(05:07:38.63,-68:47:45.9) with age ${\rm log(age)}\sim{7.9}$ (\citealt{2000AcA....50..337P}); AGN [Veron+2010]1.4418.1930 at 7.9\arcsec\, (R.A., Dec.)${\rm _{J2000.0}}$=(05:07:36.5,-68:47:51) from the X-ray source position with z=0.53 and V=19.81 mag \citep{2010A&A...518A..10V}.\\
\textbf{Src 14:} [Massey2002]119691 with $V=14.12$ mag, and $B-V=0.02$
mag at 1.34\arcsec\, from X-ray source position; 2MASS
J05202986-6931556 with $J=14.372\pm0.059$ mag, $H=14.211\pm0.070$ mag,
$K=14.276\pm0.099$ mag at 0.69\arcsec\, from the X-ray source position. This is the same counterpart as the one given by Edge \etal (2004), though the photometry listed there (which is taken from Coe \etal 2001) is a bit different (average OGLE  magnitudes: $B=14.39$ mag, and $V=14.45$ mag, thus $B-V=-0.06$ mag).\\
\textbf{Src 43:} [OGLE-II]05440524-7100512 with (R.A., Dec.)${\rm
  _{J2000.0}}$=(05:44:05.24,-71:00:51.2) is a Type-1/2 object
(i.e. Type-1 star with luminosity jumps in its light curve) in the
work of Sabogal \etal 2005, located at 5.31\arcsec\, from the X-ray source; [MACHO]15.10314.14 variable star 5.73\arcsec\, from X-ray position; 2MASS J05440522-7100507 at 5.31\arcsec\, from X-ray source.\\

\noindent
$\star$ Although this object (a known B-type supergiant; B3I(a) from \citealt{2009AJ....138..510F}) is the brightest match within the 2$\sigma$ search radius around the well studied source LMC X-1, the known optical counterpart of this BH-HMXB \citep{Cowley et al. 1995} is another object (i.e., the nearest and bluest match) within the 1$\sigma$ search radius (shown in bold in Table \typeout{\ref{tableOpt}}2).

    }
 
 \end{minipage}

\end{sidewaystable*}

\clearpage
\begin{sidewaystable*}
\hspace{-3.5cm}{\bf Table 4.} Additional bright ($V\lesssim20$ mag and $B-V\lesssim0.2$ mag) optical counterparts of HMXBs with positional uncertainty larger than 10\arcsec\label{tableMatchesOut}\\
\centering
  \begin{tabular}{@{}lcccccccccccccc@{}}
\hline
Src ID & R.A. & Decl. & U & errU & B & errB & V & errV & I & errI & B-V & err(B-V) & Offset & C/part\\
            & (h\, m\, s) & (\degr\, \arcmin\, \arcsec)  &  \multicolumn{10}{c}{(mag)} & (\arcsec)& \\[0.2cm]
$[1]$ & $[2]$ & $[3]$ & $[4]$ & $[5]$ & $[6]$ & $[7]$ & $[8]$ & $[9]$ & $[10]$ & $[11]$ & $[12]$ & $[13]$ & $[14]$ & $[15]$\\
\hline
4   &   \textbf{04 57 15.39}   &  \textbf{-66 12 22.7}  &  \textbf{15.452}  &  \textbf{0.027}  & \textbf{15.633} &  \textbf{0.030} &  \textbf{15.630}  &  \textbf{0.021}  & \textbf{15.459}  &  \textbf{0.022}  &  \textbf{0.003}  &  \textbf{0.037}  &  \textbf{22.10} & n \\[0.05cm]    
\hline
15      &  \textbf{05 23 16.94}   &   \textbf{-70 04 00.0}  &  \textbf{15.803}   &  \textbf{0.041}   &   \textbf{15.358}  &  \textbf{0.035}  &  \textbf{15.025}  &  \textbf{0.052} &  \textbf{14.402} &  \textbf{0.045}  & \textbf{0.333}  & \textbf{0.063} &  \textbf{15.88}  &  n,A \\[0.05cm]
\hline
17      & 05 27 06.47   &  -70 04 40.2    &  18.370    &  0.075   &     18.692   &    0.054   &   18.580   &    0.122   &    18.826  &    0.078  &   0.112  & 0.133 &  21.08  &    ...    \\[0.05cm]
          &  \textbf{05 27 10.14}  &  \textbf{-70 04 53.4}  &  \textbf{19.211}  &  \textbf{0.084}  & \textbf{18.782}  &  \textbf{0.052}  & \textbf{18.626}  &  \textbf{0.074} &  \textbf{18.372}  &  \textbf{0.061}  & \textbf{0.156} & \textbf{0.090} &  \textbf{13.23} & n,A \\[0.05cm]
          & 05 27 10.95   &  -70 04 45.2    &  18.656    &  0.081    &    18.661   &    0.054   &   18.381   &    0.060   &    18.068  &    0.077  &   0.280  & 0.081 &  21.48  &    ...    \\[0.05cm]
\hline        
19      & 05 29 22.81	& -69 52 16.5	& 19.168	& 0.109	& 17.771	& 0.094	& 16.711	& 0.051	& 15.511	& 0.076   &  1.060   & 0.107    & 16.88     &    ...    \\[0.05cm]
           & 05 29 22.97	& -69 52 00.0	& 18.984	& 0.093	& 18.476	& 0.076	& 18.366	& 0.058	& 18.434	& 0.355  &  0.110   & 0.096   &  18.69 &    ...    \\[0.05cm]
         & 05 29 22.99	& -69 52 21.5	& 20.236	& 0.153	& 19.063	& 0.154	& 18.817	& 0.064	& 17.918	& 0.135  & 0.246  & 0.167  &  18.34	&    ...    \\[0.05cm]
        &   05   29 23.18	   & -69 52 07.9	& 18.463	& 0.076	& 18.546	& 0.062	& 18.805	& 0.180	& 18.506	& 0.131  & -0.259   & 0.190  &  14.38	 &    ...    \\[0.05cm]
  & 05 29 24.76 & -69 51 59.3 & 20.326  & 0.209     & 19.385 & 0.109 & 18.357 & 0.096 & 17.153 & 0.093 & 1.028  & 0.145  & 13.07  &    ...    \\[0.05cm]
  & 05 29 24.87 & -69 52 24.7 &    ...    &    ...     & 20.247 & 0.153 & 18.798 & 0.122 & 17.907 & 0.296 & 1.449  & 0.196  & 14.73  &    ...    \\[0.05cm]
  & 05 29 24.87 & -69 52 29.3 & 20.271  & 0.186     & 19.355  & 0.114  & 18.635  & 0.115  & 17.624  & 0.079 &   0.720   &  0.162  &  19.07  &    ...    \\[0.05cm]
        &  \textbf{05 29 24.93}	& \textbf{-69 52 19.7}	& \textbf{16.389}	& \textbf{0.193}	& \textbf{16.490}	& \textbf{0.083}	& \textbf{16.508}	& \textbf{0.044}	& \textbf{16.608}	& \textbf{0.094}    &    \textbf{-0.018}  & \textbf{0.094}  &  \textbf{10.08}	   &  n,A \\[0.05cm]
       &  05 29 25.83 &  -69 52 27.0 &  18.797 &  0.114 &  18.871 &  0.260 & 17.725 &  0.144 &  16.924 &  0.166 & 1.146  &  0.297 &  16.05       &    ...    \\[0.05cm]
        & 05 29 26.62	& -69 51 53.3	&    ...   	&    ...    & 19.011	& 0.095	& 18.912	& 0.120	& 18.870	& 0.144  & 0.099   & 0.153    &  18.11 &    ...    \\[0.05cm]
          & 05 29 26.88	& -69 51 58.1	& 18.528	& 0.088	& 18.612	& 0.079	& 18.504	& 0.088	& 18.473	& 0.106  & 0.108    & 0.118   &  13.85   &    ...    \\[0.05cm]
        & 05 29 26.96	& -69 52 27.3	& 20.121	& 0.152	& 19.585	& 0.096	& 19.747	& 0.110	& 19.251	& 0.150 & -0.162  & 0.146  &  17.16 &     ...    \\[0.05cm]
       & 05 29 27.10	& -69 52 23.2	& 20.038	& 0.167	& 18.257	& 0.072	& 16.916	& 0.053	& 15.213	& 0.053   &  1.341  & 0.089   &  13.67	 &    ...    \\[0.05cm]
           & 05 29 28.33	& -69 52 12.1	& 19.875	& 0.168	& 18.834	& 0.099	& 18.431	& 0.073	& 17.396	& 0.147 & 0.403  &  0.123  &  12.60	&    ...    \\[0.05cm]
         & 05 29 28.66	& -69 52 05.8	& 18.907	& 0.097	& 18.417	& 0.042	& 18.514	& 0.041	& 18.323	& 0.091  & -0.097   & 0.059   &   15.16  &    ...    \\[0.05cm]
        & 05 29 28.82	& -69 52 00.7	& 19.430	& 0.113	& 18.834	& 0.069	& 18.508	& 0.066	& 17.947	& 0.083  & 0.326    & 0.095   &   18.25  &    ...    \\[0.05cm]
       &  05 29 29.04 &  -69 52 09.8 &  19.196 &  0.146 & 18.245 &  0.066 &  17.543 &  0.064 &  16.582 &  0.064 & 0.702 &  0.092 & 16.25      &    ...    \\[0.05cm]
  & 05 29 28.20 & -69 52 05.2 &    ...    &    ...     & 19.256  & 0.099  & 18.013  & 0.066  & 16.625  & 0.081 &  1.243    &  0.119  &  13.20  &    ...    \\[0.05cm]
  & 05 29 28.33 & -69 52 12.1 & 19.875 & 0.168      & 18.834  & 0.099  & 18.431  & 0.073  & 17.396  & 0.147 &  0.403    &  0.123  &  12.60  &    ...    \\[0.05cm]
\hline
23       & \textbf{05 30 48.60}  &  \textbf{-66 06 35.5} &	   ...    &    ...    & \textbf{18.270} & \textbf{0.129}	& \textbf{18.082} & \textbf{0.052} &    ...    &    ...    & \textbf{0.188} & \textbf{0.139} &   \textbf{21.72} & n,A \\[0.05cm]
           & \textbf{05 30 48.79} & \textbf{-66 06 35.5}  & \textbf{17.879} & \textbf{0.089} & \textbf{17.719} & \textbf{0.051} & \textbf{18.165} & \textbf{0.029} & \textbf{17.811} & \textbf{0.107} & \textbf{-0.446}  &  \textbf{0.059}  &  \textbf{22.13} &  n,A \\[0.05cm]
\hline
28  & \textbf{05 32 24.83} & \textbf{-71 06 56.0}	& \textbf{18.060} & \textbf{0.046} & \textbf{18.354} & \textbf{0.032} & \textbf{18.354} & \textbf{0.034} & 	\textbf{18.413} & 	\textbf{0.051} &      \textbf{0.000} &   \textbf{0.047} &  \textbf{37.48}  & n  \\[0.05cm] 
\hline      
29 &  05 32 20.71 & -65 35 17.7 & 20.035 & 0.116 & 19.644 & 0.028 & 18.745 & 0.030 & 17.789 & 0.037 & 0.899 &  0.041 &  29.79  &    ...    \\[0.05cm]
      &  05 32 20.90 & -65 35 25.9 & 20.377 & 0.281 & 19.812 & 0.059 & 19.540 &  0.035 & 19.692 & 0.097 &  0.272 &  0.069 &  32.09  &    ...    \\[0.05cm]
\hline
\end{tabular}
\end{sidewaystable*}

\begin{sidewaystable*}
\hspace{-8.5cm}{\bf Table 4}{\it \,\,--  continued}\\
\centering
  \begin{tabular}{@{}lcccccccccccccc@{}}
\hline
Src ID & R.A. & Decl. & U & errU & B & errB & V & errV & I & errI & B-V & err(B-V) & Offset & C/part\\
            & (h\, m\, s) & (\degr\, \arcmin\, \arcsec)  &  \multicolumn{10}{c}{(mag)} & (\arcsec)& \\[0.2cm]
$[1]$ & $[2]$ & $[3]$ & $[4]$ & $[5]$ & $[6]$ & $[7]$ & $[8]$ & $[9]$ & $[10]$ & $[11]$ & $[12]$ & $[13]$ & $[14]$ & $[15]$\\
\hline
      &  05 32 21.98 & -65 35 15.9 & 20.239 & 0.133 & 19.877 & 0.063 & 18.912 & 0.040 & 18.105 & 0.095 & 0.965 &  0.075 &  21.72  &    ...    \\[0.05cm]
      &  05 32 23.53 & -65 35 33.3 & 20.661 & 0.160 & 20.012 & 0.038 & 19.043 & 0.037 & 18.127 & 0.043 & 0.969 &  0.053 &  26.65  &    ...    \\[0.05cm]
      & \textbf{05 32 24.24} & \textbf{-65 34 59.4} &    ...   	&    ...    & \textbf{19.550}  & \textbf{0.042	} & \textbf{19.490}	& \textbf{0.037} & \textbf{19.640} & \textbf{0.109}  & \textbf{0.060}   &   \textbf{0.056}  & \textbf{11.61}	 &  n,A \\[0.05cm]
      &  05 32 27.65 & -65 34 49.0 & 19.789 & 0.123 & 19.722 & 0.049 & 18.837 & 0.066 & 17.919 & 0.060 & 0.885 &  0.082 &  24.74  &    ...    \\[0.05cm]
      & \textbf{05 32 27.72}  & \textbf{-65 35 02.4}  & \textbf{20.149}  & \textbf{0.203}  & \textbf{19.639}  & \textbf{0.047}  & \textbf{19.281}  & \textbf{0.038}  & \textbf{19.181}  & \textbf{0.078}  & \textbf{0.358}  & \textbf{0.060}  & \textbf{16.37}	 &  n,A \\[0.05cm]
      &  05 32 28.33 & -65 35 33.8 & 19.628 & 0.090 & 18.717 & 0.028 & 17.600 & 0.032 & 16.590 & 0.030 & 1.117 &  0.043 &  31.08  &    ...    \\[0.05cm]
      &  05 32 28.99 & -65 35 29.9 & 20.504 & 0.138 & 20.191 & 0.043 & 19.231 & 0.032 & 18.415 & 0.142 & 0.960 &  0.054 &  31.00  &    ...    \\[0.05cm]
      &  05 32 29.19 & -65 34 59.5 &    ...    &    ...    & 18.893 & 0.034 & 17.470 & 0.065 &    ...    &    ...    & 1.423 & 0.073 & 25.91  &    ...    \\[0.05cm]
\hline
35   &  \textbf{05 35 56.76} & \textbf{-65 30 22.8} & \textbf{17.558} & \textbf{0.048} & \textbf{17.825} & \textbf{0.031} & \textbf{17.882} & \textbf{0.027} & \textbf{18.060} & \textbf{0.051} & \textbf{-0.057} &  \textbf{0.041} &   \textbf{21.57} &   n,A \\[0.05cm]
       &  \textbf{05 35 51.28} & \textbf{-65 30 22.8} & \textbf{18.630} & \textbf{0.054} & \textbf{18.822} & \textbf{0.083} & \textbf{18.818} & \textbf{0.036} & \textbf{19.061} & \textbf{0.120} & \textbf{0.004} & \textbf{0.090}  &    \textbf{19.25}    &  n,A \\[0.05cm]\hline
39   &  \textbf{05 41 20.10}  &  \textbf{-69 36 22.9} & \textbf{11.307} &  \textbf{0.067} &  \textbf{11.790}  &   \textbf{0.093} &  \textbf{11.790}  &  \textbf{0.051} &  \textbf{11.910} &  \textbf{0.130}  &  \textbf{0.000}  &  \textbf{0.106}  &  \textbf{12.52} & k,A \\[0.05cm]
\hline
45   & \textbf{05 46 44.77} & \textbf{-68 51 43.4} &  \textbf{17.844} &   \textbf{0.070} &  \textbf{17.994} &   \textbf{0.038}  & \textbf{17.890}  &  \textbf{0.059} &  \textbf{17.934} &   \textbf{0.061} &  \textbf{0.104} & \textbf{0.070} & \textbf{19.41}  & n,A \\[0.05cm]
      & \textbf{05 46 44.97} & \textbf{-68 50 53.3}  & \textbf{16.231}  &  \textbf{0.055}  & \textbf{16.905}  &  \textbf{0.050}  & \textbf{16.887}  &  \textbf{0.033}  & \textbf{17.165} &   \textbf{0.131} &  \textbf{0.018} & \textbf{0.060}  & \textbf{56.67}   & n,A \\[0.05cm]
      & \textbf{05 46 54.37} & \textbf{-68 51 06.9}  & \textbf{17.819}  &  \textbf{0.067}  & \textbf{18.175}  &  \textbf{0.042}  & \textbf{18.115}  &  \textbf{0.035}  & \textbf{18.138}  &  \textbf{0.053} &  \textbf{0.060} &  \textbf{0.055} & \textbf{51.82}   & n,A \\[0.05cm]
\hline
\end{tabular}

 \begin{minipage}{195mm}
\footnotetext{

\noindent
Notes --  Similarly to Table \typeout{\ref{tableOpt}}2, the most likely counterpart of each X-ray source is indicated in bold. Only X-ray sources with a probable optical counterpart in
larger than 10\arcsec\, distances are shown in this table. For
example, although HMXB \#3 has an 11.1\arcsec\, positional
uncertainty, it is not listed in this table, since its most likely
optical counterpart is a match found using a search radius of
10\arcsec (i.e. listed in Table \typeout{\ref{tableBright}}3). Three dots indicate that an entry was not available in the MCPS catalog (see also general notes of Tables \typeout{\ref{tableOpt}}2 and \typeout{\ref{tableBright}}3). Eleven sources have bright matches ($V\lesssim20$ mag and $B-V\lesssim0.2$ mag) in larger than 10\arcsec\, distances from the X-ray position. Two of those sources (HMXB\#4 and \#28 with 14.5\arcsec\, and 24.3\arcsec\, positional uncertainties, respectively) have an unambiguous counterpart even at those large distances. The remaining 9 sources have matches usually in both Tables \typeout{\ref{tableOpt}}2 and \typeout{\ref{tableBright}}3.\\
    }
 
 \end{minipage}

\end{sidewaystable*}

\clearpage
\begin{sidewaystable*}
\hspace{-8.cm}{\bf Table 5.} Final classification of sources listed as HMXBs in the literature\label{finalclass}\\
\centering
  \begin{tabular}{@{}lllll@{}}
\hline
\multicolumn{2}{c}{X-ray Source} & \multicolumn{2}{c}{XRB Classification} & Notes\\
ID  & Name  &  Literature  & This Work & \\[0.2cm]
$[1]$ & $[2]$ & $[3]$ & $[4]$ & $[5]$\\
\hline
 1  & Swift J045106.8-694803, LXP187		&  HMXB/Be-XRB? (10), Be-XRB (70) & Be-XRB &   ...   \\[0.05cm]
\hline
2 & Swift J045558.9-702001    & Be-XRB (72) & Be-XRB &   ...   \\[0.05cm]
\hline
3$\star$  & RX J0456.9-6824	& HMXB? (37,38)  & Be-XRB? &   ...    \\[0.05cm]
\hline
4$\star$  & RX J0457.2-6612     & HMXB? (37,38)  &  Be-XRB? &   ...    \\[0.05cm]
\hline
5  & IGR J05007-7047 (23), CXOU J050046.0-704436 (20)   & HMXB (15), Be-XRB? (24)   & Be-XRB? &    ...   \\[0.05cm]
\hline
6  & RX J0501.6-7034,  CAL 9, 		          & Be-XRB (62)   & Be-XRB &    ...    \\[0.05cm]
     & 2E 0501.8-7038, 1E 0501.8-7036               &                            &                &                  \\[0.05cm]
\hline
7  & RX J0502.9-6626, CAL E, LXP4.10             & Be-XRB (62)  & Be-XRB &   ...     \\[0.05cm]
\hline
8$\star$  & RX J0507.6-6847, RX J050736-6847.8      & HMXB/Be-XRB? (36)   & Be-XRB? &   ...     \\[0.05cm]
\hline
9       & LXP169, XMMU J050755.4-682505   & eclipsing Be-XRB  (73)   & Be-XRB & we do not attempt to identify the eclipsing nature of this system    \\[0.05cm]
\hline
10$\star$  & RX J0512.6-6717      & HMXB? (38)   & Be-XRB? &   ...    \\[0.05cm]
\hline
11  & Swift J0513.4-6547 (12), LXP27.3		 & HMXB (12), Be-XRB? (45), Be-XRB (76)   & Be-XRB &    ...    \\[0.05cm]
\hline
12 & RX J0516.0-6916             				& HMXB? (25)   & Be-XRB? & see Appendix \ref{newspec}  \\[0.05cm]
\hline
13 & XMMU J052016.0-692505 	                   & likely WD/Be-XRB (8)   & Be-XRB & we do not attempt to classify the compact object    \\[0.05cm]
\hline
14 & RX J0520.5-6932, LXP8.04             		&  Be-XRB (4)   & Be-XRB & see Appendix \ref{appendixTOOblue}   \\[0.05cm]
\hline
15$\star$ & RX J0523.2-7004     & HMXB (38)   & Be-XRB? & see Appendix \ref{equalprob}    \\[0.05cm]
\hline
16$\star$ & RX J0524.2-6620     & HMXB (38)   & Be-XRB? &   ...     \\[0.05cm]
\hline
17$\star$ & RX J0527.1-7005     & HMXB (38)   & not an HMXB & see Appendix \ref{aboveRGB}  \\[0.05cm]
\hline
18$\star$ & RX J0527.3-6552     & HMXB? (37,38)   & Be-XRB? &   ...    \\[0.05cm]
\hline
19$\star$ & RX J0529.4-6952     & HMXB (37,38)   & Be-XRB? & see Appendix \ref{equalprob2} \\[0.05cm]
\hline
20 &   XMMU J052947.4-655639 (6), LXP69.2,   & Be-XRB (5)   & Be-XRB &   ...    \\[0.05cm]
      &  RX J0529.8-6556 (5), RX J0529.7-6556    &                         &                 &                 \\[0.05cm] 
\hline
\end{tabular}
\end{sidewaystable*}

\begin{sidewaystable*}
\hspace{-8.5cm}{\bf Table 5}{\it \,\,--  continued}\\
\centering
  \begin{tabular}{@{}lllll@{}}
\hline
\multicolumn{2}{c}{X-ray Source} & \multicolumn{2}{c}{XRB Classification} & Notes\\
ID  & Name  &  Literature  & This Work & \\[0.2cm]
$[1]$ & $[2]$ & $[3]$ & $[4]$ & $[5]$\\
\hline
21 & 	XMMU J053011.2-655122,  LXP272,            & Be-XRB? (6)  & Be-XRB? &   ...    \\[0.05cm] 
     & RX J0530.1-6551                                                &                         &                   &                  \\[0.05cm] 
\hline
22 & Swift J053041.9-665426 (58), LXP28.8     &  Be-XRB? (59), Be-XRB (26)  & Be-XRB &   ...    \\[0.05cm]
\hline
23$\star$ & RX J0530.7-6606     & HMXB (38)   & Be-XRB? & see Appendix \ref{equallyBRIGHTcparts}   \\[0.05cm]
\hline
24 & RX J0531.2-6607,  EXO 0531.1-6609,                                    & Be-XRB (47)   & Be-XRB  & see Appendix \ref{diffXRAYsrcs}  \\[0.05cm]
      & LXP13.7, XMMU J053113.3-660705,                                     &                           &                   &                                                                  \\[0.05cm]
      & EXO 053109-6609.2 (46), XMMU J053113.1-660707,       &                            &                  &                                                                  \\[0.05cm]
      & 1RXS J053111.0-660657 (19),  IGR J05305-6559 (69)      &                           &                  &                                                                   \\[0.05cm]
\hline
25 & XMMU J053115.4-705350 			& HMXB (48, 71) &  SG-XRB?  & see Appendix \ref{newSGXRB?}  \\[0.05cm]
\hline
26$\star$ & XMMU J053118.2-660730 	& HMXB? (48)   & Be-XRB? & see Appendix \ref{diffXRAYsrcs}  \\[0.05cm]
\hline
27 & RX J0531.5-6518,            		                   & HMXB (49), BeXRB? (1)  & Be-XRB? &   ...    \\[0.05cm] 
      & 1RXS J053137.1-651759 (19)    		&                                               &                   &                 \\[0.05cm]
\hline
28$\star$ & RX J0532.3-7107, CAL 50 (19),        & HMXB? (37,38)  & Be-XRB? &    ...   \\[0.05cm]
      & 1RXS J053219.5-710806 (19)                                               &      	                        &                   &                \\[0.05cm]
\hline
29$\star$ & RX J0532.4-6535,              & HMXB? (49)   & not an HMXB  & see Appendix \ref{RGBcparts}  \\[0.05cm]
      & 1RXS J053226.0-653505 (19)                             &      	                   &                         &                                                               \\[0.05cm]
\hline
30 & RX J0532.5-6551, XMMUJ053232.4-655139,  & SG-XRB (51)   & SG-XRB  &  see Appendix \ref{appendixSGXRB} \\[0.05cm]
      &  1RXS J053224.1-655112 (19)                            &                           &                   &                                                                        \\[0.05cm]
\hline
31 & 2A 0532-664, LMC X-4, LXP13.5,                &  HMXB (79)  & HMXB  & see Appendix \ref{appendixUband}  \\[0.05cm]
      &  RX J0532.8-6622 (19), CAL 49,                  &                 &              &                                                                       \\[0.05cm]
      & 1RXS J053246.1-662203 (38),    		 &                  &              &                                                                       \\[0.05cm]
      &  4U 0532Ð66 (57), RASS 232 (49)              &                  &              &                                                                       \\[0.05cm]
\hline
32 & Swift J053321.3-684121 (66)                        & HMXB? (66) & SG-XRB  &   ...   \\[0.05cm]
\hline
33 & RX J0535.0-6700             				  & Be-XRB (1)  & Be-XRB  &   ...    \\[0.05cm]
\hline
34 & 1A 0535-668, RX J0535.6-6651, LXP0.07,      & Be-XRB (52)   & Be-XRB  &   ...    \\[0.05cm]
      & CAL G (19),  1RXS J053539.0-665158 (19)    &                           &                  &                 \\[0.05cm]  
\hline
\end{tabular}
\end{sidewaystable*}

\begin{sidewaystable*}
\hspace{-8.5cm}{\bf Table 5}{\it \,\,--  continued}\\
\centering
  \begin{tabular}{@{}lllll@{}}
\hline
\multicolumn{2}{c}{X-ray Source} & \multicolumn{2}{c}{XRB Classification} & Notes\\
ID  & Name  &  Literature  & This Work & \\[0.2cm]
$[1]$ & $[2]$ & $[3]$ & $[4]$ & $[5]$\\
\hline
35$\star$ & RX J0535.8-6530,                           & HMXB? (49)                               & not an HMXB   & see Appendix \ref{RGBcparts}  \\[0.05cm]
                  & 1RXS J053555.0-653039 (19)     &                                                       &                            &                                                             \\[0.05cm]
\hline
36 & 1H 0538-641, LMC X-3,                                 & BH-HMXB (18)                       &   ...               & there is no MCPS coverage  \\[0.05cm]
     & XMMU J053856.7-640503 (19),   		 &                                                                           &                &   \\[0.05cm] 
      & 1RXS J053855.6-640457 (19),	  		 &                                                                           &                &   \\[0.05cm]                         
      & 1RXS J053855.5-640457 (19),	  		 &                                                                           &                &   \\[0.05cm]                         
      & RX J0538.9-6405, CAL 70 (19),  		 &                                                                           &                &   \\[0.05cm]               		 
      & 3A 0539-641, 4U 0538-64 (19),  		 &                                                                           &                &   \\[0.05cm]               		 
\hline
37 & 3A 0540-697, LMC X-1,                           	& BH-HMXB (53,54)                     & HMXB  &  we do not attempt to classify the compact object;  \\[0.05cm]
   & 1RXS J053938.8-694515 (19)    		        &                                     &       & counterpart compatible with having an OB spectral type  \\[0.05cm]
\hline
38 & IGR J05414-6858 (13), LXP4.42		&  Be-XRB? (14), Be-XRB (61)   & Be-XRB  &   ...     \\[0.05cm]
\hline
39 & RX J0541.4-6936             				& SG-XRB? (2)   & SG-XRB  &  see Appendix \ref{stillSGXRB}  \\[0.05cm]
\hline
40 & XMMU J054134.7-682550, LXP60.8	& HMXB? (48), Be-XRB? (60)  & Be-XRB?  &   ...    \\[0.05cm]
\hline
41 & RX J0541.5-6833, RX J0541.6-6832 (40)     & HMXB? (2)  & HMXB  &  see Appendix \ref{simplyHMXB} \\[0.05cm]
\hline
42$\star$ & RX J0543.9-6539 (38)  &  HMXB? (37)  & not an  HMXB  & see Appendix \ref{RedClump}  \\[0.05cm]
\hline
43 & 1SAX J0544.1-7100,  LXP96.1,      		& Be-XRB (4)  & Be-XRB  &   ...    \\[0.05cm]
      & RX J0544.1-7100 (38),                                 &                        &                 &                 \\[0.05cm] 
      & AX J0548-704, AX J0544.1-7100 (40),     &                        &                 &                  \\[0.05cm]  
      & 1WGA J0544.1-7100                                    &                        &                 &                  \\[0.05cm]   
\hline
44 & H 0544-665   (41)                                          & Be-XRB (55,56)  & Be-XRB  &    ...   \\[0.05cm]
\hline
45$\star$ & RX J0546.8-6851, RX J0547.0-6852 (38),       & HMXB? (38) & Be-XRB?  & see Appendix \ref{NOTtypicalclass}  \\[0.05cm]
                                           & 2E 0547.2-6852, 1RXS J054655.3-685142     &                        &                    &   \\[0.05cm]
\hline
\end{tabular}

\begin{minipage}{180mm}
\footnotetext{

\noindent
Notes -- References given in parenthesis are similar to those listed in Table \typeout{\ref{tableCensus}}1.

\noindent
$\star$ New counterparts for these X-ray sources are identified in the present work.\\

 }
 \end{minipage}

\end{sidewaystable*}

\clearpage
\begin{table}
\hspace{-0.5cm}{\bf Table 6.} The young X-ray source population of the LMC\label{tblPOP}\\
\centering
  \begin{tabular}{@{}lcc@{}}
\hline
XRB type & \multicolumn{2}{c}{Total number of sources}	\\
                  & Literature & This work\\[0.2cm]
$[1]$ & $[2]$ & $[3]$\\
\hline
Be-XRBs                                               & 21     &   33  \\ 
SG-XRBs                                              &   2      &   4  \\
HMXBs (w/o further classification)   & 20     &  3   \\
BH-HMXB                                             &  2        &   ...  $\bullet$   \\
HMXBs (all)                                          &  45     &   40   \\
X-ray pulsars                                        & 14      &  14$\star$    \\
\hline
\end{tabular}

\begin{minipage}{80mm}
\footnotetext{

\noindent
Notes -- Each class includes confirmed and candidate systems. There is
one source (LMC X-3) without MCPS coverage, thus we do not classify
this system.\\
$\bullet$ With the present work we do not obtain info about the nature of the compact object, so we cannot classify any source as a BH-HMXB. For the one known BH-HMXB that there is MCPS coverage (LMC X-1; see Table \typeout{\ref{finalclass}}5), we only classify this X-ray source as an HMXB.\\
$\star$ We did not perform any timing analysis, so the number of known
X-ray pulsars is the one found in the literature. 
    }
 
 \end{minipage}

\end{table}

\clearpage
\begin{sidewaystable*}
\hspace{-5.cm}{\bf Table 7.} Compilation of metallicity values of B-type stars in the Magellanic Clouds\label{tblFeH}\\
\centering
  \begin{tabular}{@{}llcclccc@{}}
\hline
Galaxy & Source & R.A. & Decl. & Spectral &  [Fe/H] &  Z/\zsun$\dag$ &  Z$\ddag$ \\
 & Name & \multicolumn{2}{c}{(J2000.0)} & Type &  &   &   \\
 &  &(h\, m\, s)  & (\degr\, \arcmin\, \arcsec) &  & (-dex) &  &  \\[0.2cm]
$[1]$ & $[2]$ & $[3]$ & $[4]$ & $[5]$ & $[6]$ & $[7]$ & $[8]$\\
\hline
LMC    &    N11-024 (6)                              & 04 55 32.93 (6)           & -66 25 27.7 (6)                 &   B1 Ib (6)             & $-0.39\pm0.16$$\bullet$ (5)     &      $0.41\pm0.15$    &      $0.005\pm0.002$    \\[0.13cm] 
\cline{2-8}
           &    N11-017 (6)                                & 04 56 17.57 (6)           & -66 18 18.5 (6)                 &   B2.5 Iab (6)       & $-0.30\pm0.16$$\bullet$ (5)     &      $0.50\pm0.18$    &      $0.007\pm0.002$    \\[0.13cm]
\cline{2-8}
           &    N11-016 (6)                                & 04 56 20.59 (6)           & -66 27 14.0 (6)                 &   B1 Ib (6)             & $-0.24\pm0.28$$\bullet$ (5)     &      $0.58\pm0.37$    &      $0.008\pm0.005$    \\[0.13cm] 
\cline{2-8}
           &    N11-002 (6)                                & 04 56 23.51 (6)           & -66 29 51.7 (6)                 &   B3 Ia (6)             &  $-0.16\pm0.16$$\bullet$ (5)    &      $0.69\pm0.25$    &      $0.009\pm0.003$    \\[0.13cm]  
\cline{2-8}
           &    N11-014 (6)                                & 04 56 48.02 (6)           & -66 20 09.8 (6)                 &   B2 Iab (6)           &  $-0.27\pm0.17$$\bullet$ (5)    &      $0.54\pm0.22$    &      $0.007\pm0.003$    \\[0.13cm] 
\cline{2-8}
           &    N11-012 (6)                                & 04 56 51.15 (6)           & -66 31 48.3 (6)                 &   B1 Ia (6)             &  $-0.40\pm0.27$$\bullet$ (5)    &      $0.40\pm0.25$    &      $0.005\pm0.003$    \\[0.13cm]
\cline{2-8}
           &    N11-001 (6)                                & 04 57 08.85 (6)           & -66 23 25.1 (6)                 &   B2 Ia (6)             &  $-0.05\pm0.17$$\bullet$ (5)    &      $0.89\pm0.36$    &      $0.012\pm0.005$    \\[0.13cm]
\cline{2-8}
           &    N11-009 (6)                                & 04 57 17.68 (6)           & -66 26 31.5 (6)                 &   B3 Iab (6)           &  $-0.26\pm0.14$$\bullet$ (5)    &      $0.55\pm0.17$    &      $0.007\pm0.002$    \\[0.13cm]
\cline{2-8}
           &    N11-015 (6)                                & 04 57 22.08 (6)           & -66 24 27.5 (6)                 &   B0.7 Ib (6)          &  $-0.29\pm0.26$$\bullet$ (5)    &      $0.51\pm0.31$    &      $0.007\pm0.004$    \\[0.13cm] 
\cline{2-8}
           &    N11-110 (6)                                & 04 57 37.11 (6)          & -66 23 44.7 (6)                  &   B1 III (6)             & $-0.38\pm0.16$$\bullet$ (5)     &      $0.42\pm0.16$     &      $0.006\pm0.002$    \\[0.13cm]
\cline{2-8}
           &    N11-036 (6)                                & 04 57 41.00 (6)          & -66 29 56.4 (6)                  &   B0.5 Ib (6)          & $-0.33\pm0.23$$\bullet$ (5)     &      $0.47\pm0.25$    &      $0.006\pm0.003$    \\[0.13cm] 
\cline{2-8}
           &    PS 34-16 (3)                               &  05 04 32.5 (3)	           & -66 24 47 (3)                     &   1 dwarf               &  $\sim -0.17\pm0.06$ (3)                                    &      $\sim0.68$            &      $\sim0.009$              \\[0.13cm] 
           &    NGC 1818/D1                            & 05 04 32.5 (14)	          & -66 24 51.0 (14)                &  B1 V (14)             & $-0.16\pm0.30$ (1)                                             &      $0.69\pm0.48$    &      $0.009\pm0.006$    \\[0.13cm]
\cline{2-8}
           &    NGC 2004/B15	(6)                    & 05 29 11.46 (6)          & -67 15 24.40 (6)                & B1.5 II (6)             & $-0.45\pm0.40$ (1)                                              &      $0.35\pm0.33$    &      $0.005\pm0.004$    \\[0.13cm]
\cline{2-8}
           &    NGC 2004/B30 (6)                    & 05 30 11.03 (6)	          & -67 22 57.10 (6)                & B0.2 Ib (6)            & $-0.42\pm0.30$ (1)                                              &      $0.38\pm0.26$    &      $0.005\pm0.004$    \\[0.13cm] 
\cline{2-8}
            & BRU 217 (1)                                 &  05 34 51.76  (11,7)   &  -69 22 44.7 (11,7)          & B Ib (1), O/B0 (11)  & $-0.55\pm0.30$ (1)                                              &      $0.28\pm0.20$    &      $0.004\pm0.003$    \\[0.13cm]
\cline{2-8}
            & BRU 231 (1)                                 & 05 35 17.44   (11,7)   &  -69 19 54.2 (11,7)          &  B Ib (1), B1/2 (11)     & $-0.46\pm0.30$ (1)                                              &      $0.35\pm0.24$    &      $0.005\pm0.003$    \\[0.13cm] 
\cline{2-8}
            & LH 10-3270 (20)                          & 05 57 21.2 (20)	 & -66 25 01 (20)	           & B1 V (20)        & -0.9 (16)                                                                & 0.13                   &    0.002                      \\
\hline
\hline
  SMC & AV 175 (1)                                     & 00 56 38.07 (12)        & -72 36 35.3 (12)               & B1 I (12)               & $-0.86\pm0.40$ (1)                                              &      $0.14\pm0.13$    &      $0.002\pm0.002$    \\[0.13cm] 
\cline{2-8}
            &  NGC346-11 (2)                           &  00 57 29.50	(7)      & -72 16 00.5 (7)                 & B9 II (6)                & $\sim-0.7$$\star$ (9)                   &       $\sim0.20$           &      $\sim0.003$     \\[0.13cm]
\cline{2-8}
            &  NGC346-039 (6)                         & 00 57 47.42 (6)	& -72 16 40.44	(6)                & B0.7 V (6)            & $-0.56\pm0.17$$\bullet$ (5)      &      $0.28\pm0.11$     &      $0.004\pm0.001$    \\[0.13cm] 
\cline{2-8}
            & AV 218 (1)                                     & 00 59 04.34 (12)        & -72 19 40.8 (12)               & B1 I (12)               & $-0.51\pm0.30$ (1)                                              &      $0.31\pm0.22$    &      $0.004\pm0.003$    \\[0.13cm]
\hline
\end{tabular}
\end{sidewaystable*}

\begin{sidewaystable*}
\hspace{-8.5cm}{\bf Table 7}{\it \,\,--  continued}\\
\centering
  \begin{tabular}{@{}llcclccc@{}}
\hline
Galaxy & Source & R.A. & Decl. & Spectral &  [Fe/H] &  Z/\zsun$\dag$ &  Z$\ddag$ \\
 & Name & \multicolumn{2}{c}{(J2000.0)} & Type &  &   &   \\
 &  &(h\, m\, s)  & (\degr\, \arcmin\, \arcsec) &  & (-dex) &  &  \\[0.2cm]
$[1]$ & $[2]$ & $[3]$ & $[4]$ & $[5]$ & $[6]$ & $[7]$ & $[8]$\\
\cline{2-8}
            &  NGC346-637 (2)                         & 00 59 14.60$\rhd$ (10) & -72 12 01.45$\rhd$ (10)    &  B0 V (10)   & $\sim-0.8$$\star$ (9)   &   $\sim0.16$   &     $\sim0.002$  \\[0.13cm]            
\cline{2-8}
            &  NGC346-037 (6)                         & 00 59 18.84 (6)	& -72 06 29.27 (6)                & B3 III (6)               & $-0.67\pm0.21$$\bullet$ (5)      &      $0.21\pm0.11$     &      $0.003\pm0.001$    \\[0.13cm] 
\cline{2-8}
            &  NGC346-021 (6)                         & 00 59 19.58 (6)	& -72 14 50.47	(6)                & B1 III (6)               & $-0.36\pm0.18$$\bullet$ (5)      &      $0.44\pm0.19$     &      $0.006\pm0.002$    \\[0.13cm] 
\cline{2-8}
            &  NGC346-044 (6)                         & 00 59 26.46 (6)	& -72 13 11.78	(6)                & B1 II (6)                & $-0.51\pm0.18$$\bullet$ (5)      &      $0.31\pm0.13$     &      $0.004\pm0.002$    \\[0.13cm] 
\cline{2-8}
            & AV 304 (2,4)                                  & 01 02 21.47 (7)          & -72 39 14.7 (7)                 &  B0.5 V (8,4)        & $\sim-0.8$$\star$ (9)                   &       $\sim0.16$          &      $\sim0.002$    \\[0.13cm]
            &                                                         &                                       &                                             &                               & $-0.79\pm0.19$ (4)                                               &      $0.16\pm0.07$    &      $0.002\pm0.001$    \\[0.13cm]
            &                                                         &                                       &                                             &                               & -1.2 (16)                                                                   & 0.06                   &    0.001                      \\
\hline
\hline
SMC    & SK 194 (14) &	01 45 03 (14)     & -74 31 32 (14)	            & B9 Ia (15)            & $\sim-1.35$ (14)  &   $\sim0.04$          & $\sim0.001$     \\[0.13cm]
\cline{2-8}
Wing$\boxplus$   &  IDK-D2 (13)                                  & 02 06 59.06 (13)      & -74 16 55.9  (13)               & B2 (13)                 &  $\sim-1.1$$\star$ (9)                  &       $\sim0.08$           &     $\sim0.001$       \\
\cline{2-8}
           & DI 1239 (19)	                        & 02 30 40.81 (19)   & -74 04 45.3 (19)	          & B1 (19)           & -1.6 (16)                                                                   & 0.03                   &    $\sim4\times10^{-4}$                      \\[0.13cm]
\hline
\hline
Magellanic     & DI 1388 (18)	                        & 02 57 11.94 (18)   & -72 52 54.61 (18)	          & B0 V (14), O8 (13) & -1.9 (16)                                                                   & 0.01                  &    $\sim10^{-4}$                     \\[0.13cm]
\cline{2-8}
Bridge$\boxplus$                         & DGIK 975 (18)	                         & 04 19 58.63 (18)   & -73 52 25.80 (18)	          & B2 (17)           & -1.7 (16)                                                                   & 0.02                   &    $\sim10^{-4}$                      \\
\hline
\end{tabular}

\begin{minipage}{195mm}
\footnotetext{

\noindent
$\dag$ Using Z/\zsun ${\rm =10^{[Fe/H]}}$.\\
$\ddag$ With the solar chemical composition of \citet{Asplund et
  al. 2009}, the mass fractions of H, He, and metals in the
present-day photosphere become X $=$ 0.7381, Y $=$ 0.2485, and Z $=$
0.0134, i.e. the solar metallicity has been revised from the canonical
2\% value first recommended in the literature in the 90's. In the
present study we use the revised value (\zsun$=$ 0.0134).\\
$\bullet$ Defined as the difference between the weighted mean absolute abundances of \cite{2007A&A...471..625T} and the Fe abundance in the present-day solar photosphere (the latter is equal to $7.50\pm0.04$ from \citealt{Asplund et al. 2009}).\\
$\star$ Heavy element depletion ([M/H]).\\
$\rhd$ Original equatorial B1950.0 coordinates were transformed to J2000.0 using NASA/IPAC Extragalactic Database (NED; https://ned.ipac.caltech.edu/forms/calculator.html).\\
$\boxplus$ The SMC Wing is the area defined as ${\rm -75.5\degree \lesssim Decl. \lesssim -73.5\degree}$ and $\rm{1.6^{h} \lesssim R.A. \lesssim 2.6^{h}}$, while for the Magellanic Bridge it is $\rm{R.A. \gtrsim 3^{h}}$ (see Fig.1 in \citealt{2008ApJ...678..219L}).\\

\noindent
{\bf References:} (1) \cite{2000A&A...353..655K}; (2)
\cite{1993A&A...277...10R}; (3) \cite{2002A&A...396...53R}; (4)
\cite{2006ASPC..348..136P}; (5)  \cite{2007A&A...471..625T}; (6)
\cite{2006A&A...456..623E}; (7) \cite{2006AJ....131.1163S}; (8)
\cite{1982PASP...94...31C}; (9) \cite{1995IrAJ...22...17R}; (10)
\cite{1989AJ.....98.1305M}; (11)
ftp://ftp.lowell.edu/pub/bas/starcats/brunet.lmc (version:  15 Feb
2008; originally in \citealt{1975A&AS...21..109B}); (12) ftp://ftp.lowell.edu/pub/bas/starcats/azv.dat (version:  28 Aug 2008); (13) \cite{1990AJ.....99..191I}; (14) \cite{2009PASP..121..634B}; (15) \cite{1977A&AS...30..261A}; (16) \cite{2008MNRAS.385.2261D}; (17) \cite{2010ApJS..187..149A}; (18) \cite{2002ApJ...578..126L}; (19) \cite{1999A&A...348..728R}; (20) \cite{1992AJ....103.1205P}.\\

    }

 \end{minipage}

\end{sidewaystable*}

\clearpage
\begin{sidewaystable*}
\hspace{-1.5cm}{\bf Table 8.} Compilation of age and metallicity values of young ($<100$ Myr) star clusters in the Magellanic Clouds\label{tblAgeFeH}\\
\centering
  \begin{tabular}{@{}llccccccccc@{}}
\hline
Galaxy & Cluster & \multicolumn{2}{c}{Age} & \multicolumn{2}{c}{[Fe/H]} & \multicolumn{2}{c}{Z/\zsun} &  \multicolumn{2}{c}{Z} &  Ref.$\spadesuit$\\
 &  & Value & Error & Value & Error & Value & Error & Value & Error & \\[0.2cm]
 &  &  \multicolumn{2}{c}{(Myr)} & \multicolumn{2}{c}{(-dex)} &  &  &  &  & \\[0.2cm]
$[1]$ & $[2]$ & $[3]$ & $[4]$ & $[5]$ & $[6]$ & $[7]$ & $[8]$ & $[9]$ & $[10]$ & $[11]$\\
\hline
LMC &  LH47/48         &   2.00           &      ...      &   -0.40                   &       ...           &   0.40             &       ...      &  0.005           &      ...       & (1) \\[0.05cm] 
          &  NGC1850B   &   3.98           &   0.92         &   -0.12                   &    0.03             &   0.76             &    0.05         &  0.010           &   0.001       & (2) \\[0.05cm] 
          &  NGC1948      &   7.08           &   2.45         &  -0.40                    &      ...           &   0.40             &       ...      &  0.005           &      ...       & (3) \\[0.05cm] 
          &  NGC1858      &   7.94           &      ...      &   -0.40                    &      ...           &   0.40             &       ...      &  0.005           &      ...       & (4) \\[0.05cm] 
          &  LH72              &   8.91           &   5.13         &   -0.60                    &      ...           &   0.25             &       ...      &  0.003           &      ...       & (5) \\[0.05cm] 
          &  R136              &   10.00          &      ...      &   -0.40                     &      ...          &   0.40             &       ...      &  0.005           &      ...       & (6) \\[0.05cm] 
          &  LH52/53        &   10.00          &     ...      &   -0.40                    &      ...           &   0.40             &       ...      &  0.005           &      ...       & (7)  \\[0.05cm] 
          &  NGC2027     &   11.48          &   3.70        &   -0.40                    &      ...           &   0.40             &       ...      &  0.005           &      ...       & (8)  \\[0.05cm] 
          &  NGC1955     &   15.49          &   5.35        &   -0.40                    &      ...           &   0.40             &       ...      &  0.005           &      ...       & (8)  \\[0.05cm] 
          &  NGC2004     &   15.49          &   5.35        & $\sim$-0.40          &    ...             &   0.40             &       ...      &  0.005           &      ...       & (8)  \\[0.05cm] 
          &  LH77             &   15.85          &   5.11        &   -0.40                    &      ...           &   0.40             &       ...      &  0.005           &      ...       & (8)  \\[0.05cm] 
          &  SL503           &   15.85          &   8.03        &   -0.40                     &      ...          &   0.40             &       ...      &  0.005           &      ...       & (8)  \\[0.05cm] 
          & NGC1818      &  17.78           & 16.38       & $\sim$-0.40           &     ...            & $\sim$0.40  &      ...       & $\sim$0.005 &      ...       &   (18,19)   \\[0.05cm] 
          &                         &   44.67           &  5.14          &     ...                    &       ...          &     ...          &      ...       &      ...           &      ...       &  (18)  \\[0.05cm] 
          &  NGC2100     &  $<$40         &     ...         &   $>$-0.03             &      ...           &   $>$0.93     &      ...       &   $>$0.013     &      ...       & (13)  \\[0.05cm] 
          &                         &   15.85           &      ...         &   -0.32                  &  0.03           &  0.48             &   0.03         &  0.006             &   0.000      &     (17,16)    \\[0.05cm] 
          &  SL218           &   50                  &  10             &   -0.40                &  0.15            &   0.40            & 0.14           &  0.005             &  0.002        & (20)  \\[0.05cm] 
          &  NGC1711     &   50.12            &  5.77         &   -0.57                 &  0.17               &  0.27             &  0.11          &  0.004             &  0.001        & (10) \\[0.05cm] 
          &  NGC1850A  &   50.12            &  11.54       &   -0.12                 &  0.03               &  0.76             &  0.05          &  0.010             &  0.001        & (2)  \\[0.05cm] 
          &  NGC1863     &   50                &  10             &  -0.40                  &  0.15               &  0.40             &  0.14          &  0.005             &  0.002        & (20)  \\[0.05cm] 
          &  BRHT4b        &   100              &  25              &  -0.40                 & 0.15                &  0.40            & 0.14            &  0.005            & 0.002        & (20)  \\[0.05cm] 
          &  NGC1838     &   100              &  25              &   -0.40                & 0.15                &  0.40            &  0.14           & 0.005             &  0.002       & (20)  \\[0.05cm] 
          &  NGC1866     &   100              &     ...        &  -0.43                    &  0.18               &  0.37           &  0.15            &  0.005            &  0.002       & (11)  \\[0.05cm] 
          &  NGC2136     &   100.00        &  23.03        &  -0.55                &  0.23               &  0.28            &  0.15           &  0.004            &  0.002       & (10)  \\[0.05cm] 
          &  NGC2164     &   100              &     ...       &  -0.20                     &  0.20               &   0.63           &  0.29           &  0.008            &  0.004       & (12)  \\[0.05cm] 
          &  NGC2214     &   100              &    ...        &  -0.20                     &  0.20               &  0.63            &  0.29           &  0.008            &  0.004       & (12)  \\[0.1cm] 
\hline
\end{tabular}
\end{sidewaystable*}

\begin{sidewaystable*}
\hspace{-8.5cm}{\bf Table 8}{\it \,\,--  continued}\\
\centering
  \begin{tabular}{@{}llccccccccc@{}}
\hline
Galaxy & Cluster & \multicolumn{2}{c}{Age} & \multicolumn{2}{c}{[Fe/H]} & \multicolumn{2}{c}{Z/\zsun} &  \multicolumn{2}{c}{Z} &  Ref.$\spadesuit$\\
 &  & Value & Error & Value & Error & Value & Error & Value & Error & \\[0.2cm]
 &  &  \multicolumn{2}{c}{(Myr)} & \multicolumn{2}{c}{(-dex)} &  &  &  &  & \\[0.2cm]
$[1]$ & $[2]$ & $[3]$ & $[4]$ & $[5]$ & $[6]$ & $[7]$ & $[8]$ & $[9]$ & $[10]$ & $[11]$\\
\hline
SMC &  L66                &   15                &  10            &   -0.70                     &    ...             &   0.20            &    ...        &  0.003             &     ...      &    (21,14)   \\[0.05cm] 
          &  L51                &   15                &  10            &   -0.70                     &    ...             &   0.20            &     ...       &  0.003             &    ...      &    (21,14)    \\[0.05cm] 
          &  L39                &   15                &  10            &   -0.70                     &    ...             &   0.20            &    ...        &  0.003             &    ...      &    (21,14)    \\[0.05cm] 
   &  L72$\equiv$NGC376   & 25      &  10             &   -0.70                   &     ...            &   0.20            &    ...        &  0.003             &    ...       & (22)  \\[0.1cm] 
   &  L49$\equiv$NGC299   & 25   &  $+$6$/-$5  &   -0.70                  &    ...               &   0.20            &     ...     &  0.003             &     ...       & (14)  \\[0.1cm] 
          &  NGC 330      &   25                 &  15            &   -0.82                  & 0.10             &   0.15            &  0.03          &  0.002             &  0.000       & (8) \\[0.05cm] 
          &                         &   30                 & 2               &   -0.82                  &   0.11           &   0.15            &  0.04          &  0.002             &  0.001       & (24,15)  \\[0.05cm]  
          &                         &   31.62           &     ...           &  -0.22                   &    ...              &   0.60            &    ...        &  0.008           &     ...       &   (25)   \\[0.05cm] 
          &                         &   17.5$\bullet$  & 7.5$\bullet$  &    $<$-0.22            &    ...             &  $<$0.60       &    ...        &  $<$0.008   &     ...       &  (26)    \\[0.05cm] 
          &                         &   29$\star$    & 19$\star$       &       ...                 &    ...             &    ...            &     ...       &     ...             &     ...       &  (26)   \\[0.05cm] 
          &  L63                &   45                 &     ...        &   -0.70                  &     ...            &   0.20             &     ...       &  0.003             &     ...       & (14)  \\[0.1cm] 
          &                         &  110  & $+$30$/-$20   & -0.70        &     ...            & 0.20              &     ...       &  0.003              &     ...      & (14)  \\[0.1cm] 
          &                         &   50.12            & $<$34.62  &      ...              &      ...           &     ...           &      ...       &      ...           &      ...       & (23)  \\[0.05cm] 
          &  HW8              &   50                 &  20           &   -0.70            &    ...              &   0.20            &    ...        &  0.003             &    ...       &    (21,14)    \\[0.05cm] 
          &  L50$\equiv$NGC306  &   80                    &  20                  &   -0.70               &    ...              &   0.20            &    ...        &  0.003             &    ...       & (14)  \\[0.05cm] 
\hline
\end{tabular}

\begin{minipage}{185mm}
\footnotetext{

\noindent
Notes -- The age and [Fe/H] values (Columns 3 and 5, respectively) are taken from the references listed in Column 11. The values of the remaining columns have been derived using Z/\zsun ${\rm =10^{[Fe/H]}}$, and \zsun$=0.0134$ (\citealt{Asplund et al. 2009}). For few clusters, e.g., LMC systems NGC2100 and NGC1818, the values of age and [Fe/H] are taken from two different references. In these cases, we list both references in Column 11: the first one corresponds to the age and the second one to the [Fe/H] value.\\

$\bullet, \star$ In the original reference, \cite{1995A&A...293..710C} give an age range (10--25 Myr and 10--48 Myr, respectively), while here we chose to list a mean age value along with its error (defined as half this age range).\\

\noindent
$\spadesuit$ {\bf References:} (1) \cite{1995ApJ...452..210O}; (2) \cite{1994ApJ...435L..43G}; (3) \cite{1996A&A...315..125W}; (4) \cite{1994A&A...284..447V}; (5) \cite{1997ApJ...475..545O}; (6) \cite{1995ApJ...448..179H}; (7) \cite{1995ApJ...446..622H}; (8) \cite{1998AJ....115.1934D}; (10) \cite{2000A&A...360..133D}; (11) \cite{1995A&A...294..648H}; (12) \cite{1991A&A...250..324S}; (13) \cite{2011ApJ...735...55C}; (14) \cite{2008MNRAS.389..429P}; (15) \cite{1999A&A...345..430H}; (16) \cite{1994A&A...282..717J}; (17) \cite{1991ApJS...76..185E}; (18) \cite{2009MNRAS.396.1665L}; (19) \cite{2000A&A...353..655K}; (20) \cite{2003MNRAS.343..851P}; (21) \cite{2005A&A...440..111P}; (22) \cite{2007MNRAS.377..300P}; (23) \cite{2006A&A...452..179C}; (24) \cite{2002ApJ...579..275S}; (25) \cite{1999AcA....49..157P}; (26) \cite{1995A&A...293..710C}
    }

 \end{minipage}

\end{sidewaystable*}

\clearpage
\begin{sidewaystable*}\small
\hspace{-8cm}{\bf Table 9.} Formation efficiency for different XRB types\label{FRates}\\
\centering
\begin{tabular}{@{}lcccccccc@{}}
\hline
XRB type &  N(XRBs) & \multicolumn{3}{c}{Major Star-Formation Burst} & {\color{black} N(MCPS)} &  XRB Formation & SFR required for &  {\color{black} N(XRBs)/M$_{\star}$} \\
 \multicolumn{2}{c}{} &   Age & Duration & Observed Mean Rate &  &  Efficiency & production of 1 system & \\
  \multicolumn{2}{c}{} &  \multicolumn{2}{c}{(Myr)} &  ($10^{-6}$\mdot/${\rm (arcmin)^{2}}$) & & (systems/(\mdot)) & ($10^{-3}$\mdot) & {\color{black} ($10^{-2}$\msun$^{-1}$)}\\[0.2cm]
$[1]$ & $[2]$ & $[3]$ & $[4]$ & $[5]$ & {\color{black} [6]} & $[7]$ & $[8]$ & {\color{black} $[9]$}\\
\hline
 all HMXBs      &  40   & 6.3  & $>$22     & $333.9_{-24.9}^{+32.2}$  & {\color{black} 38}  & {\color{black} $21.9_{-3.8}^{+4.1}$}    &  {\color{black} $45.7_{-8.0}^{+8.5}$} &  {\color{black} $0.6\pm0.1$}\\[0.25cm]
                          & ...    & 50.1$\dag$ & $>$123 & $84.6_{-7.7}^{+9.6}$     & ...       & ... &  ...  &  {\color{black} ...} \\[0.25cm]
 \hline\\
 all HMXBs w/o WD or BH & 38   & 6.3  & $>$11  & $318.7_{-23.7}^{+31.0}$   & {\color{black} 36}  & {\color{black} $23.0_{-4.1}^{+4.4}$}  &  {\color{black} $43.5_{-7.8}^{+8.2}$} &  {\color{black} $0.7\pm0.1$}\\[0.25cm]
                                               & ...   & 50.1$\dag$  & 108    & $79.3_{-7.3}^{+9.1}$         & ...      &  ...  &  ...  &  {\color{black} ...}   \\[0.25cm]
 \hline\\
  confirmed Be-XRBs &  16   & 12.6  & 20  &  $161.1_{-10.6}^{+12.9}$  & {\color{black} 16}  &  {\color{black} $43.1_{-11.2}^{+11.3}$} & {\color{black} $23.2_{-6.0}^{+6.1}$}  &  {\color{black} $0.3\pm0.1$}\\[0.25cm]
 \hline\\
  confirmed NS/Be-XRBs & 15    & 12.6  &  16  & $148.3_{-9.7}^{+12.0}$   & {\color{black} 15}   &  {\color{black} $46.8_{-12.5}^{+12.7}$} & {\color{black} $21.4_{-5.7}^{+5.8}$}  &  {\color{black} $0.3\pm0.1$}\\[0.25cm]
 \hline\\
  candidate Be-XRBs  & 17    & 6.3    & $>$10  &  $121.6_{-15.1}^{+19.4}$ & {\color{black} 16}   & {\color{black} $60.7_{-16.5}^{+17.6}$}  & {\color{black} $16.5_{-4.5}^{+4.8}$}   &  {\color{black} $0.8\pm0.2$}\\[0.25cm]
                                        & ...   & 100$\dag$  & 133      &  $48.6_{-4.4}^{+5.4}$ & ...  & ...   &  {\color{black} ...}  \\[0.25cm]
 \hline\\
 X-ray pulsars &  14  & 12.6  & 32  &  $113.3_{-9.5}^{+11.5}$ & {\color{black} 14}   &  {\color{black} $61.3_{-17.2}^{+17.5}$}   &  {\color{black} $16.3_{-4.6}^{+4.7}$}  &  {\color{black} $0.3\pm0.1$}\\[0.25cm]
 \hline\\
 SG-XRBs        &  4    &  6.3 & $>$11  & $81.4_{-10.2}^{+12.6}$ & {\color{black} 4}    &  {\color{black} $85.3_{-44.0}^{+44.7}$} & {\color{black} $11.7_{-6.0}^{+6.1}$}  &  {\color{black} $0.3\pm0.2$}\\[0.25cm]
 \hline\\
 BH-HMXB       &   1 & 6.3   &  $>$16 & $15.1_{-6.3}^{+5.6}$ & {\color{black} 1}   & {\color{black} $458.4\,\,\,(\leq947.8)$} & {\color{black} $2.2\,\,\,(\leq4.5)$} &  {\color{black} $0.5_{-0.5}^{+0.6}$}\\[0.25cm]
\hline\\
  LMC WD/Be-XRB  &   1 &  25.1 & 29 & $17.8_{-3.2}^{+3.4}$ & {\color{black} 1}   & {\color{black} $390.3\,\,\,(\leq788.0)$} & {\color{black} $2.6\pm2.6$}  &  {\color{black} $0.2\pm0.2$}\\[0.25cm]
 SMC WD/Be-XRB  &   1   &  42.2  &  110$\ddag$ & $28.6_{-13.6}^{+43.6}$ & {\color{black} 1}   & {\color{black} $242.7\,\,\,(\leq685.2)$}   & {\color{black} $4.1\,\,\,(\leq11.6)$}  &  {\color{black} $0.0_{-0.0}^{+0.1}$} \\[0.25cm]
\hline
\end{tabular}

\begin{minipage}{190mm}
\footnotetext{

\vspace{-0.5cm}
\noindent
Notes -- The duration of the major star-formation burst (Column 4) refers to the total duration of the star-formation episode and it includes the period before it reached its maximum SFR (see Figs. \ref{fig1SFHmean}, \ref{fig2SFHmean}). {\color{black} The number of MCPS subregions (Column 6; each 12\arcmin$\times$12\arcmin\, [HZ09]) denotes the total area used for the derivation of the average star-formation history of regions with different XRB types (as these are given in Columns 1 and 2).} The formation efficiencies presented here (Column 7) are derived as the ratio of the total number of XRBs for each class of objects to the maximum SFR (for more details see Section \ref{sectionFRs}). {\color{black} M$_{\star}$ is the total stellar mass produced on the specific SF burst episode.  The errors in the N(XRBs)/M$_{\star}$ quantity (listed in Column 8) reflect the upper and lower limits of the SFR in Figs. \ref{fig1SFHmean}, \ref{fig2SFHmean}.}\\

\noindent
$\dag$ This peak is not associated with the particular XRB population (see Section \ref{SFH-youngXRBs}). This non-symmetric SF burst was approximated by a Gaussian.\\
$\ddag$ Double-peak star-formation burst episode.

    }

 \end{minipage}

\end{sidewaystable*}

\clearpage
\begin{table}
\hspace{-1.5cm}{\bf Table 10.} Main-sequence lifetimes of OB-type stars\label{tblLIFE}\\
\centering
\begin{tabular}{@{}lcc@{}}
\hline
Spectral Class &  M &  ${\rm t_{MS}}$\\
  & (${\rm M_{\odot}}$) & (Myr)\\[0.2cm]
$[1]$ & $[2]$ & $[3]$\\
\hline
O7.5 V    &   22.9   &  4.0  \\[0.05cm]
O8 V       &    20.8   &  5.1 \\[0.05cm]
O8.5 V    &   18.8    &  6.5 \\[0.05cm]
O9 V       &   17.1    &  8.3  \\[0.05cm]
O9.5 V    &   15.6   &  10.4   \\[0.05cm]
B0 V       &    14.6   &  12.3   \\[0.05cm]
B0.5 V    &   13.2    &  15.8   \\[0.05cm]
B1 V       &    11.0   &  24.9   \\[0.05cm]
B2 V       &     8.6    &  46.1   \\[0.05cm]
B3 V       &    6.1     &  108.8   \\[0.05cm]
B4 V       &    5.1     &  170.2   \\[0.05cm]
B5 V       &    4.4     & 246.2   \\[0.05cm]
\hline
\end{tabular}

\begin{minipage}{80mm}
\footnotetext{

\noindent
Note -- The spectral classes and masses are taken from Table 1.1 of
Williams, Stephen J., "Optical Spectroscopy of Massive Binary Stars"
(2011), Physics and Astronomy Dissertations, Paper 49,
http://digitalarchive.gsu.edu/phy\_astr\_diss/49, while their
main-sequence lifetimes are estimated using equation (1) reported in Section \ref{yngPULSARS}.

    }
    
 \end{minipage}

\end{table}

\clearpage   



\appendix

\section{NOTES ON SPECIFIC SOURCES}\label{appendixA}

\subsection{RX J0516.0-6916 (source ID  \# 12)}\label{newspec}
   The optical counterpart of this source has been spectroscopically classified by \citet{Cowley et al. 1997} as approximately a B1-type star. However, RX J0516.0-6916 has been listed in the literature as a candidate HMXB (i.e. HMXB? in Tables \typeout{\ref{tableCensus}}1 and \typeout{\ref{finalclass}}5). Based on its $B$ and $V$ magnitudes and its position on the $V,B-V$ color-magnitude diagram shown in Fig. \ref{VBVall}, we have identified here that this source may well be a Be-type star that when Cowley \etal observed it in 1997 did not show any emission (possibly observed during a disk-loss episode). We thus classify this source as a candidate Be-XRB.

\subsection{RX J0520.5-6932 (source ID  \# 14)}\label{appendixTOOblue}
None of the optical matches listed within 5\arcsec\, from the X-ray source position (in Table \typeout{\ref{tableOpt}}2) has been presented in the literature. Although most of the reported coordinates for the optical match of RX J0520.5-6932 are compatible with the identified matches in this work, their photometric magnitudes are totally different, especially the $B-V$ color. In particular, RX J0520.5-6932 has one bright, very blue match with $V=16.004\pm0.471$ mag, and $B-V=-1.692\pm0.479$ mag. We note that there are very few sources spectroscopically classified with these $B$ and $V$ magnitudes (also note that this very blue match is outside the $B-V$ color range shown in Fig. \ref{VBVall}). Although the identified counterpart has been spectroscopically classified as an O8-9e star (\citealt{1994PASP..106..843S}, \citealt{2001MNRAS.324..623C}), thus bright, blue, magnitudes are expected, perhaps the deblending photometric algorithm used in the MCPS survey resulted in erroneous $B$ and/or $V$ magnitudes in this case, given the very contradictory values present in the literature (also, in the MCPS catalog in particular, the errors in the $B$ and $V$ magnitudes for this object is of the order of 0.5 mag). As noted in Table \typeout{\ref{tableBright}}3, there is a blue source (ID 119691 in the catalog of \citealt{2002ApJS..141...81M}) with $V=14.12$ mag, and $B-V=0.02$ mag at 1.34\arcsec\, from the X-ray source position. This is also source 66519 in LMC field (SC) 6 from the OGLE-II survey \citep{2000AcA....50..307U} with $V=14.461$ mag and $B-V=-0.054$ mag. In the latest OGLE-III catalog \citep{2012AcA....62..247U} there are 2 sources in Field 100.1 both within only $\sim$1\arcsec, of the X-ray position: source 5424 at 1.0\arcsec\, with $I=14.362$ mag, and $V=14.449$ mag, and source 138 at 1.1\arcsec\, with $I=14.358$ mag, and $V=14.446$ mag. Thus, most probably the MCPS, but also prior works, have blended together multiple point sources, resulting in erroneous $B$ and/or $V$ magnitudes. This source has been classified in the past as a Be-XRB and given the uncertainty of its photometric properties we decided not to revise its classification.

\subsection{RX J0523.2-7004 (source ID  \# 15)}\label{equalprob}
This source has two equally probable matches (based on their similar chance coincidence probabilities; Fig. \ref{figCCtests}), thus we cannot propose either as its optical counterpart. They are located at 15.88\arcsec\, and 3.62\arcsec\, from the X-ray source position (90\% error radius 9.9\arcsec), and they are both bright ($V\sim15$ mag; the nearest one is redder by $\sim$0.5 mag). Based on their photometric properties, we classify this source as a candidate Be-XRB.

\subsection{RX J0527.1-7005 (source ID  \# 17)}\label{aboveRGB}
Based on the brightness of the identified matches for source RX J0527.1-7005, which has a positional uncertainty of 11.9\arcsec, we propose as the most likely counterpart the source at 4.91\arcsec\, with $V=15.308\pm0.050$ mag and $B-V=1.208\pm0.066$ mag (see Table \typeout{\ref{tableOpt}}2). This optical match is perhaps too red for an HMXB (as previously classified; \citealt{1999A&AS..139..277H}), although few Be and B[e] III-V sources have been found by \citet{2012MNRAS.425..355R} to have even redder $B-V$ colors (see Fig. \ref{VBVall}). Nevertheless, following a conservative approach we opted here to exclude this source from the list of HMXBs in the LMC. We also note that outside the 1$\sigma$ search radius (i.e. at 13.23\arcsec) there is another optical match with $V=18.626\pm0.074$ mag and $B-V=0.156\pm0.090$ mag (from Table \typeout{\ref{tableMatchesOut}}4), which falls at the faint end of the Be-stars locus in the $V,B-V$ color-magnitude diagram (see  Fig. \ref{VBVall}).

\subsection{RX J0529.4-6952 (source ID  \# 19)}\label{equalprob2}
Within its ${\rm r_{90}}$ ($\sim$9.8\arcsec), this source has two equally probable matches (based on their similar chance coincidence probabilities; Fig. \ref{figCCtests}). We propose as the most likely counterpart the brightest one, located at 10.08\arcsec\, from the X-ray source position. This match falls well within the locus of Be stars, thus we classify this source as a candidate Be-XRB.

\subsection{RX J0530.7-6606 (source ID  \# 23)}\label{equallyBRIGHTcparts} 
This X-ray source has a positional uncertainty of 11.4\arcsec\, (see Table \typeout{\ref{tableCensus}}1), and we have identified two optical sources (in Table \typeout{\ref{tableMatchesOut}}4) as equally likely matches due their similar chance coincidence probabilities (Fig. \ref{figCCtests}), photometric properties (i.e. position on the $V,B-V$ color-magnitude diagram) and offsets from the X-ray position. Nevertheless, we choose as the most probable match the one that falls nearer to the locus of Be stars in Fig. \ref{VBVall} (i.e. the one with an offset of 21.72\arcsec, while the one at 22.13\arcsec\, is too blue and it is presented in this color-magnitude diagram as a less probable match) and we classify this source as a candidate Be-XRB.

\subsection{RX J0531.2-6607 (source ID  \# 24) and XMMU J053118.2-660730 (source ID  \# 26): Two different X-ray sources}\label{diffXRAYsrcs}
We treat sources with IDs \# 24 and \# 26, which have been both
detected by \xmm (for more details see Table \typeout{\ref{tableCensus}}1), as
two different sources since they are $\sim$38\arcsec\, away from each
other. For RX J0531.2-6607 we find an OB counterpart within
5.71\arcsec\, from the X-ray position, in agreement with previous
studies (\citealt{2002A&A...385..517N} assigned a B0.7Ve spectral type
to this counterpart, so we retain its classification as a Be-XRB
system). Similarly, we find a bright OB star within 3.21\arcsec\, from
source with ID \# 26, which has $V=16.873\pm0.023$ mag and $B-V=-0.069\pm0.031$ mag (see Table \typeout{\ref{tableOpt}}2). Based on this new identification, we classify XMMU J053118.2-660730 as a candidate Be-XRB (\citealt{2005A&A...431..597S} list this as an uncertain HMXB).

\subsection{XMMU J053115.4-705350 (source ID  \# 25): a new candidate supergiant XRB}\label{newSGXRB?}
This source has been classified by \citet{2005A&A...431..597S} as a likely HMXB candidate with $R=13.6$ mag and $V-R=0.1$ mag. Here, we identify its counterpart as the MCPS source with $V=14.051\pm0.025$ mag and $B-V=-0.042\pm0.043$ mag located at 1.64\arcsec\, from the X-ray position (Table \typeout{\ref{tableOpt}}2; we note that this is a unique match within a search radius of 5\arcsec). \citet{2012MNRAS.425..355R} assign a B0 Ie spectral type to this source (object [RP2006]444 in their catalog) and list it as an HMXB. We go one step further and classify this source as a new candidate SG-XRB given its luminosity class. Although the optical counterpart is a B-type emission-line object (i.e. a Be-type star), thus XMMU J053115.4-705350 could be classified as a Be-XRB, on the other hand, being a supergiant contradicts the definition of Be stars, which explicitly excludes luminosity class I-II objects.

\subsection{RX J0532.4-6535 (source ID  \# 29) and RX J0535.8-6530  (source ID  \# 35)}\label{RGBcparts}
These are two sources for which all identified brighter than $V\sim20$
mag optical matches have similarly large chance coincidence
probabilities (Fig. \ref{figCCtests}). Following the same conservative
approach as for source RX J0527.1-7005 (source ID \# 17), we chose to
exclude both of them from the list of HMXBs in the LMC. Nevertheless, we note that these sources have other faint matches that lie along the main sequence (see Tables \typeout{\ref{tableBright}}3 and \typeout{\ref{tableMatchesOut}}4), and in particular the ones for source RX J0535.8-6530 lie on the faint end of the Be-star locus (see Fig. \ref{VBVall}).

Moreover, we note that RX J0532.4-6535 is erroneously listed in \citet{2005A&A...442.1135L} (but not in their Tables 1 and 2) as a supergiant system, citing for this classification \citet{1999A&A...344..521H}. However, \citet{1999A&A...344..521H} classify RX J0532.4-6535 as an uncertain HMXB and based on its low variability they suggest that this source is an X Per type HMXB detected during the persistent quiescent state. Even if we look for an OB counterpart out to $\sim35$\arcsec\, (about twice the ${\rm r_{90}}$ listed above) from the X-ray position, we find none.

\subsection{RX J0532.5-6551 (source ID  \# 30): one of the few identified LMC supergiant XRBs}\label{appendixSGXRB}
RX J0532.5-6551 has been detected by \citet{1995A&A...303L..49H} (this is also ROSAT PSPC source \# 184 with (R.A., Dec.)${\rm _{J2000.0}}$ = (05:32:32.1,-65:51:42) and $r_{90}=3.3$\arcsec\, in \citealt{1999A&A...344..521H}). Based on the optical magnitude of the associated source Sk-65 66, \citet{1995A&A...303L..49H} suggested that this may be a supergiant system. Later, \citet{2002A&A...385..517N} classified source Sk-65 66 as a B0 II type star and confirmed that RX J0532.5-6551 was the first wind-fed SG-XRB in the LMC. Independently, \citet{2001PASP..113.1130J} derived a B0.5 II spectral type for Sk-65 66 (with (R.A., Dec.)${\rm _{J2000.0}}$ = (05:32:32.6,-65:51:42) and $B=12.9$\,mag, $V=13.1$\,mag). In the MCPS catalog, this counterpart has $B=12.474\pm0.138$ mag and $V=13.048\pm0.076$ mag and it has a 3.85\arcsec\, offset from the X-ray position, thus we adopt the classification of this system as a SG-XRB.

\subsection{LMC X-4 (source ID  \# 31)}\label{appendixUband}
The counterpart of this well-studied X-ray source presented here is the same as the one reported in the literature, i.e. a very bright blue object with $B\simeq V\sim 14$ mag (the MCPS source listed in Table \typeout{\ref{tableOpt}}2 has $B-V=-0.227\pm0.041$ mag). The striking difference between this study and earlier works though is the reported $U$ magnitude. The MCPS catalog lists this source with a $U$ magnitude equal to 15.449$\pm$0.070 mag, resulting in $U-B$ color of 1.509$\pm$0.076 mag. This is significantly redder than any previous reports, although significant variability has also been reported (e.g., \citealt{1976IAUC.3017....3P}, \citealt{1977IAUC.3039....3B}, \citealt{1977A&A....59L...9C}, \citealt{1984A&A...140..251I}). Given the spectral type of the donor star of this RLOF-fed X-ray pulsar (O8 III; \citealt{2002A&A...385..517N}), we classify this system as an HMXB.

\subsection{RX J0541.4-6936 (source ID  \# 39)}\label{stillSGXRB}
\citet{2005A&A...442.1135L} list this source as a B2 type SG-XRB
following \citet{2000A&AS..143..391S}. However,
\citet{2000A&AS..143..391S} only classify RX J0541.4-6936 (ROSAT HRI
source 328 with (R.A., Dec.)${\rm _{J2000.0}}$ =
(05:41:22.2,-69:36:29) and ${\rm r_{90}=6.6}$\arcsec) as a candidate
HMXB and associate it with a source from Sanduleak 1970. This object
is source Sk -69 271 ((R.A., Dec.)${\rm _{J2000.0}}$ =
(05:41:20.14,-69:36:23.0)) $\equiv$ CPD-69 500 at 12.33\arcsec\, from
the Xray source position, which has photometric parameters typical of OB supergiants ($V=11.790\pm0.051$ mag and $B-V=0.000\pm0.106$ mag from the MCPS catalog). Recently, \citet{2009AJ....138..510F} classified source Sk -69 271 as a B4 III-I type star (with (R.A., Dec.)${\rm _{J2000.0}}$ = (05:41:20.04,-69:36:22.4) and $V=11.80$\,mag). RX J0541.4-6936 is also listed in the XMM-Newton Serendipitous Source Catalogue 2XMMi-DR3 as 2XMM J054123.0-693633 (source ID 53010). 

\begin{figure}
\centering
\includegraphics[height=8cm]{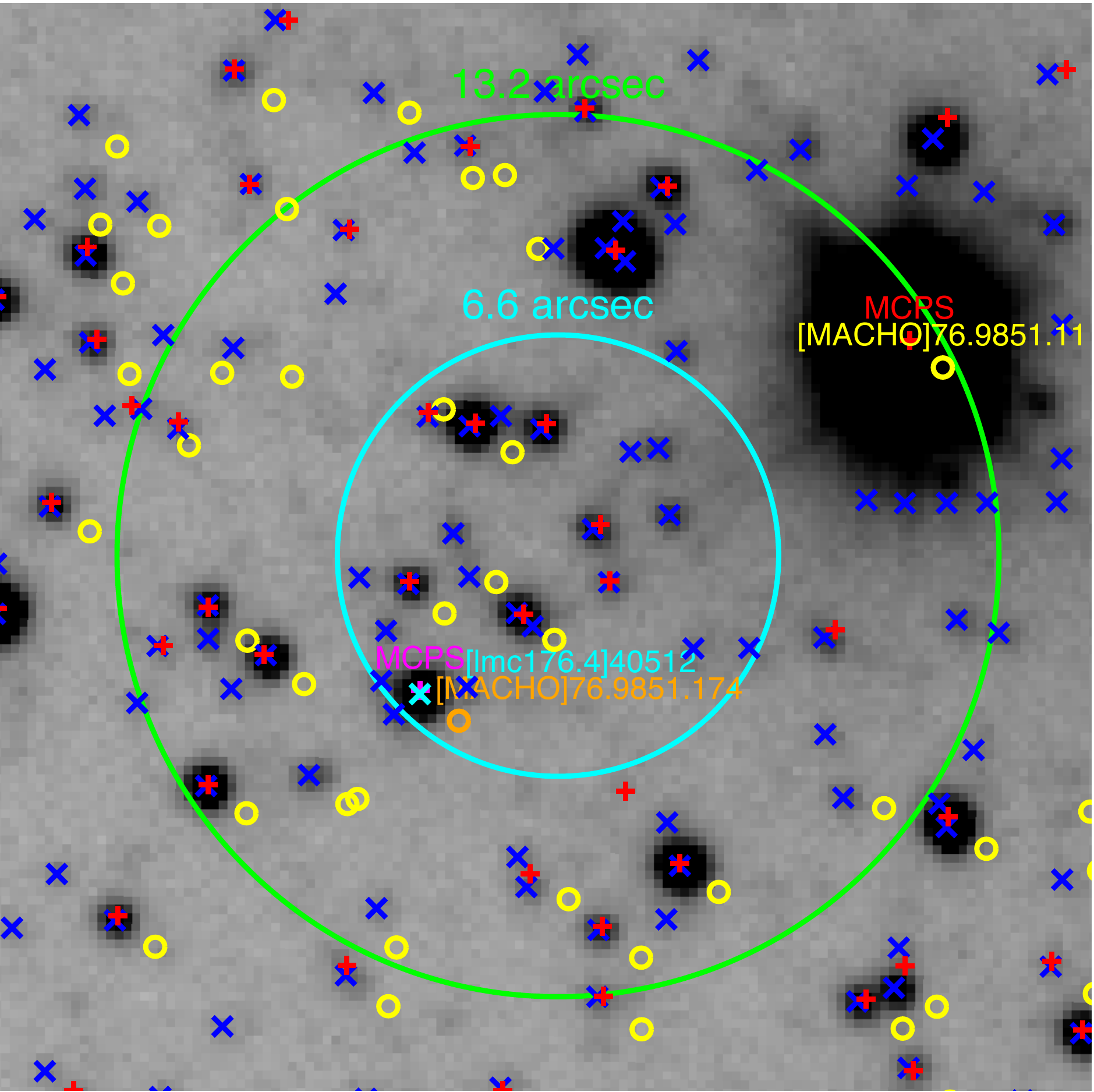}
\caption{Finding chart of RX J0541.4-6936 (source ID \# 39; $\sim30$\arcsec\, in each side with North up and East on the left) from the OGLE-III $V$-band image (Field ID 176.4; the WCS was tied to 2MASS). OGLE-III, MCPS and MACHO sources are shown with blue "x" points, red crosses and yellow circle points, respectively. The 1$\sigma$ (6.6\arcsec) and 2$\sigma$ (13.2\arcsec) positional uncertainties around source RX J0541.4-6936 are shown in cyan and green circles, respectively. In the literature, the proposed counterpart of this source is MACHO object 76.9851.11 (a supergiant star). In this work, we have also identified a closer Be-star like match: its OGLE-III position is shown with a cyan "x" point (marked by field and source ID), while this is also MACHO object 76.9851.174 (shown with an orange circle point). For a detailed discussion see Appendix \ref{stillSGXRB}.}\label{FindingChart}
\end{figure}

Although there are several other sources within its error circle (Fig. \ref{FindingChart}), some of which fall in the locus of B-type stars, the chance coincidence probability for the initially proposed counterpart is much lower than that for the other candidate sources, and therefore we adopt the previously reported classification.

\subsection{RX J0541.5-6833 $=$ RX J0541.6-6832 (source ID  \# 41)}\label{simplyHMXB}
This source has photometric properties compatible with OB III-V stars (see Fig. \ref{VBVall}). In addition, its most likely counterpart has been classified as a B0 III star \citep{2002ApJS..141...81M}, thus we only classify this source as an HMXB and not as a (candidate) Be-XRB.

\subsection{RX J0543.9-6539 (source ID \# 42)}\label{RedClump}
This source has a positional uncertainty (${\rm r_{90}}$) of
8.6\arcsec, and within a search radius of 10\arcsec\, there are 3
matches from the MCPS catalog. Based on their very large chance
coincidence probabilities (Fig. \ref{figCCtests}), and similarly to
X-ray sources with IDs \#23, \#29, and \#35, we exclude this source from the list of HMXBs in the LMC.

\subsection{RX J0546.8-6851 (source ID \# 45)}\label{NOTtypicalclass}
The adopted optical counterpart of this source is located at
56.67\arcsec\, from the X-ray source position (${\rm
  r_{90}}\sim48.7$\arcsec; Table \typeout{\ref{tableCensus}}1). It is well within
the locus of Be stars, thus we classify this source as a candidate
Be-XRB system. We also note there are 3 additional fainter matches located at 5.93\arcsec, 19.41\arcsec, and 51.82\arcsec\, from the X-ray source position.


\bsp	
\label{lastpage}
\end{document}